\documentclass[10pt]{article}
\usepackage{geometry}

\title{Massless scalar free Field in 1+1 dimensions II: Net Cohomology and Completeness of Superselection Sectors}
\author{Fabio Ciolli \\
 \small{Dipartimento di Matematica, Universit\`a di Roma ``Tor Vergata''}\\
    \small{Via della Ricerca Scientifica I-00133, Roma,  Italy}  \\
          \small{\texttt{ciolli@mat.uniroma2.it}}}

\usepackage{latexsym, amssymb, amsmath}
\def\1{{\mathbf 1}}
\def\0{{\mathbf 0}}
\def\Zb{{\cZ_b}}
\def\dG{{\widehat{\cG}}}
\def\dS{{\partial \cS}}
\def\duS{{\partial^{-1} \cS}}
\def\duzS{{\partial^{-1}_0 \cS}}
\def\duqS{{\partial^{-1}_q \cS}}%
\def\dulS{{\partial^{-1}_l\cS}}
\def\durS{{\partial^{-1}_r\cS}}
\def\fzu{{f_0\oplus f_1 }}
\def\gzu{{g_0 \oplus g_1 }}
\def\cA{{\cal A}}
\def\cB{{\cal B}}
\def\cC{{\cal C}}
\def\cD{{\cal D}}
\def\cE{{\cal E}}
\def\cF{{\cal F}}
\def\cG{{\cal G}}
\def\cH{{\cal H}}
\def\cI{{\cal I}}
\def\cJ{{\cal J}}
\def\cK{{\cal K}}
\def\cL{{\cal L}}
\def\cM{{\cal M}}
\def\cN{{\cal N}}

\def\cP{{\cal P}}
\def\cQ{{\cal Q}}
\def\cR{{\cal R}}
\def\cS{{\cal S}}
\def\cT{{\cal T}}
\def\cU{{\cal U}}

\def\cW{{\cal W}}

\def\cZ{{\cal Z}}

\def\bC{{\mathbb C}}

\def\bN{{\mathbb N}}

\def\bR{{\mathbb R}}

\def\bZ{{\mathbb Z}}

\def\bbN{{\mathbf{N}}}%
\def\a{\alpha}
\def\b{\beta}
\def\g{\gamma}        											\def\G{\Gamma}
\def\d{\delta}        											\def\D{\Delta}
\def\ep{\varepsilon}

\def\e{\eta} 
          
\def\i{\iota}

\def\l{\lambda}       \def\L{\Lambda}
\def\m{\mu}

\def\x{\xi}
\def\p{\pi} 
\def\r{\rho}
\def\s{\sigma}
    
\def\S{\Sigma}
\def\t{\tau}
			
       \def\F{\Phi}

\def\c{\chi}
\def\o{\omega}        \def\O{\Omega}

\def\sF{{\mathsf F}}

\def\to{\rightarrow}

\def\spam{{\mathrm{spam\,}}}
\def\supp{{\mathrm{supp\,}}}
\def\loc{{\mathrm{loc\,}}}
\def\Loc{{\mathrm{Loc\,}}}
\def\ad{{\mathrm{ad\,}}}
\def\repb{{\mathrm{Rep^\bot\,}}}
\def\open{{\mathrm{Open\,}}}
\def\sub{{\mathrm{Sub\,}}}
\def\rest{{\upharpoonright}}

\def\nbot{\vbox {\hbox to 10pt {$\not$\hskip -1.3pt $\bot$\hfil}}}
\newtheorem{thm}{Theorem}[section]

\newtheorem{dfn}[thm]{Definition}
\newtheorem{lemma}[thm]{Lemma}
\newtheorem{prop}[thm]{Proposition}
\newtheorem{cor}[thm]{Corollary}

\newtheorem{rem}[thm]{Remark}
\newtheorem{axi}[thm]{Axiom}
\newcommand{\baxi}{\begin{axi}}
\newcommand{\eaxi}{\end{axi}}
\newcommand{\bassum}{\begin{assum}}
\newcommand{\eassum}{\end{assum}}
\newcommand{\bdfn}{\begin{dfn}}
\newcommand{\edfn}{\end{dfn}}
\newcommand{\blemma}{\begin{lemma}}
\newcommand{\elemma}{\end{lemma}}
\newcommand{\bprop}{\begin{prop}}
\newcommand{\eprop}{\end{prop}}
\newcommand{\bthm}{\begin{thm}}
\newcommand{\ethm}{\end{thm}}
\newcommand{\bcor}{\begin{cor}}
\newcommand{\ecor}{\end{cor}}
\newcommand{\prf}{\noindent{\it Proof. }}
\newcommand{\qed}{\ \hfill $\square$ \\\\}
\newcommand{\bconj}{\begin{conjecture}}
\newcommand{\econj}{\end{conjecture}}
\newcommand{\brem}{\begin{rem}}
\newcommand{\erem}{\end{rem}}
\newcommand{\beq}{\begin{equation}}
\newcommand{\eeq}{\end{equation}}
\newcommand{\beqn}{\begin{equation*}}
\newcommand{\eeqn}{\end{equation*}}
\newcommand{\barr}{\begin{array}}
\newcommand{\earr}{\end{array}}
\newcommand{\beqa}{\begin{eqnarray}}
\newcommand{\eeqa}{\end{eqnarray}}
\newcommand{\beqan}{\begin{eqnarray*}}
\newcommand{\eeqan}{\end{eqnarray*}}

\newcommand{\bdes}{\begin{description}}
\newcommand{\edes}{\end{description}}
\newcommand{\bitem}{\begin{itemize}}
\newcommand{\eitem}{\end{itemize}}
\numberwithin{equation}{section}
\begin{document}
\maketitle\noindent

\begin{abstract}
As an application of Roberts' cohomology (net cohomology), we prove the completeness of the DHR sectors of the local observables of the model in the title, detailed in \cite{fabio3}. 
This result is achieved via the triviality of the net $1$-cohomology, with values in the local fields, enhancing 
the Roberts' methods to the case of anyonic Weyl nets, not satisfying the split property. 
We take advantage of using different causal index sets for the nets involved. 
The presence of anyonic commutation relations is treated introducing the notion of nets graded by a generic group, and the related properties of graded locality and graded duality. 
As a further result, we obtain the description of twisted and untwisted sectors of the model as two symmetric subcategories of a $W^*$-braided category, whose objects are the same as the dual category of the compact Abelian group of the gauge symmetry.
The work also furnish some hints for the analysis of the sector structure of generic models on different spacetimes.
\end{abstract}
\tableofcontents
%
%
\section{Introduction}
The purpose of this paper is twofold: complete the superselection sectors analysis of the massless scalar free field in 1+1 dimensions begun in \cite{fabio3}, and give a concrete example of Roberts' net cohomology at work on a low dimensional spacetime.

The physical motivation for studying this conformal model can be traced back to the original paper of Streater and Wilde \cite{sw70} and to the solution of its simpler counterpart in $1+3$ dimensions given in \cite{bdlr92}.

\smallskip
The results obtained in \cite{fabio3} shown that the model fits the main lines of the celebrated higher dimensional DHR paradigm \cite{dhr69I, dhr69II}. In fact, from a twisted crossed product underlying the Weyl algebras, we proved that given the observables net of the model $\cA_\cI$ on the index set $\cI$ of the bounded open intervals of the  time zero real line, in a regular vacuum representation on a separable (physical) Hilbert space $\cH_a$, there exist a (putative) fields net $\cF_\cI$, in a non-regular representation, and a compact Abelian group $\cG$ such that $\cA_\cI=\cF_\cI^\cG\rest\cH_a$, the global gauge symmetry, being isomorphic to the Bohr compactification of the additive group $\bR^2$.
 
The non-regular representation of the fields net turns out to be given on a (unphysical) non-separable Hilbert space that carries two families of uncountably many charged sectors of the observables, well distinguished in the time zero formulation, as twisted (those of solitonic origin) and untwisted. 

\smallskip
The above results also focus the relations between the model and the more recent achievements of the Conformal Algebraic Quantum Field Theory, both for the rational (i.e.\ finite number of sectors) and irrational (infinite sectors) case, see for example the papers of Kawahigashi, Longo, M\"uger and Xu \cite{{klm01}, {mug01}, kl04,lx04}.\smallskip

\smallskip
The present article addresses the question  of the completeness of the DHR sectors of the observables net, that is if they are \emph{all} implemented by the Weyl operators in $\cF_\cI$. 
A positive answer is given in terms of the triviality of $Z^1(\cF_\cI)$, the category of the $1$-cohomology of $\cI$ with values in the net $\cF_\cI$. 
The proof bases on a major outcome of the Roberts' net cohomology: premising that the category of the DHR representations of the net $\cA_\cI$ is  equivalent to (a full subcategory of) $Z^1(\cA_\cI)$, if $\cA_\cI=\cF_\cI^\cG$ and $Z^1(\cF_\cI)$ is trivial, then no nontrivial sectors exist for the net $\cA_\cI$ but the ones carried by the net $\cF_\cI$, see \cite{rob90, rob04} for details.
This cohomological procedure, albeit with model specific peculiarities, is a well established tool in the theory of DHR sectors; for example in \cite{cdr01, cc05} it turns out to have central relevance for the description of the superselection of subsystems.

\smallskip
It is useful to recall that the net cohomology dates back to the pioneering paper by Roberts \cite{rob76}. From the early beginnings as in recent advances (see for example Brunetti, Roberts and Ruzzi in \cite{rob04, ruz05, rr06, br08} and the explicit massive examples of  \cite{bfm}) it responds to the challenge of extending the operator algebraic approach to superselection rules in different physical circumstances: curved spacetimes with nontrivial topologies; topological or electromagnetic charges.\smallskip

\smallskip
Another relevant issue of the present paper is the treatment of field nets on the line, or 1+1-dimensional spacetime, satisfying anionic commutativity rules. Similarly to the standard notion of the Bose/Fermi nets in higher dimensional spacetime, we introduce nets graded by a (a priori) generic discrete group, and give definitions of graded locality and graded duality, generalizing the $\bZ_2$-twisted analogous of the Bose/Fermi case. This allows to describe the main properties of the observable nets in non vacuum representations, and gives a net-oriented version of the categorial treatment of graded categories of M\"uger to the conformal orbifold theories, see \cite{mug05}.

\smallskip
The main results and other byproducts of this paper are presented as follow.

In Section \ref{s: Prop of net} we review the abstract definition and properties of a net on a causal index set. In particular, we focus on the problem of extending a net to a larger index set. Then we analyze the characteristic of the useful index set on the real line and introduce the abstract notion of a net with generic grading group, acting exclusively on one of the two disconnected components of the complements of the elements of the index set. Here we also underline the relation with the geometric spatial inversion of the spacetime.

We then define the graded structural properties of the net, and give the major example of grading, the one obtained by represented crossed product of the grading group. 
In this construction only translation covariance is used and the M\"{o}bius group covariance plays almost no role.  
We conclude describing the concerns for the gauge symmetry group and the twisted and untwisted superselection sectors of the involved nets, and discussing the closeness to the categorial approach of \cite{mug05}. 

\smallskip
In Section  \ref{s: net coom model rep} we approach the question of the triviality of net $1$-cohomology. We start from the standard method established by Roberts, that requires: 
1) the (quasi-)split property for the net, ensuring the independence of the algebras of spacelike separated regions and the existence of a useful conditional expectation; and 
2) a proper cohomological condition, requiring that the intersection of the algebras of all paths of regions having the same two end regions equals the algebra generated by the two end regions, see Subsection \ref{ss:Nets} for details.

As a first step we generalize the triviality result for a net (Bosonic, $\bZ_2$-graded or with generic grading) when the (quasi-)split property is missing but the conditional expectation exists; then, for nets defined from a Weyl algebras even if non-separably represented, we show the existence of such a conditional expectation without the help of the Tomiyama slice map of the split case.    

\smallskip
In the first part of Section \ref{s:completenes SW} we discuss the structural properties of the nets of the model in the title, both for the index set $\cI$ and for its enlargement $\cD:=\cI\cup\cI_2$, where $\cI_2$ is the set of couples of open bounded intervals with disjoint closures.
We obtain the strong condition that all the superselection data of the observables are just encoded in the nets $\cA$ on $\cI$ and $\cD$.
For the net $\cF$, we easily show $\bR$-graded locality and $\bR$-graded duality on $\cI$ and $\cD$, and the cited cohomological condition on $\cD$.
Finally, we show the triviality of $Z^1(\cF)$ on $\cD$ and, by Roberts' general theorems about the change of 1-cohomology when the enlarging or restricting the index set (see \cite{rob04}), also on $\cI$. 

\smallskip
We hence answer the completeness question and describe the category of the sectors of $\cA$ as a braided tensor $W^*$-category, equivalent as a tensor category (i.e.\ disregarding the symmetry) to the dual of the gauge group $\cG$, containing the twisted and untwisted sectors as two symmetric subcategories. 
\section{Properties of Nets on  Causal Index sets}\label{s: Prop of net}
In this section we review the standard definition of nets in Algebraic Quantum Field Theory (AQFT), discussing in detail their main properties and the dependence on their index set. 

\smallskip
In particular, we focus on the procedure to obtain nets of von Neumann algebras from nets of symplectic spaces, in order to grasp the relevant features for the discussion of  superselection sector theory of the concrete models.

Then we specialize to the 1+1-dimensional Minkowski spacetime and to the formulation of a field theory on the Cauchy surface of its time zero real line.
Here the notion of a net with grading group is introduced, first in all generality and then the main example of represented crossed product. This allows to manage with the anyonic commutations rules of fields, the basic properties of locality and duality of the relative nets and the concerns with the DHR superselection sectors determined on their subnets by twisted automorphisms.   
\subsection{Nets on index sets}\label{ss:Nets}
We recall some basic notions associated to a partially ordered set  $\cP:=(\cP, \subseteq)$, a poset for short, used to formulate a cohomological and a net theory on $\cP$.

\smallskip
For given a poset $\cP$, its \emph{simplicial set} $\S_{*}(\cP)$ is defined by the collection of all the sets of \emph{standard $n$-simplices} $\S_{n}(\cP)$ on $\cP$.
Recall that the \emph{standard $n-$simplex} is defined by 
\beqn
\D_n:=\left\{ (\l_0,\dots\l_n)\in\bR^{n+1}\,:\,\l_0+\dots\l_n=1,\, \l_i \in [0,1] \right\}\,,\quad n\in \bN.
\eeqn
The standard $n-$simplex $\D_n$ itself  is a poset respect to the inclusion map of subsimplices and the set of singular $n$-simplices $\S_{n}(\cP)$ of $\cP$ is formed by the inclusion preserving maps $f:\D_n\to \cP$.

To be explicit, we have $\S_0(\cP):=\{a:=(|a|, a) \in \cP\times \cP: a\subseteq |a|\}$, i.e.\ any $0$-simplex is just an element in $\cP$ and an element $|a|\in \cP$ (that may be thought equal to $a$) called the \emph{support} of $a$; the set $\S_1(\cP):=\{b:=(|b|, \partial_0 b,\partial _1 b) \in \cP\times \S_0(\cP)\times \S_0(\cP): \partial_0 b\,,\, \partial _1 b\subseteq |b|\}$,  a $1$-simplex $b$ is thus a triple of element in $\cP$, where $|b|\in \cP$ is called the support of $b$. Similarly,  
$\S_2(\cP):=\{c:=(|c|, \partial_0 c,\partial _1 c, \partial _2 c) \in \cP\times \S_1(\cP)\times \S_1(\cP)\times \S_1(\cP): \partial_0 c\,,\, \partial _1 c\,,\,\partial _2 c\subseteq |c|\}$.

\smallskip
Recall that a poset $\cP$ is \emph{directed}  if for given $o_1,\,o_2 \in \cP$ there exits $o\in\cP$ such that $o_1,\,o_2\subseteq o$.
\emph{Connectedness} of $\cP$ means, for given $o_0, o_1\in \S_0(\cP)$, the existence of a path $p$, i.e.\ a $n$-uple of elements $b_1,b_2,\dots, b_n$ in $\S_1(\cP)$ such that $\partial_1 p:=\partial_1 b_1 =o_0,\, \partial_0 p:=\partial_0 b_n =o_1$ and $\partial_1 b_{i+1} =\partial_0 b_i$, with $i=1,\dots,n$. 
A directed poset $\cP$ is a trivial example of a connected poset. 

\smallskip
For two given posets $\cP\subseteq \cL$, the poset $\cP$ is said to be \emph{cofinal} in $\cL$ if for every $a\in \cL$ there exists a $o\in \cP$ such that $a\subseteq o$. 

\smallskip
The following result can be proved easily
\blemma\label{l:cofinal}
Let $\cP\subseteq\cL$ an inclusion of posets with $\cP$ cofinal in $\cL$. If $\cL$ is connected (respectively directed) then $\cP$ is connected (respectively directed). 
\elemma
The poset $\cL$ is said to be \emph{generated} by the poset $\cP\subseteq \cL$ if for every $a\in \cL$ there exist $o_i\in \cP$ such that $a=\bigvee_i o_i$, with $\bigvee$ the supremum on the index set $\cL$. 
For example, if $\cP$ is a base of neighborhoods for a topological space $M$, by definition it generates $\open(M)$, the set of non-void, open subsets of $M$ ordered by inclusion.

\smallskip
A poset with \emph{disjointness relation} is a triple $(\cP,\subseteq,\bot)$ where $(\cP,\subseteq)$ is a partially poset and $\bot$ is a binary relation  that, for $o_1,\,o_2,\,o_3\in \cP$, satisfies 
\bdes
\item[a)]   $o_1\bot o_2 \Rightarrow o_2\bot o_1$, symmetry;
\item[b)]   $o_1\subseteq o_2$ and  $o_2\bot o_3 \Rightarrow o_1\bot o_3$, i.e.\ $o_1^\bot\supseteq o_2^\bot$, notation as in equation (\ref{e: disjunto}), order-reversing; 
\item[c)]   given $o_1\in\cP$ there exists $o_2\in \cP$ such that $o_1 \bot o_2$, existence of a disjoint element.
\edes
We briefly denote such a triple by $\cP$ and call it a \emph{causal index set}, as the disjointness relation $\bot$ assumes a causal interpretation when referring to an index set on a spacetime.

\smallskip
We pass now to recall the definition and the properties of a net on a index set.
A \emph{net $\bbN$ on index set $\cP$} is defined by an inclusion preserving (i.e.\ isotonic) map

\bigskip
\noindent \textbf{1.} {\sc Isotony.} 
\quad $\bbN: o \longmapsto \bbN (o),\qquad o \in \cP$.\\

\noindent 
If necessary, when different index sets are in use, we shall denote the net $\bbN$ on index set $\cP$ by $\bbN_\cP$, and $\cP$ is referred to as \emph{the index set} of the net $\bbN_\cP$. Moreover, an index set is always assumed to be directed, if not otherwise stated.
The image of a net $\bbN$ is in a generic category, also furnished with inclusion order and a disjointness relation. The reference examples are the following, where the arrows are the inclusion map of the objects sets: 
\bitem
\item
$(\sub(V),\subseteq,\bot_{\s_V})$ where $(V,\s_V)$ is a symplectic space, and $\sub(V)$ is the set of its symplectic subspaces ordered by inclusion. $\bot_{\s_V}$ is the disjointness relation on $\sub(V)$, defined by $V_1\bot_{\s_V}V_2$ iff $\s_V(v_1,v_2)=0$, for all $v_1\in V_1,\,v_2\in V_2$. 
\item
$(\sub(\cW(V,\s_V)),\subseteq,\bot_{\s_V})$ the functorially defined category of Weyl subalgebras of $\cW(V,\s_V)$ from the preceding example, where $\bot_{\s_V}$ means commuting Weyl subalgebras.
\item
$(\sub(\cR),\subseteq,\bot)$ for given a von Neumann algebra $\cR$ on the Hilbert space $\cH$, the set of its von Neumann subalgebras, ordered by inclusion and disjointness defined by commutant, i.e.\ $\cM\bot\cN$ iff $\cM\subseteq\cN'$, for $\cM,\,\cN \in \sub(\cR)$. In particular we may take $\cR=\cB(\cH)$. An equivalent definition for a $C^*$-algebras and its subalgebras is possible.
\eitem
To give classical examples of nets and fix the notation, we recall the standard construction of Weyl algebras models in AQFT.
  
For a given spacetime $M$, we fix an index set $\cP\subseteq\open (M)$ with the causal disjointness relation deriving from the metric of $M$.
Then, a linear space $V$ of sufficiently regular test functions on $M$ is given, together with a symplectic form $\s_V$, that defines the disjoint relation $\bot_{\s_V}$ for elements of $V$.
For the elements of $V$, it have also to be possible to define a notion of \emph{localization} on the elements of $\cP$; for instance arising from their support as functions, or from the one of their derivatives. 
From these ingredients, a net $V_\cP$ with range category a subcategory of $(\sub(V),\subseteq,\bot_{\s_V})$ is hence defined by injectivity.
\footnote
{Note that the symplectic space net $V_\cP$ itself can also be used as index set, with (causal) disjointness relation $\bot_{\s_V}$, see \cite[Sec. 23]{rob04}.
}

The Weyl functor, recalled in \cite[Subsection 2.1]{fabio3}, preserves the net structure on $\cP$, so that a net of Weyl algebras  $\cW_\cP: o \longmapsto \cW(V(o), \s_V)$ is  functorially defined. Moreover, once a reference or defining representation $\p_n$ of the algebra $\cW$ has been fixed, for instance the vacuum representation, there exists a canonically associated net of von Neumann algebras  
\beq\label{e:nets von weyl}
 \cN_\cP :o\longmapsto \cN(o):= \p_n (\cW (V(o),\s))'', \qquad o \in \cP.
\eeq
More features about Weyl algebras nets, for example isomorphisms and twisted crossed products, are defined in Section 2 of \cite{fabio3}.

\smallskip
We now list the basic properties of nets. The second piece of general structure associated to a net is the existence of the upper bound for the set of all elements in the image category, i.e.\

\bigskip
\noindent \textbf{2.} {\sc Maximal Element.} 
For $\cP\subseteq \open(M)$ a directed index set, there exists  \emph{the maximal element} $\bbN^\cP(M):=\bigvee_{o\in \cP}\bbN_\cP(o)$.\\

\noindent
For a net of symplectic subspace $V_\cP$, the symbol $\bigvee$ has to be understood as the generated symplectic space and the maximal element is $(V^\cP(M), \s_V):=\bigvee_{o\in \cP} (V(o), \s_V)\subseteq (V, \s_V)$. 
For a net of von Neumann subalgebras $\cN_\cP$ in the representation $\p_n$, with $\cP\subseteq\open(M)$ the von Neumann algebra $\cN^\cP(M)$ equals 
the weak closure in $\cB(\cH_n)$ of the \emph{quasilocal C*-algebra $\cN$} of the net, indicated with the same symbol of the net as usual, i.e.\ it coincides with the commutant of the self-intertwiners of $\p_n$. In symbols  $\cR=(\p_n, \p_n)'\subseteq \cB(\cH_n)$ and $\cN^\cP(M)=\cR$.

\smallskip
By far the most relevant property of a net $\bbN_\cP$ is  the property of disjointness preserving, called \emph{locality} for physical reasons, and written as 

\bigskip
\noindent \textbf{3.} {\sc Locality.} 
\quad $\bbN(o_1)\,\bot\, \bbN(o_2)$, for $o_1,o_2\in \cP$ and $o_1 \bot o_2$.\\

\noindent
For example, the net of von Neumann algebras  $\cN_\cP$  defined in equation (\ref{e:nets von weyl}) turns out to be local if the underlying symplectic subspace net $V_\cP$ is. 

\smallskip
The property of \emph{additivity} for the net $\bbN_\cP$  is satisfied if for any element $o\in \cP$ and any cover of it by $o_i$ elements in $\cP$, i.e.\ such that $\bigvee_i o_i =o\in \cP$, we have

\bigskip
\noindent \textbf{4.} {\sc Additivity.}
\quad $\bbN(o)=\bigvee_{i} \bbN(o_i)$.\\

\noindent
Actually we are interested in \emph{extensions of nets on larger index sets}.
For given an index set $\cL$ such that $\cP\subseteq\cL$ and with compatible causal disjointness relation $\bot$, such an extension of a net $\bbN_\cP$ is a net $\bbN_{\cL}$ having the same range category, so that the following elementary restriction requirement is fulfilled
\beq\label{e: extension}
\bbN_{\cL} (o) \,=\,\bbN_\cP (o),\qquad o \in \cP.
\eeq
The extension mainly used is the one \emph{by additivity}, defined as follow
\beq\label{e:N(o)}
\bbN^\cP_{\cL}: a\longmapsto \bbN^\cP_{\cL} (a):=\bigvee\{\bbN_\cP(o_i),\,o_i\subseteq a,\,o_i\in\cP \},\qquad a\in \cL.
\eeq
Here the superscript $\cP$ means \emph{definition by additivity on the index set} $\cP$.
It is to observe, see \cite[Section 7]{rob04}, that if $\cP$ generates $\cL$, the set $\cP$ is cofinal in $\cL$ and the net $\bbN_\cP$ is additive, then $\bbN^\cP_{\cL}$ is a canonical extension of the net $\bbN_\cP$, in the sense that it turns out to be the minimal definable one and is the unique additive extension of $\bbN_\cP$ to a net on $\cL$, i.e.\ we have
\beq
\bbN^\cP_{\cL}(a)=\bigvee_i \bbN_\cP(o_i),\qquad a=\bigvee_i o_i,\quad o_i \in \cP,\quad a\in \cL.
\eeq 
Moreover, if $\cP$ is cofinal in $\cL$ and $\bbN_\cL$ is any extension of the net $\bbN_\cP$, then for the maximal elements always holds $\bbN^\cP(M)=\bbN^\cL(M)$.

\smallskip
We recall now (also with a slight generalization) the well known Haag duality property for a von Neumann algebra net on $\cP$. 

\smallskip
The $\cP$-\emph{dual net} of a local net $\bbN_\cP$ is defined as the (possibly larger) net $\bbN^d_\cP$, on the same index set $\cP$ and the same image category, by
\beq\label{e:dual net}
o \longmapsto \bbN^d_\cP (o):=\bigcap \{\bbN_\cP(o_i)',\, o_i\bot o,\, o_i\in \cP\}\cap\bbN^\cP(M),\qquad o \in \cP.
\eeq
Locality of the net $\bbN_\cP$ can also be expressed by the net relation as $\bbN_\cP\subset \bbN_\cP^d$ so that the net $\bbN_\cP$ is said to satisfies \emph{$\cP$-duality} iff 

\bigskip
\noindent \textbf{5.} {\sc Duality.}
\quad $\bbN_\cP\,=\,\bbN_\cP^d$.\\

\noindent
Notice that for nets of von Neumann or Weyl algebras we give the definition of $\cP$-duality in terms of commutants of algebras. A similar definition is also possible for nets of symplectic subspaces. 

\smallskip
In general, given a net $\bbN_\cP$ and index sets $\cP\subseteq \cL$, \emph{the dual net of the additively extended net} $\bbN_{\cL}^\cP$ is denoted by $\bbN^{\cP d}_{\cL}$. It is a (possibly non-trivial) enlargement of $\bbN_\cP$, both in the sense that $\bbN_\cP$ sits as a subnet in $\bbN^{\cP d}_{\cL}\rest\cP$, if it satisfies locality, and that it is defined on the larger index set $\cL$. 
Explicitly, an object of $\bbN^{\cP d}_{\cL}$ for $a\in \cL$ is given by
\beqa\label{e:dual ex}
\bbN^{\cP d}_{\cL}(a)&:=&\bigcap \{\bbN_{\cL}^\cP(a_i)',\,a_i\bot a,\, a_i \in \cL\}\cap(\bbN_{\cL}^\cP)^{\cL}(M)\nonumber\\
&=&\bigcap \{\bbN_\cP(o_j)',\,o_j\bot a,\, o_j \in \cP\}\cap\bbN^\cP(M).
\eeqa
Note that such an enlargement is not an extension, since the restriction requirement of an extension expressed in equation (\ref{e: extension}), is not fulfilled in general.

\smallskip
A similar enlargement of the net $\bbN_\cP$ is also given reversing the order in the above definition: the net $\bbN^{d \cP}_{\cL}$ is called \emph{the additively extended net of the dual net} $\bbN^d_\cP$. It is defined on the generic element, $a\in \cL$, by
\beqa\label{e:ex dual}
\bbN^{d \cP}_{\cL}(a)&:=&\bigvee \{\bbN_{\cP}^d(o_i),\,o_i\subseteq a,\, o_i \in \cP\}.
\eeqa
We have $\bbN^{d \cP}_{\cL}\subseteq \bbN^{\cP d}_{\cL}$ and the dual net $\bbN^{\cP d}_{\cL}$ turns out to be more interesting from the point of view of superselection theory. However, the following holds
\blemma\label{l: algebraic dual}
If $\bbN_\cP$ is a local net on an index set $\cP$ and $\cP\subseteq\cL$ we have:
\bdes
\item[i)]
the extended dual and the dual extended nets coincide on $\cP$ with the dual net, i.e.\ 
\beqn
\bbN^{d \cP}_{\cL}\rest\cP\,=\, \bbN^{\cP d }_{\cL}\rest\cP\,=\,\bbN^d_\cP\,;
\eeqn
\item[ii)]  
if the net $\bbN_\cP$ is $\cP$-dual then the nets $\bbN^{d \cP}_{\cL}$ and $\bbN^{\cP}_{\cL}$ are equal; if also the $\cL$-duality for net $\bbN^\cP_{\cL}$ holds, the three above defined nets coincide, i.e.\ 
\beqn
\bbN^{\cP}_{\cL}= \bbN^{d \cP}_{\cL}= \bbN^{\cP d}_{\cL}\,.
\eeqn
\edes
\elemma
\prf
Trivial from the definitions.
\qed
Observe that it is not possible to derive in general the $\cL$-duality of the net $\bbN^\cP _{\cL}$ from the $\cP$-duality of the net $\bbN_\cP$, i.e.\ the heritability of duality to an extended net, see also \cite[Section 30]{rob04}. We shall however obtain such a result for a specific model in Section \ref{ss:Other properties of fields subnets}.

\smallskip
We end this section recalling two useful widely used notations
\bitem
\item
for an element $o\in\cP$, the \emph{causally disjoint set} $o^\bot$ is the sieve 
\footnote{
A \emph{sieve} in $\cP$ is a subset of elements $\cS\subset\cP$ such that if $o\in \cS$ and $o_1\subset o$ then $o_1\in \cS$.  
}
of elements in $\cP$ disjoint from $o$, i.e.\
\beq\label{e: disjunto}
o^\bot:=\{o_1\in \cP: o_1 \bot o\}\subset \cP\,;\\
\eeq
\item
for  a topological space $M$, and and element $o$ in a causal index set $\cP\subset\open (M)$, 
a subspace $o'\in\open(M)$ called \emph{the causal complement of $o$}, is defined  by 
\beq\label{e: disjoit}
o':=\cup\{o_1\in \cP: o_1 \bot o\}=\cup\{o_i\in o^\bot\}\,.
\eeq
\eitem
Observe that $o'$ is not an element of $\cP$ and that both $o^\bot$ and $o'$ may be used to rewrite in a simpler way the properties of nets listed above. 
For instance, if $\bbN_\cP$ is a net of (von Neumann) algebras over the index set $\cP$ in the spacetime $M$, then the duality property may be expressed in a shorthand form as
\beq\label{e:dual casual}
\bbN_\cP(o)=\bbN_\cP (o')' \,\cap\, \bbN^{\cP}(M)\,,\qquad o \in \cP.
\eeq
\subsection{Index sets for the real line}\label{ss: SW nets}
In this subsection we fix the notation and the main results about the index sets of the real line that will be used in the sequel.

\smallskip 
Let $M$ be the 1+1-dimensional Minkowski spacetime and denote by $\bR$ its time zero axes (or when indicated, the chiral left/right line of the light cone). Let $\open(\bR)$ denotes the set of the open non-void subsets of $\bR$, partially ordered under inclusion.  The causal disjointness relation $\bot$ of $M$ restricted to $\bR$ coincides with the set theoretic disjointness relation, i.e.\ for $o,a\in\open(\bR)$ $o\bot a\, \iff \,o\cap a=\emptyset$. Moreover, for $o,\, a\in \open(\bR)$, we write $o<a$ iff $x<y$ for any $x\in o$ and $y \in a$; if $o,\, a\in \open(M)$ then $o<a$ means that $o$ sits in the left causal spacelike complement of $a$. 

The relation $<$ on the set $\open(M)$, that implies the relation $\bot$ and only satisfies point \textbf{b)} and \textbf{c)} of a causal disjointness relation, is in fact derived from the usual orientation on $\bR$ and will be will be called the \emph{orientation relation} on $\open(M)$. Similarly is defined the opposite orientation $>$ on $\open(M)$.

\smallskip
To recall standard notation about subsets of $\open(\bR)$, considering at first \emph{non-empty, bounded, open intervals}, we denote the generic one by $I$. The causal complement of $I$ is denoted by $I'$ and is the union of two open connected components; we may write $I'=J_l\cup J_r$ where $J_l:=\cup_{I_i<I}\,I_i$ and $J_r:=\cup_{I<I_i}\,I_i$. We call $J_l$ and $J_r$ \emph{the left} and \emph{the right complement} of $I$ respectively, and we have $J_l<J_r$. Observe that if we denote by $J_{ll}:=\bR\setminus\overline{J_r}$ and by $J_{rr}:=\bR\setminus\overline{J_l}$, then $I$ equals $J_{ll}\cap J_{rr}$. 
In general, the \emph{the left} and \emph{the right open halflines} in $\open(\bR)$ are used, usually denoted by $J_l$ and $J_r$ respectively and defined as above.  
Given two non-empty, bounded, open intervals $I_1, I_2$ with $I_1\cap I_2=\emptyset$, we call a non-empty bounded \emph{double interval} the union $E=:I_1 \cup I_2\in \open(\bR) $ and usually suppose $I_1<I_2$.
It is hence useful to introduce the following notation for the above families of subsets of $\open(\bR)$:
\beq\label{e:sets}
\barr{ll}
\cI&:=\,\{ \textrm{non-empty  bounded open intervals of }\bR  \}\,,\\
\cI_2&:=\,\{ \textrm{non-empty  bounded open double-intervals of }\bR \}\,,\\
\cD&:=\,\cI \cup \cI_2\,,\\
\cJ&:=\,\cJ_l \cup \cJ_r=\{ \textrm{left or right open half-lines of } \bR \}\,.
\earr
\eeq
Furnished with the inclusion as partial order and with the causally disjointness relation inherited from $\open (M)$, all these sets may be used as index sets for nets, in the sense of Subsection \ref{ss:Nets}.

The set $\cI$ is a base of neighborhoods for the topology on the time zero line $\bR$ of open (non-empty) path connected elements, so that it generates the other three. 

\smallskip
Also to fix notation, if $J_r\in \cJ_r$ is the generic right half-line, we have $J_r^\bot:=\{J\in \cJ_l : J\bot J_r\}$ and we put $J_r'=\bigcup_{J\in J^\bot_r} J= \bR\setminus \bar{J_r}:=J_l$. Vice versa $J_l'=J_r$.
The causal complement  of $E\in \cI_2$ is $E' = J_{l}\cup I_3 \cup J_{r}$, with $J_l<I_1<I_3<I_2<J_r$, the subset $I_3\in \cI$ being the largest bounded open interval between $I_1$ and $I_2$. Moreover $J_{l}\in \cJ_l$ and $J_{r}\in \cJ_r$, and we have $J_l \cup J_r = I'$ for  $I:=I_1\cup \bar{I_3}\cup I_2\in \cI$. Hence the choice of any double interval gives a relevant five piece decomposition of $\bR$, corresponding to the four piece one of the circle (discussed for example in \cite{klm01}), the one-point compactification of $\bR$.

\smallskip
Note that the elements in $\cI$ and $\cJ$ are linearly path connected and the elements in $\cI_2$ are not. 
Some properties of the above index sets as posets are summarized in the following easily proved result
\blemma\label{l:prop index}
The index sets $\cI,\,\cI_2$ and $\cD$ are directed and connected. The index set 
$\cJ$ is  neither directed nor connected: it is the union of the two directed and connected components $\cJ_l$ and $\cJ_r$. The set $\cI$ is cofinal in $\cI_2$ and $\cD$ but not in $\cJ$; the set $\cI_2$ is cofinal in $\cD$ but not in $\cJ$.
\elemma
For every element $o\in \cP$, the sieve $o^\bot$ may have more than one connected component. The number of these connected components is related to the number of the connected components of the graph of the relation $\bot$ on $\cP$, defined by 
\beq\label{e: graph bot}
\cG_\cP^\bot:=\{(o_1,o_2)\in \cP\times \cP:\, o_1\bot o_2\}\,
\eeq
with order inherited from $\cP\times \cP$ and the product ordering, see \cite[Section 23]{rob04} for details.

\smallskip
If the index set $\cP$ is connected as a poset, and for the index set of the list (\ref{e:sets}) above this is the case of  $\cP=\cI,\cI_2$ and $\cD$, the number of the connected components of $o^\bot$ is two and equals the one of $\cG_\cP^\bot$, see \cite[Lemma 23.6]{rob04}. 

\smallskip
The case of half-lines is more insidious: the index set $\cJ$ is not connected and is the union of the two connected components $\cJ_l$ and $\cJ_r$, as said in Lemma \ref{l:prop index}. For any $J_1\in \cJ_l$ and $J_2\in \cJ_r$ with $J_1\bot J_2$, there is no path from the couple $(J_1, J_2)$ to the couple $(J_2,J_1)$ in $\cG^\bot_\cJ$, just because $\cJ$ is not connected. Hence also $\cG_\cJ^\bot$ has two connected components, namely $\{(J_l,J_r)\in \cJ_l \times \cJ_r:\,J_l\bot J_r\}$ and $\{(J_r,J_l)\in \cJ_r \times \cJ_l:\,J_l\bot J_r\}$ whose union generates $\cG_\cJ^\bot$.  

\smallskip
It is possible to introduce two \emph{oriented} versions of the graph of the relation $\bot$ on $\cP$, i.e.\ the graph of the orientation relations $<$ and $>$, by
\beq\label{e: graph bot orient}
\cG_\cP^<:=\{(o_1,o_2)\in \cP\times \cP:\, o_1 < o_2\},\qquad\textrm{and}\qquad
\cG_\cP^>:=\{(o_1,o_2)\in \cP\times \cP:\, o_1 > o_2\}\,,
\eeq
with product ordering, i.e.\ the one inherited from $\cG_\cP^\bot$.
For $\cP$ any of the index sets above, the graphs $\cG_\cP^<$ and $\cG_\cP^>$ are connected, and may be easily identified with the two connected component of $\cG_\cP^\bot$.

\smallskip
We define by $o^\bot_l :=\{o_1\in\cP : o_1<o\}$ and $o^\bot_r :=\{o_1\in\cP : o<o_1\}$ \emph{the left} and \emph{the right} components of $o^\bot$.
If $o\in \cP$ and we fix by convention the first entry in $\cG_\cP^<$ and $\cG_\cP^>$, we have $o_r^\bot\subset \cG_\cP^< $ and $o_l^\bot\subset \cG_\cP^> $.  
In practice, we may think that the elements in $o_r^\bot$ and $o_l^\bot$ connect $o$ to the right and left infinity respectively. 
In general it holds $o^\bot_l\cup o^\bot_r \subseteq o^\bot$. 

In particular, if $I\in \cI$, the equality holds. If $J\in \cJ_r$, then  $J^\bot_l =J^\bot\subsetneq \cJ_l$ and $J^\bot_r =\emptyset$. Similarly for $J\in \cJ_l$. For a double interval $E\in \cI_2$, it only holds $E^\bot_l\cup E^\bot_r \subsetneq E^\bot$. 

\smallskip
With the same convention, and a slight abuse of notation, we call $\cG_\cP^< $ and $\cG_\cP^>$ \emph{the right} and \emph{the left} component of $\cG_\cP^\bot$ respectively.

\smallskip
Finally, we denote by $S= S^{-1}$ the space inversion about $0\in\bR$, thus $S: x\mapsto -x$. All the index sets $\cP$ in the above list (\ref{e:sets}) are closed under the action of $S$, i.e.\ $So\in \cP$ for every $o\in \cP$, and because $So_2 <So_1$ for every  $o_1<o_2$ in $\cP$, the action of $S$ on $\cG^\bot_\cP$ realizes the \emph{flip} between the two oriented components $\cG_\cP^<$ and $\cG_\cP^>$, i.e.\  $(o_1, o_2)\in\cG_\cP^<$ iff  $S(o_1,o_2):=(So_1,So_2)\in \cG_\cP^>$. 

\smallskip
If $J\in \cJ_l$, the connected component of its causal complement is $\cJ_r$, and $S$ flips it to $\cJ_l$ itself. Similarly, if $I\in \cI$, the left component $I'_l$ of its complement is in the connected component $\cJ_l$ and is mapped  by $S$ to an element of $\cJ_r$, connected to the right component $I'_r$ of the causal complement of $I$. 

\smallskip
These arguments, that may be extended to the index sets of the whole 1+1-dimensional Minkowski space $M$, will be used in Lemma \ref{l: stat net implement} to implement some isomorphisms between the algebras of the net, also related to the representation of the geometric covariance group of the net itself.
\subsection{Nets with a grading group}\label{ss:K graded net}
In the following three subsections we present a generalization of the locality and duality properties of non local nets defined on low dimensional spacetimes, to show how the commutation relations of these nets may be relevant for the description of the DHR sectors of their fixed point subnets.

\smallskip
From now on, we shall use mainly the index sets $\cP=\cI, \cD$ if not differently stated, and will refer to $\cJ$ only as an ancillary index set. In fact, we will show that the index sets $\cI$ and $\cD$ are sufficient to determine the superselection sectors of interest.

\smallskip
It is worth underlining that only translation covariance have a role in the sequel, according to the fact that we want to describe (sharply localized transportable) DHR sectors, hence M\"{o}bius group covariance plays almost no role.   

\smallskip
The Bose-Fermi alternative accounts for the normal commutation rules of a field net defined on an index set $\cP$, of a $1+3$-dimensional spacetime, for fields localized on disjoint elements of $\cP$. 
This may be formalized in AQFT using $\bZ_2$-graded non local nets. 
In particular, it is possible to associate to any $\bZ_2$-graded net $\cN_\cP$, another net  $\cN^t_\cP$ called \emph{the twisted net}.
\footnote{\label{f:tnet}
Remember that the twisted net of a net $\cF_\cP$ is defined locally for any $o\in \cP$ by $\cF_\cP^t(o):=\{\a_V(F),\, F\in \cF_\cP(o)\}$ where $\a_V$ is a global Klein automorphism implemented by an operator $V$. For example $V=(I+ik)/\sqrt{2}$, where $k$ is the operator giving a $\bZ_2$-grading of the Hilbert space of the representation $\o$ of $\cF$, in the sense that $k\o(F)=\o(\a_k(F))k$ for $F\in \cF_\cP(o)$ and $\a_k$ is the automorphism acting locally and giving the Bose-Fermi alternative, i.e.\ such that  $\a_k(F)=\pm F_\pm$, where $F=F_+ + F_-$ is the decomposition of any $F\in\cF_\cP(o)$ in its Bose and Fermi parts, see for example \cite[Section 4.1.1]{rob90} for details.   
}
This procedure replaces the Bose-Fermi alternative and allows to introduce, similarly to the case of a local net, the properties of \emph{twisted locality} and \emph{twisted duality} for (mainly von Neumann algebras) nets

\bigskip
\noindent \textbf{3\hspace{0,5pt}a.} {\sc Twisted Locality.}
\quad $\bbN_\cP(o_1)\,\bot\, \bbN^t_\cP (o_2)$, if $o_1\,,o_2\in \cP$ with $o_1 \,\bot\, o_2$.\\ 

\noindent \textbf{5\hspace{0,5pt}a.} {\sc Twisted Duality.}
\quad $\bbN_\cP(o)=\bbN_\cP^{td}(o):=\bigcap_{o_1\bot o}\bbN_\cP^t(o_1)'\cap \bbN^\cP(M)$, for $o\in \cP$.\\

\noindent 
We are initially interested in the generalization of these properties in the case of  1+1-dimensional spacetime and for a generic (Abelian, discrete) group replacing $\bZ_2$. 

\smallskip
Recall that given a group $K$, a von Neumann algebra $\bbN$ is said to be \emph{$K$-graded} iff for any $k\in K$, a linear weakly closed subspace $\bbN_k$ is defined and called the \emph{$k$-grade homogeneous subspace} of $\bbN$, such that 
\bdes
\item[a)]   $\bbN=\bigvee_{k \in K} \bbN_k\,$;
\item[b)]   $\bbN_k \cap\bbN_h =0\,,\qquad k\neq h,\;h,k\in K$;
\item[c)]   $\bbN_h \cdot \bbN_k\subseteq\bbN_{h k}\,,\qquad k,h \in K$.   
\edes
Notice that an homogeneous subspace $\bbN_k$ is a subalgebras of $\bbN$ iff $k=e$, for $e$ the identity element of $K$. 
In the sequel we shall denote by $K$ an Abelian group, furnished with the discrete topology.   
We hence give the following
\bdfn\label{d:graddnetdef0}
A net of von Neumann algebras $\cF_\cP$ on the index set $\cP$, is said to be \emph{graded by a group $K$}, or for short \emph{a $K$-graded net}, if it is graded as net, i.e.\ 
\bdes
\item[a)]
for every $o\in \cP$ the von Neumann algebras $\cF_\cP(o)$ is $K$-graded; and
\item[b)]
for every $o\,,o_1\in \cP$ with $o\subseteq o_1$ the inclusion $\cF_\cP(o)\subseteq \cF_\cP(o_1)$ is grade preserving, i.e.\ denoting for any $k\in K$ by $\cF_\cP(o)_k$ the $k-$homogeneous subspace of the algebras $\cF_\cP(o)$, it holds 
$\cF_\cP(o)_k\subseteq \cF_\cP(o_1)_k$.
\edes
\edfn
Albeit different group constructions may give a graded net, for example \emph{twisted} crossed products, we shall concentrate in Subsection \ref{ss:cross prod graded net} to the less general case of crossed products, also used to treat the Streater and Wilde model, after having established the main features of graded nets in the present section.  

\smallskip
We begin fixing the purely algebraic aspects of \emph{nets graded by a couple of groups}. 
For example, when the index set $\cP$ has two different connected components of $\cG^\bot_\cP$, i.e.\ for every index set in the list (\ref{e:sets}), we shall use nets graded by the direct product of two isomorphic groups. 

\smallskip
If we are given with two distinguished nets $\cF_{1\cP}$ and $\cF_{2\cP}$, represented on the same Hilbert space and graded by the groups $K_1$ and $K_2$ respectively, we easily obtain another net, graded by $K_1\times K_2$ and \emph{generated by product} in an essentially unique way. This is possible through the product of the homogeneous subspaces of $\cF_{1\cP}$ and $\cF_{2\cP}$, disregarding of the order of the elements, given for $(k_1,k_2)\in K_1\times K_2$ and $o\in \cP$, by the following linear space
\beq\label{e:spam prod}
\cF_1(o)_{k_1} \cdot \cF_2(o)_{k_2}:=\spam \big\{F_{k_1}\,F_{k_2}\,, F_{k_1}\in \cF_1(o)_{k_1}\,, F_{k_2}\in\cF_2(o)_{k_2} \big\}=\cF_2(o)_{k_2} \cdot \cF_1(o)_{k_1}\,.
\eeq 
We define the $K_1\times K_2$-homogeneous subspaces of the product net through the following equation 
\beq\label{e:k12 homog} 
\cF(o)_{(k_1,k_2)}:= \cF_1(o)_{k_1} \cdot \cF_2(o)_{k_2} \,,\quad (k_1,k_2)\in K_1\times K_2\,, \quad o\in \cP\,.
\eeq 
A result in this direction is hence given by the following
\blemma\label{l: net bi grad}
Let $K_1$ and $K_2$ be two (Abelian, discrete) groups. Then 
\bdes
\item[i)]
a net $\cF_\cP$ of von Neumann algebras on $\cP$, is $K_1\times K_2$-graded iff there exist a couple of nets $(\cF_{1\cP},\,\cF_{2\cP})$ graded by $K_1$ and $K_2$ respectively, such that $\cF(o) = \cF_1(o) \bigvee \cF_2(o)$ and the equation (\ref{e:k12 homog}) holds for any $o\in \cP$; 
\item[ii)]
a net $\cF_\cP$ is uniquely defined by a couple of nets $(\cF_{1\cP},\cF_{2\cP})$  iff  $\cF_1$ is a $\cF_{2, e_2}$-bimodule graded net, i.e.\ for every $k_1\in K_1$ and $o\in \cP$ it holds
\beq\label{e:bimod}
\cF_2 (o)_{e_2} \cdot \cF_1(o)_ {k_1}\,=\, \cF_1(o)_ {k_1}\cdot \cF_2 (o)_{e_2}\subseteq \cF_1 (o)_ {k_1} \,.  
\eeq
Similarly exchanging $\cF_1$ by $\cF_2$ and $K_1$ by $K_2$.
\edes
\elemma
\prf 
i) for $(\Rightarrow)$ we define the homogeneous subspaces of a net $\widetilde{\cF}_{1,\cP}$ by  (similarly for $\widetilde{\cF}_{2,\cP}$)
\beq\label{e:omogcouple}
\widetilde{\cF}_1 (o)_{k_1} \,:=\, \cF (o)_{(k_1, e_2)}\,,\qquad o\in \cP\,.
\eeq
This gives two nets graded by $K_1$ and $K_2$ respectively, that generate $\cF_\cP$ by definition. 
The identity subnets of $\widetilde{\cF}_{1,\cP}$ and $\widetilde{\cF}_{2,\cP}$ both equal $\cF_{\cP,(e_1, e_2)}$, i.e.\ it holds 
\beq\label{e:equal identity}
\widetilde{\cF}_1(o)_{e_1}\,=\,\widetilde{\cF}_2(o)_{e_2}\,=\,\cF(o)_{(e_1, e_2)}\,, \qquad o\in\cP\,.
\eeq
$(\Leftarrow)$ 
We define through the equation (\ref{e:k12 homog}) the $K_1\times K_2$-homogeneous subspaces. Then, for any $o\in \cP$, it remains defined a von Neumann algebra $\cF_\cP(o)$ graded by $K_1\times K_2$ and generated by $\cF_1(o)$ and $\cF_2(o)$. In fact, 
it is easy to see, using the definition of equation (\ref{e:k12 homog}), that  
\beqn
\bigvee_{(k_1,k_2)\in K_1\times K_2}\,\cF(o)_{(k_1,k_2)}=\bigvee_{(k_1,k_2)\in K_1\times K_2}\,\cF_1(o)_{k_1}\cdot \cF_2(o)_{k_2}=
\cF_1(o)\bigvee\cF_2(o)=:\cF(o)\,.
\eeqn
This also satisfies the point \textbf{a)} of the definition of a graded algebra. 
Point \textbf{b)} comes from the same property for the graded subspaces of $\cF_1$ and $\cF_2$: for $(k_1,k_2), \,(h_1,h_2)\in K_1\times K_2$ it holds 
\beqn
\cF(o)_{(k_1,k_2)}\bigcap\cF(o)_{(h_1,h_2)}=\cF_1(o)_{k_1}\cdot \cF_2(o)_{k_2} \bigcap 
				\cF_1(o)_{h_1}\cdot \cF_2(o)_{h_2} = 0 \quad \textrm{if} \quad (k_1,k_2)\neq(h_1,h_2)\,.
\eeqn
Point \textbf{c)} follows from
\beqan
\cF(o)_{(k_1,k_2)}\cdot\cF(o)_{(h_1,h_2)}&=&\big(\cF_1(o)_{k_1}\cdot \cF_2(o)_{k_2}\big)\cdot \big(\cF_1(o)_{h_1}\cdot \cF_2(o)_{h_2}\big)\\
&=&\cF_1(o)_{k_1}\cdot \cF_1(o)_{h_1}\cdot \cF_2(o)_{k_2}\cdot \cF_2(o)_{h_2} \subseteq \cF_1(o)_{k_1 h_1}\cdot\cF_2(o)_{k_2 h_2} \\
&=& \cF(o)_{(k_1 h_1, k_2 h_2)}\,.
\eeqan
Hence, the $K_1\times K_2$-algebras are well defined, so that point \textbf{a)} of the Definition \ref{d:graddnetdef0} of graded net is satisfied and easily also point \textbf{b)}. Hence $\cF_\cP$ is a net graded by $K_1\times K_2$.\\
ii)
We may work with the net itself or with any fixed algebras at $o\in \cP$.\\ 
$(\Rightarrow)$ 
If the couple $(\cF_1,\cF_2)$ is unique, using the definition of equation (\ref{e:omogcouple}) and the relation (\ref{e:k12 homog}), we obtain 
\beqn
\cF_{1,k_1}:= \cF_{(k_1,e_2)}= \cF_{1,k_1}\cdot\cF_{2,e_2}=\cF_{2,e_2}\cdot\cF_{1,k_1}\,,   
\eeqn
hence $\cF_1$ is a $\cF_{2,e_2}$-bimodule graded net and similarly for $\cF_2$.\\
$(\Leftarrow)$ 
Because the identity operator $I$ has $e_2$-grade, we have $I\in \cF_{2,e_2}$. Moreover, $\cF_{1,k_1}$ is a $\cF_{2,e_2}$-bimodule, then for any $k_1\in K_1$ it holds
\beqn
\cF_{1,k_1} \cdot \bC \subseteq \cF_{(k_1,e_2)}= \cF_{1,k_1}\cdot\cF_{2,e_2} \subseteq \cF_{1,k_1}\,,
\eeqn
hence $\cF_{1,k_1}=\cF_{(k_1,e_2)}$ and similarly $\cF_{2,k_2}=\cF_{(e_1,k_2)}$. The same holds for any other couple of nets $(\widetilde{\cF}_1,\widetilde{\cF}_2)$ defining the $K_1\times K_2$-graded net $\cF$, so that these couples coincide.
\qed
From the unicity condition of the above lemma, i.e.\ the bimodule conditions of equation (\ref{e:bimod}) for $\cF_1$ and $\cF_2$, it follows that the identity subnets $\cF_{1,e_1}$ and $\cF_{2,e_2}$ coincide. Hence we have an uniquely defined identity net, see also equation (\ref{e:equal identity})
\beq\label{e: identity bigraded}
\cF_{(e_1,e_2)} \,=\, \cF_{1,e_1}\cdot \cF_{2,e_2}\,=\,\cF_{1,e_1}\,=\,\cF_{2,e_2}\,.
\eeq
We consider now the space inversion map $S$, introduced at the end of Subsection \ref{ss: SW nets}: 
if we suppose that the index set $\cP$ is closed under the discrete action of $S$, for a net $\cF_\cP$ graded by two copies of the same group, i.e.\ by the product group $K\times K$, we can give the following
\bdfn\label{d:graddnetdef}
A net $\cF_\cP$ is said to have a \emph{$K\times K$-grading flipped by $S$} if there exists a group $K$ such that $\cF_\cP$ is $K\times K$-graded, and $S$ acts covariantly on the homogeneous subspaces of $\cF_\cP$, flipping the elements in the couples of $K\times K$. This means that for $U$ a (faithful) unitary representation of $S$ on the same Hilbert space of $\cF_\cP$, it holds
\footnote{
Remember that if the net  $\cF_\cP$ (or the nets $\cF_l$ and $\cF_r$, see Proposition \ref{o:properties}) satisfies the Reeh-Schlieder property, so that the vacuum vector $\O$ is cyclic and separating for every $\cF(o)$ with $o\in \cP$, then the representation of $S$ may be obtained using the modular conjugation of the  
half-line $(-\infty, 0)$, see for example \cite{gl} for details.
}  
\beq\label{e:flipped graded}
U(S)\ \cF(o)_{(h,k)}\ U(S)\,=\, \cF(So)_{(k,h)}\,,\qquad \,o\in \cP,\quad (h,k)\,,(k,h)\in K\times K\,.
\eeq   
\edfn
Notice that in this case the net itself have to be $S$-covariant, in fact it holds
\beqn
\ad U(S)\big(\cF(o)\big)=\ad U(S)\,\big(\bigvee_{(h,k)\in K\times K}\cF(o)_{(h,k)}\big)=\bigvee_{(k,h)\in K\times K}\cF(So)_{(k,h)}=\cF(So)\,.
\eeqn 
Moreover,  $S$ acts covariantly on the diagonal homogeneous subspaces of $\cF_\cP$, in particular on the identity homogeneous component subnet $\cF_{\cP,(e,e)}$.  

\smallskip
If $\cF_\cP$ is a net with $K\times K$-grading flipped by $S$, because the space inversion $S$ flips between the left $\cG_\cP^<$ and the right $\cG_\cP^>$ oriented component of the graph $\cG_\cP^\bot$, we prefer to replace the subscripts $1$ and $2$ that label the two graded subnets generating $\cF_\cP$, by  the labels $l$ and $r$ for \emph{left} and \emph{right} graded subnet. 
The motivation of this labeling shall be more clear in the subsequent two subsections an on the specific models. 

\smallskip
We easily obtain from point i) of Lemma \ref{l: net bi grad}, that a net $\cF_\cP$ has a $K\times K$-grading flipped by $S$ 
iff the two subnets $\cF_{l\cP}$ and $\cF_{r\cP}$, graded by the same group $K$ and generating $\cF_\cP$, are \emph{homogeneously conjugated by the space inversion operator $S$}, i.e.\ if for $U$ a (faithful) unitary representation of $S$ on $\cH$, it holds
\beq\label{e:S action}
U(S)\ \cF_l(o)_k\ U(S)\,=\, \cF_r(So)_k,\qquad \,o\in \cP,\quad k\in K\,.
\eeq
In fact it holds: $\ad U(S) \big(\cF(o)_{(h,k)} \big)=\cF_r(So)_h\cdot \cF_l(So)_k = \cF(So)_{(k,h)}$.

If $\cF_l$ is inversion covariant, i.e.\ if $\ad U(S)\big(\cF_l(o)\big)=\cF_l(So)$ for any $o\in \cP$, and homogeneously conjugated by $S$ to the net $\cF_r$, then the two nets are equal. 

\smallskip
More relevant is the following observation. 
Being $S=S^{-1}$, the equation (\ref{e:S action}) holds also exchanging $\cF_l$ and $\cF_r$, and for every $o\in \cP$ it furnish a unitary isomorphism between the algebras $\cF_l(o)$ and $\cF_r(So)$. 
For an element $o\in \cP$ invariant under the action of $S$, i.e.\ $So=o$, the action of the space inversion implements the following \emph{local} unitary isomorphism
\beq\label{e:invariance local}
\ad U(S) \big(\cF_l(o) \big)\,=\,\cF_r(o)\,.
\eeq
This occurs only for the elements in $\cP=\cI,\cD$ that are symmetric about the point $x=0$. In general, for elements $o\in \cP$ non-invariant under the action of $S$, the isomorphism in equation (\ref{e:invariance local}) may be obtained by the action of $S$ combined with an element of the geometric covariance group $\G$, as given in the following
\blemma\label{l: stat net implement}
Let $\cF_{c\cP}$ with $c=l,r$ be two nets on $\cP$, graded by the (Abelian, discrete) group $K$, homogeneously conjugated by the space inversion $S$ and with geometric covariance group $\G$. If $U$ denotes a (strongly continuous, faithful) unitary representation of $S$ that extends to $\G$, on the common Hilbert space of the nets $\cF_{l\cP}$ and $\cF_{r\cP}$, then there exist a family $\{\Xi_o\}_\cP$ of implemented isomorphic involutions, i.e.\ satisfying $\Xi_o=\Xi_o^{-1}$, such that  
\beq\label{e: o isom grad}
\Xi_o(\cF_c(o))=\cF_{Sc}(o)\,,\qquad  o \in \cP\,.
\eeq
In particular
\bdes
\item[i)]
if $\cP=\cI$ and $\G$ contains the translations, for any bounded open $I\in \cI$ there exists a translation $\t_t$ with $t\in \bR$, such that $\Xi_I =\ad U(\t_t S)$; 
\item[ii)]
if $\cP=\cD$ and $\G$ contains the translations, for any open double interval $E=I_1\cup I_2\in \cI_2$ there exists a translation $\t_t$ with $t\in \bR$, such that for $i=1,2$ it holds $\t_t S (I_i)\cap I_i\in \cI$. Moreover, if $\G$ is the M\"{o}bius group, for any open double interval $E=I_1\cup I_2\in \cI_2$ there exists a M\"{o}bius transformation $g$ such that $\Xi_E =\ad U(gS)$. 
\edes  
\elemma
\prf
The proof is similar to the case with $So=o$ of equation (\ref{e:invariance local}).  The isomorphism $\Xi_o$ is implemented by an \emph{orientation-reversing element}, given by $S$ and an element $g\in \G$, such that $gSo=o$. Being $\cF_l$ and $\cF_r$ homogeneously conjugated by $S$, from equation (\ref{e:S action}) we have
\beq
\Xi_o\big(\cF_r(o)\big)\,=\, \ad U(gS)\big(\cF_r(o)\big)\,=\,\ad U(g)\big(\cF_l(So) \big) \,=\,\cF_l(gSo)\,=\,\cF_l(o)\,.
\eeq		
Easily, through the same isomorphism is obtained $\Xi_o(\cF_l(o))=\cF_r(o)$. We compute the proper elements of the geometrical group in the various cases for completeness.\\
i) 
if $I=(\a, \b)$ and $t=\a+\b$, then $SI =\t_t\, I$, for $\t_t\,I:=\{y=x+t\in \bR,\, x\in I\}$ the translation of $I$ by $t$. The required isomorphism is hence $\Xi_I=\ad U(\t_t S)$, for $\t_t S:x\mapsto {-x+ \a+\b}$.\\
ii)
The first part is obtained by point i), using the cofinality of $\cI$ in $\cD$. Explicitly, consider $I_1<I_2$, $I_1=(\a,\b)$ and $I_2=(\g,\d)$. The translation $\t_t$ is obtained as in i), with $t=\a+\d$ and is such that $\t_t S(\a)=\d$ and $\t_t S(\d)=\a$. The automorphism $\Xi_E$ is given by the adjoint action of $U(g S)$, with $\t_t$ as above and $g$ is a M\"{o}bius transformation such that $g(\a)=\a$, $g(\d)=\d$,  $g(\t_t S(\b))=g(-\b+\a+\d)=\g$ and  $g(\t_t S(\g))=g(-\g+\a+\d)=\b$. Such a $g: x\mapsto \frac{ax+b}{x+c}$ (where $a=c+\a+\d$, $b=-\a\d$ and $c=\frac{\a\d-\b\g}{(\d-\g)-(\b-\a)}-(\a+\d)$), is uniquely determined and it equals the identity of the group $\G$ if the length of $I_1$ and $I_2$ is equal. 
\qed
\brem\label{r: xi and cofinality}
In the applications it may be enough to use the first part of point ii) of Lemma \ref{l: stat net implement}, where only the translation covariance of the net $\cF$ is required. 
Because of the cofinality of $\cI$ in $\cD$, this means that it is often suffices to use the family $\Xi_\cI$ also for a net defined on the index set $\cD$. In particular, for $E\in \cD$, we replace $\Xi_{E}$ by $\Xi_{I}$, for $I$ the minimal open interval containing $E$.
The non-trivial intersection in the same point ii), suffices to define the algebras $\cF_{Sc}(o)$, from  $\Xi_o\big(\cF_c(o)\big)$, for $c=l,r$. Hence if $\cF_c$ is a field net,  the charges localized in this intersection, continue to have the same localization after the action of $\Xi_o$.    
This is true at least in the case of graded nets obtained as crossed product, see the comments before Lemma \ref{l:local grading}.
\erem
In addiction we have
\blemma\label{l: symmetries of xi}
In the notation of Lemma \ref{l: stat net implement}, for $c=l,r$ it holds
\bdes
\item[i)]
if $I\in \cI$ and $gS(I)=I$ then for any $I_c\in (I^\bot)_c$ there exists $I_{Sc}\in (I^\bot)_{Sc}$ such that $gS(I^\bot_c)=I^\bot_{Sc}$ and for the isomorphisms $\Xi_I$, we have for any $s=l,r$
\beq\label{e:aitom on compl}
\Xi_I\big( \cF_s(I_c)\big) \,=\,\cF_{Ss}(I_{Sc})\,, \qquad \textrm{where}\qquad I_{Sc}=gS I_c\,;
\eeq
\item[ii)]
if $E=I_1\cup I_2\in \cI_2$ with $I_1\cap I_2=\emptyset$ and $E'=J_l \cup I_3\cup J_r$ with $J_l<I_1<I_3<I_2<J_r$, then $\Xi_E \big(\cF_c(I_i)\big)=\cF_{Sc}(I_i)$ for  $i=1,2,3$, and the result of the equation (\ref{e:aitom on compl}) holds for $I=(J_l\cup J_r)' $  and for any $I_c\in J_c$, with $I_{Sc}\in J_{Sc}$. 
\edes
\elemma
\prf
Easily obtained from the action of the elements in the geometric covariance group $\G$ and $S$ on the elements of the index sets, and of the algebras isomorphisms $\Xi_o$ defined in the above Lemma \ref{l: stat net implement}. 
\qed
\brem\label{r:sympl xi}
Also for a net of symplectic subspaces $V_\cP$ or abstract Weyl algebras it is possible to give the notion of grading by a (Abelian, discrete)  group and covariance under the action of $S$. 
In particular, if  $V_\cP$ has a geometric symmetry group $\G$ as above, a family of symplectic automorphisms similar to the algebraic ones $\{\Xi_o\}_\cP$ above also exists. When used, these will be denoted by $\{\xi_o\}_\cP$ and trivially obtained by the action of $S$ and $\G$ on the elements of  $V_{l\cP}$ or $V_{r\cP}$, the left or right net of symplectic subspaces that generate the net $V_\cP$.
For example, if $o\in \cP$ is invariant under the action of $S$, for every for $c=l,r$ and $F \in V_{c\cP}(o)$ we have $\xi_o (F)(x)=F(-x)\in V_{Sc\cP}(o)$; e.g.\ if $I=(\a,\,\b)$ is an open interval, we have $\xi_I (F)(x)=F (-x+\a+\b)$ for the translation covariance, and $\xi_I(V_{c\cI}(I))=V_{Sc\cI}(I)$.
\erem
The construction of nets graded by a crossed product in the following Subsection \ref{ss:cross prod graded net} will gives the left and right graded nets explicitly on $\cP=\cI, \cD$, and represent the main example of grading used on models below.
\subsection{Graded locality and graded duality}\label{ss: graded prop}
A major outcome of the Definition \ref{d:graddnetdef} of a net with double grading flipped by $S$, is the following  generalization of the locality property for nets, that we state for nets of von Neumann algebras

\bigskip
\noindent \textbf{3\hspace{0,5pt}b.} {\sc Graded Locality.}
A net $\cF_\cP$ on the index set $\cP=\cI, \cD$, with $K\times K$-grading flipped by $S$, is said to satisfy the \emph{$K$-graded locality} if for every $o_l,\,o_r\in \cP$ with $o_l<o_r$, we have
\beq\label{e:K graded locality}
\cF_l(o_l)\;\bot\; \cF_r(o_r)\,.
\eeq
Mnemonically, we may think that the algebras $\cF_l (o_l)$ are graded on the left complement of $o_l$ and have a Bosonic behavior 
on the right one, where they have the identity grade; hence these elements commutes with the Bosonic elements localized on the right complement. Respectively, the algebras $\cF_r (o_r)$ are graded on the right complement of $o_r$ and are Bosonic on the left one.

\smallskip
The main example of graded nets will be given by the crossed products construction in Subsection \ref{ss:cross prod graded net}. In this case, we start from a local net on an index set $\cP$; then  any local algebra supported on an element of $o\in \cP$ 
will be graded through two different represented actions of the group $K$; each of them acting non trivially on the algebras supported on a connected component of the complement of $o$, but on the ones supported on the other component.   

\smallskip
Some attention is needed to introduce a notion of duality for a net with double grading flipped by $S$. In the local case, seen in the equations (\ref{e:dual net}) and (\ref{e:dual casual}), the dual net $\cF_\cP^d$ is an extension of the original net $\cF_\cP$, defined for any $o\in \cP$ by the intersection of the commutants of the algebras supported on elements contained in the causally disjoint set $o^\bot$. 
Moreover, we remember that if $\cF_\cP$ is $\bZ_2$-graded, the properties of locality  and duality are discussed using the $\bZ_2$-twisted net $\cF^t_\cP$ seen in footnote \ref{f:tnet}. 

If $\cF$ is an a net with $K\times K$-grading flipped by $S$ on the index set $\cP=\cI, \cD$, we want to define a \emph{$K$-twisted net} and a \emph{$K$-graded dual net} but, because of the graded locality condition (\ref{e:K graded locality}), this is better given only for the left $\cF_l$ and the right subnets $\cF_r$, and not for $\cF$ itself. 
For this reason, and because of the isomorphism of these two subnets given in Lemma \ref{l: stat net implement}, from now on we fix to work with the \emph{left net} $\cF_l$, if not differently stated  (see also the crossed product example in equation (\ref{e: Vin F}) and footnote \ref{f:charge cont}).

\smallskip
To define the $K$-twisted net  $\cF_l^g$ (here the ``$g$'' for \emph{graded} replaces the ``$t$'' of \emph{twisted} of the $\bZ_2$-case), we observe  from equation (\ref{e:K graded locality}) that for any $o_l,o,o_r\in\cP$ with $o_l<o<o_r$, we have at least to require for the algebra $\cF_l^g(o)$ that $\cF_l(o_l)\,\bot\, \cF^g_l(o)\,\bot\, \cF_l(o_r)$.
Such consideration has to be used in defining the graded algebras both over the elements of $o\in \cP$ that over their causal complement $o'$, necessary for the definition of the $K$-graded dual net $\cF_l^{gd}$. 
We hence give the following
\bdfn\label{d: ktwisted net}
Let $\cF_l$ be the left subnet of a $K\times K$-graded net $\cF_\cP$ over $\cP=\cI, \cD$, flipped by $S$ and satisfying $K$-graded locality. 
The $K$-twisted algebras associated to $\cF_l$ are defined 
\bdes
\item[a)]
for $I\in \cI$, by
\beqn
\cF_l^g(I):= \bigcap_{I_0\bot I} \cF_l(I_0)'\qquad \textrm{and}\qquad
\cF_l^g(I'):= \bigvee_{I_l<I}\Big(\cF_l(I_l) \bigcap \cF_l(I)'\Big)\bigvee_{I<I_r}\cF_r(I_r)\,;
\eeqn
\item[b)]
for $E=I_1 \cup I_2$, $E'=J_l \cup I_3 \cup J_r$ with $J_l=(I_1)'_l$,  $J_r=(I_2)'_r$ and $J_l\cup J_r=I'$, by
\beqn
\cF_l^g(E):= \bigcap_{I_0\bot E} \cF_l(I_0)'\qquad \textrm{and}\qquad
\cF_l^g (E'):= \cF_l^g(I')\bigvee \cF_l^g(I_3)\,.
\eeqn
\edes
The \emph{graded-dual algebra} of the algebra $\cF_l(o)$ are defined for $I\in \cI$ and $E\in \cD$ respectively by
\beqn\label{e:defgd senza xi}
\cF_l^{gd}(I):=\cF_l^g(I')'\bigcap \cF_l^\cI(\bR)\qquad \textrm{and}\qquad
\cF_l^{gd}(E):=\cF_l^g (E')'\bigcap \cF_l^\cI(\bR)\,.
\eeqn
\edfn
Several observations about these definitions are in order and we collect them in the following  
\brem\label{r:obs graded prop}
1)
For $o\in \cI, \cD$ we have $\,\cF_l^g(o')\,\bot \,\cF_l(o)\,$ and  $\,\cF_l^g(o) \,\bot \, \cF_l(o_1)$ for any $o_1\in o^\bot$.\\ 
2)
Defining the twisted algebras $\cF_l^g(I')$ in point a) above, the intersection has been required on both the complements, but on the right we are doing the minimal choice that it equals the right algebras, according to the Property \textbf{3\hspace{0,5pt}b.} of graded locality.\\   
3)
According to Lemma \ref{l: symmetries of xi} in the second equation of point a)  we may replace the algebras $\cF_r(I_r)$ by $\Xi_I\big(\cF_l(I_l)\big)$, for proper $I_l<I<I_r$. In this way, the nets  $\cF^g_l$ and $\cF^{gd}_l$ remain defined only through the net $\cF_l$ itself and the family if isomorphisms $\{\Xi_I\}_\cI$.\\ 
4)
If the element $E$ reduces to an bounded open interval, i.e.\ if $E\in \cI\subset \cD$, then the definitions in b) equals the ones of a) and so the one of the graded dual algebras.\\
5)
As in the case of the $\bZ_2$-twisted nets, we define the \emph{graded-dual algebra} of the algebra $\cF_l(o)$ by the commutants of the twisted algebras of the complement $o'$.
Explicitly we obtain   
\beqan\label{e:defgd senza xi1}
\cF_{l\cP}^{gd}(I)&=&\bigcap_{I_l<I}\Big(\cF_l(I_l)\bigcap \cF_l(I)' \Big)' \bigcap_{I<I_r}\cF_r(I_r)' \bigcap\cF_l^\cI(\bR)\,,\\
\cF_{r\cP}^{gd}(E)&=&\bigcap_{I_l<I}\Big(\cF_l(I_l)\bigcap \cF_l(I)' \Big)' \bigcap_{I<I_r}\cF_r(I_r)'\bigcap\Big(\bigvee_{I_0\bot I_3} \cF_l(I_0)\Big)\bigcap\cF_l^\cI(\bR)\,.
\eeqan
6)
If the net $\cF_\cI$ extends by additivity on the half lines, there exist some simplifications for the given definitions: for the second equation of a) and the first of b), respectively we have 
\beqan\label{e:dual addit}
\cF_l^g(I')\,&=&\, \Big(\bigvee_{I_l<I}\Big(\cF_l(I_l) \bigcap \cF_l(I)'\Big)\Big)\,\bigvee\cF_r(J_r)\,,\\
\cF_l^g(E)\,&=&\,\cF_l^g(I)\bigcap\cF_l(I_3)'\,.
\eeqan
Hence, for the graded dual algebras, we have explicitly  
\beqa\label{e:defgdd senza xi}
\cF_{l\cP}^{gd}(I)&=&\,\bigcap_{I_l<I}\Big(\cF_l(I_l)\bigcap \cF_l(I)' \Big)' \bigcap\cF_r(J_r)' \bigcap\cF_l^\cI(\bR)\,,\\
\cF_{r\cP}^{gd}(E)&=&\bigcap_{I_l<I}\Big(\cF_l(I_l)\bigcap \cF_l(I)' \Big)' \bigcap\cF_r(J_r)'\bigcap\cF_l(I'_3)\bigcap\cF_l^\cI(\bR)\,.
\eeqa
\erem
More explicit consequences of the above definition shall be presented for nets graded by crossed products in Subsection \ref{ss:cross prod graded net}.

\smallskip
Notice that the property of \emph{graded duality} may be written through the following straightforward generalization of the properties \textbf{5.} of duality  and \textbf{5\hspace{0,5pt}a.} of twisted duality

\bigskip
\noindent \textbf{5\hspace{0,5pt}b.} {\sc Graded Duality.}
A $K$-graded local net $\cF_\cP$ on $\cP=\cI, \cD$, is said to satisfy \emph{$K$-graded duality} if for any $o\in \cP$ it holds $\cF_{l\cP}(o)=\cF_{l\cP}^{gd}(o)$. Equivalently for the subnet $\cF_{r\cP}$.\\ 

\noindent
Finally, we observe that for a net defined on a multiple connected spacetime, through an index set $\cP$ with $\cG^\bot_\cP$ having more then two connected components and covariant under the action of a finite, discrete geometric symmetry group replacing $S$, the notion of a grading may result useful to describe the commutation relations of quantum fields. We reserve returning to this elsewhere.
\subsection{Graded net obtained by crossed products}\label{ss:cross prod graded net}
The main example of graded nets is the case of grading by a crossed product. 
This plays a primary role both in Section \ref{s: net coom model rep}, to compute the net cohomology of graded net derived from Weyl algebras, and in Section \ref{s:completenes SW}, to study the properties of the Streater and Wilde model.

\smallskip
In some sense, our approach may be conceived as a net-oriented version of the category-oriented one of the braided crossed $G$-categories of M\"uger in \cite{mug05}: for a net having soliton automorphisms both of them describe the presence of superselection sectors of a fixed point net, i.e.\ the presence of \emph{twisted representations}, see also \cite{fabio3} and references therein. A more precise sketch on the relation between the two approaches is given at the end of this subsection. 

\smallskip
We begin by a trivial example of crossed product net, i.e.\ a net with a \emph{global grading}: we mean here that there exists a unique representation of an Abelian, discrete group, giving the grading for all the algebras of the net.
Suppose we are given with a local net of von Neumann algebras $\cN_\cP$ and a faithful, unitary representation $V$ of an Abelian, discrete group $K$ on the same Hilbert space $\cH$ of $\cN_\cP$. 

We define a net $\cF_\cP$ through  the implemented, discrete crossed product of $\cN_\cP$ by the adjoint action $\a:=\ad V$ of $K$ on $\cN_\cP$, i.e.\  $\cF(o):=\cN(o)\rtimes_\a V(K)$, for every $o\in \cP$. 

This net is represented on the Hilbert space $\cH\otimes \ell^2(K)$ and the homogeneous subspaces of the algebra $\cF(o)$ may be defined by 
\beq\label{e:product grad}
\cF(o)_k:=\cN(o)\, V(k)\,, \qquad o\in \cP, \quad k\in K\,.
\eeq
The net $\cN_\cP$ itself may be identified with $\cF_{\cP,e}$, the identity component of $\cF_\cP$. 
If the action $\a$ is $\cP$-local on $\cF_\cP$, i.e.\ $\a_k(\cF(o))=\cF(o)$ for any $o\in \cP$, it leaves invariant any local homogeneous subspace $\cF(o)_k$, i.e.\ for any $k,h\in K$ and $o\in \cP$,  $\a_k(\cF(o)_h)=\cF(o)_h$. 
The homogeneous subspace $\cF_{\cP,k}$ defined as in equation (\ref{e:product grad}), are easily characterized by the following 
\blemma\label{l: product grad}
In the notation of equation (\ref{e:product grad}), for every $A\in \cN(o)'$ and $k\in K$ we have
\beq \label{e:eq grading}
F\in \cF_\cP(o)_k \quad\Longleftrightarrow\quad F A=\a_k(A)F\,.
\eeq
\elemma
\prf
$(\Rightarrow)$ Because of the the definition in equation (\ref{e:product grad}), we may take $F=N V(k)$ with $N\in \cN(o)$. Hence, because of the local action of $\a=\ad V$, we have 
\beqn
F A= N V(k) A = N \a_k (A) V(k) = \a_k (A) N V(k) = \a_k (A) F\,.
\eeqn
$(\Leftarrow)$
If for $F\in \cF(o)$ and $A\in \cN(o)'$ it holds $F A=\a_k(A)F$, then  
\beqn
F V(k)^{-1}\a_k(A)=F A V(k)^{-1}=\a_k(A)F V(k)^{-1} \,,
\eeqn
that means, again because of the local automorphic action, $F V(k)^{-1}\in \a_k(\cN(o))=\cN(o)$, so that $F\in \cN(o)V(k)=\cF(o)_k$.  
\qed
Using this lemma, the global grading for a crossed product net $\cF_\cP$ is achieved by the following
\bprop\label{p: product grad1}
Let $\cN_\cP$ be a local net of von Neumann algebras represented on the Hilbert space $\cH$, with a $\cP$-local automorphic action of the Abelian discrete group $K$ implemented as $\a=\ad V$ for $V$ a faithful, unitary  representation of the group $K$ on $\cH$. If for every $o\in \cP$ there exists an invertible element $Y\in \cN(o)'$ and a faithful character $\c$ of $K$ such that $\a_k(Y)=\c(k) Y$, then $\cF_\cP(o):=\bigvee_{k\in K}\cF_\cP(o)_k$ is a (global) $K$-graded net, represented on the Hilbert space $\cH \otimes \ell^2(K)$, where any homogeneous subspace $\cF_\cP(o)_k$ is defined as in the equation (\ref{e:product grad}).
\eprop
\prf
The only non trivial thing to check is the condition b) of grading, for any von Neumann algebras $\cF(o)$, with $o\in \cP$.
Note that through the above lemma, $F\in \cF(o)_k$ is equivalent to $FA= \a_k(A)F$ for every $A\in \cN(o)'$, in particular $FA =AF$ if $k=e\in K$. Choosing $A=Y$ this implies $FY=\a_k(Y)F =\c(k)YF$ for $k\neq e$, and $FY=YF$ for $k=e$.  Hence $\big(1-\c(k)\big)F=0$ using $Y^{-1}$, that gives the result because of the faithfulness of $\c$.  
\qed
The above proposition may be generalized to the case of a non Abelian group $K$, replacing the invertible element $Y$ and the faithful character $\c$, by a $K$-invariant Hilbert space inside $\cN(o)'$, carrying  a faithful representation of $K$.
\footnote{\label{f: hilbert in vna}
We recall that an \emph{Hilbert space inside a von Neumann algebra} is a norm closed subspace $H$ with the property that if $\psi,\,\psi' \in H$ then $\psi^*\psi'\in \bC$. This is a scalar product where the norm of $H$ is induced by the von Neumann algebra. If $\psi^\a$ is an orthonormal bases of $H$, the \emph{support of $H$} is the projection $\sum_\a \psi^\a \psi^{\a*}$ and turns out to be independent of the chosen basis.
}

\smallskip
Notice that if the identity component net $\cN_\cP$ is local, the net $\cF_\cP$ is not local because of the non-trivial action $\a$ of $K$. If this action is trivial then $\cF_\cP\cong \cN_\cP \otimes V(K)''$. 

\smallskip
More interesting is the case of a \emph{grading on the causal complement}, that we discuss in the sequel, starting from the following observation. Suppose that in Lemma \ref{l: product grad} the net $\cN_\cP$ satisfies the Haag duality on $\cP$, i.e.\ explicitly
\beq\label{e:duality}
\cN_\cP (o)\,=\,\cN^d_\cP (o):=\bigcap \{\cN_\cP(o_i)',\, o_i\in o^\bot,\, o_i\in \cP\}\cap\cN^\cP(M),\qquad o \in \cP.
\eeq
Then, the condition (\ref{e:eq grading}) on the commutant $\cN(o)'$ of Lemma \ref{l: product grad} characterizing the grading, is actually required for the elements of the von Neumann algebras associated to the causal complement, being in this case $\cN (o)'= \bigvee \{\cN_\cP(o_i),\, o_i\in o^\bot,\, o_i\in \cP\}$, if $\cN_\cP$ is irreducibly represented.
Consider hence that for any $o\in \cP$ the causally disjoint set  $o^\bot$ (and hence $\cG_\cP^\bot$) has only one connected component, and that there exists a faithful, unitary representation $V_{o^\bot}$ of the Abelian discrete group $K$ implementing an action $\a_{o^\bot}:=\ad V_{o^\bot}$ on any  $\cN(o_1)$ with $o_1\in o^\bot$. 
In this case, similarly to  Lemma \ref{l: product grad} and Proposition  \ref{p: product grad1}, it remains defined the following $K$-graded net on $\cP$ represented on $\cH\otimes \ell^2(K)$ 
\beq\label{e:net graded}
\cF(o)\,:=\,\cN(o)\rtimes_{\a_{o^\bot}} V_{o^\bot}(K)\,. 
\eeq
The set of the representations as above, that we denote by  $\widehat{K}_{\bot,\cP}:=\big\{\widehat{K}_\bot(o)\big\}_{o\in\cP}$,  is easily proved to be a net on $\cP$.

\smallskip 
The causal complement of the elements of the index sets on the real line we introduced in Subsection \ref{ss: SW nets}, all have two connected components, hence the construction have to be given for both of them. These two connected components were denoted by $l$ (left) and $r$ (right), abstractly related by the space inversion map $S$ through $S\, l=r$ and generically indicated by $c=l,\,r$.  
Fixed  $o\in \cP$ and the connected component $c$ of $\cG_\cP^\bot$, for the corresponding connected component $o^\bot_c\subset o^\bot$ we give the following definition, remarking its strict relation with the category-oriented one \cite[Definition 2.6]{mug05}:
\bdfn\label{d:rapo K on comlem}
Let $\cN_\cP$ be a local net of von Neumann algebras on $\cP$, represented on the Hilbert space $\cH$.
For $K$ a (Abelian, discrete) group and for $o\in \cP$, we denote by $\widehat{K}_c(o)$ the set of the faithful unitary representations of $K$ on the Hilbert space $\cH$, such that a representation $V_{o^\bot_c}\in \widehat{K}_c(o)$ defines an action $\a_{o^\bot_c}$ of $K$, and for any $k\in K$ it holds 
\beq\label{e:rapo K on comlem}
\begin{array}{lcll} 
\a_{o^\bot_c}(k)(N) &=& \ad V_{o^\bot_c}(k)(N)\,, & \qquad  N\in \cN(o_1)\,, \quad o_1\in o^\bot_c\,, \\
\a_{o^\bot_c}(k)(N) &=& N\,, 											 & \qquad  N\in \cN(o_2)\,, \quad o_2\in o^\bot_{Sc}\,.
\end{array}
\eeq
The set $\widehat{K}_{c,\cP}:=\big\{\widehat{K}_c(o)\big\}_{o\in\cP}$ is called \emph{the set of the (strongly) faithful, unitary representations of $K$, acting on the algebras of the connected component $c$ of the causal complement of $o$}. 
\edfn
For any $o_1, o_2 \in \cP$ with $o_1\subseteq o_2$, it holds $\widehat{K}_c(o_1)\subseteq\widehat{K}_c(o_2)$, hence also $\widehat{K}_{c,\cP}$ seen as a map $\cP\ni o\mapsto \widehat{K}_c(o)$, is a net on $\cP$.
For bounded subset of the real line, as the elements in $\cI$ and $\cD$, this definition provides the two nets of representations $\widehat{K}_{l,\cP}$ and $\widehat{K}_{r,\cP}$, on both the causal complements respectively.
\footnote{
For the half-lines $J_c\in\cJ_c\subset \cJ$, the definition only provides the set of the representations $\widehat{K}_{Sc}(J_c)$.
The apparent asymmetry respect to the bounded case may be restored through the compactification of the real line. We however reserve to consider this case in \cite{cr}.
}

\smallskip
Being $K$ Abelian, we shall suppose that $\widehat{K}_{l,\cP}$, $\widehat{K}_{r,\cP}$ and the net they generate is local. 
The same may be required when $\cP$ has only one connected complement, i.e.\ for the net of representations $\widehat{K}_{\bot,\cP}$, and is automatically satisfied for a global grading, i.e.\ for the trivially Abelian net of representations $o\mapsto \widehat{K}_\cP(o)\equiv\widehat{K}$. 

\smallskip
Because of the triviality requirement of the adjoint action of the elements $\widehat{K}_c(o)$ on $o^\bot_{Sc}$ stated in equation (\ref{e:rapo K on comlem}), the above representations may be uniquely defined up to local perturbations in $\cN_\cP$, in fact it holds the following
\blemma\label{l:local grading}
Using the notation of the Definition \ref{d:rapo K on comlem}, it holds
\bdes
\item[i)]
if $\cN_\cP$ be a local net on the index set $\cP$, irreducibly represented on $\cH$, for given $o\in \cP$ and for any different $V_1,V_2\in \widehat{K}_c(o)$ and $k\in K$, it holds 
\beq\label{e:op grad-c}
V_1(k)\,V_2^*(k)\,\in \bigcap \{\cN_\cP(o_i)',\, o_i\in o^\bot_c,\, o_i\in \cP\}\bigcap
\big(\cN^\cP(M)\otimes \cB(\ell^2(K))\big)\,;
\eeq
\item[ii)]
if the net $\cN_\cP$ is irreducible represented and satisfies Haag duality on $\cP$, for any
$V_1,V_2\in \widehat{K}_c(o)$, it holds 
\beq\label{e:op grad}
V_1(k)\,V_2(k)^*\,\in \cN (o)\,;
\eeq 
\item[iii)]
the results of Lemma \ref{l: product grad} and Proposition  \ref{p: product grad1} hold, so that for any connected component $c$ and choice of representation $V_{o^\bot_c}\in \widehat{K}_c(o)$, the equation 
\beq\label{e:net graded c}
\cF_c(o)\,:=\,\cN(o)\rtimes_{\a_{o^\bot_c}} V_{o^\bot_c}(K) 
\eeq
defines a unique net on $\cP$, graded by the group $K$ and represented on $\cH\otimes \ell^2(K)$.
\edes
\elemma
\prf
i)
If $\cN_\cP$ if irreducibly represented on $\cH$ we have $\cN^\cP(M)\cong \cB(\cH)$. The rest is trivial from the definition, using easy calculations as in Lemma \ref{l: product grad}.\\
ii) 
for every $k\in K$ the operator $V_1(k)\,V_2^*$ acts trivially on $\ell^2(K)$, so that from the equation (\ref{e:op grad-c}) we have 
\beqn
V_1(k)\,V_2(k)^*\,\in \bigcap \{\cN_\cP(o_i)',\, o_i\in o^\bot_c,\, o_i\in \cP\}\bigcap \cN^\cP(M)\,.
\eeqn
Then, because of the triviality condition of equations  (\ref{e:rapo K on comlem}), for $V_1, V_2\in \widehat{K}_c(o)$ and for any $k\in K$ and $o_1\in o^\bot_{Sc}$, we have $V_1(k), V_2(k)\in \cN(o_1)'$. A fortiori, again using the condition of equations (\ref{e:rapo K on comlem}), the same holds for $V_1(k)V_2(k)^*$, hence $V_1(k) V_2(k)^*\in \cN_\cP(o^\bot_c)'\cap \cN_\cP(o^\bot_{Sc})'=\cN_\cP(o)$.\\
iii)
the proofs are unchanged, because of points i) and  ii) above.
\qed
\brem\label{r:obs grading}
1)
If $\cG_\cP^\bot$  has only one connected component, for the graded net defined in equation (\ref{e:net graded}) from the elements of $\widehat{K}_{\bot,\cP}$ omitting the triviality condition on one complement, the result in the equation (\ref{e:op grad}) also holds. A fortiori this holds for trivially and globally graded nets.\\ 
2) If the grading group $K$ is not Abelian, the representations of $K$ may be carried by Hilbert spaces with support identity, inside the crossed product algebras, so the symbol $\widehat{K}_{c,\cP}$ assume the meaning of a net of (finite) Hilbert space unitary, irreducible representations in this case. 
\erem
Hence, starting from an Haag dual net $\cN_\cP$ and choosing a representation $V_c\in\widehat{K}_c(o)$ for both $c=l,r$, the above lemma defines uniquely the left and the right $K$-graded nets; namely, for any $o\in \cP=\cI, \cD$ these are defined by
\beq\label{e:net grad o}
\cF_l(o)\,:=\,\cN(o)\rtimes_{\a_l} V_l(K)\qquad \textrm{and}\qquad \cF_r(o)\,:=\,\cN(o)\rtimes_{\a_r} V_r(K)\,.
\eeq
Through this definition of the left and right nets, we can treat the $K\times K$-graded nets introduced in Subsection \ref{ss:K graded net}. Considering the nets on the index sets $\cI$ and $\cD$ of the equations (\ref{e:net grad o}), it is easy to show that  
the nets $\cF_l$ and $\cF_r$ are homogeneously conjugated by the space inversion operator $S$ iff the identity grade subnet $\cN$ is $S$-covariant and the two nets of representations $\widehat{K}_{l,\cP}$ and $\widehat{K}_{r,\cP}$ are $S$-conjugated, up to elements in $\cN_\cP$ as in Lemma \ref{l:local grading}, with the components flipped by $S$. This respectively means that for every $o\in \cP$, it holds 
\beq\label{e:conj for l r}
\ad U(S)\big(\cN (o)\big)=\cN (So)\qquad \textrm{and}\qquad  \widehat{K}_{c,\cP}(o)=\widehat{K}_{Sc,\cP}(So)\,.
\eeq
In particular, the second equation means that for $V_c \in \widehat{K}_{c,\cP}(o)$ there exists $V_{Sc} \in \widehat{K}_{{Sc},\cP}(So)$ such that $ U(S) V_c(k)=V_{Sc}(k) U(S)$, for any $k\in K$. 

Using equation (\ref{e:k12 homog}) the two subnets $\cF_{l\cP}$ and $\cF_{r\cP}$ uniquely generate by product the net $\cF_\cP$, that results $K\times K$-graded and flipped by $S$ according to the Definition \ref{e:flipped graded}. If the net $\cN_\cP$ is represented on the Hilbert space $\cH$, then the net $\cF_\cP$ is defined on the Hilbert space $\cH_f\cong\cH\otimes\ell^2(K)\otimes \ell^2(K)$ and for any $o\in \cP$ it is explicitly given by 
\beq\label{e:bi graded net prod}
\cF_\cP(o)=\big(\cN_\cP(o)\rtimes_{\a_l} V_l(K)\big)\rtimes_{\a_r} V_r(K)\,.
\eeq
For a graded net as in equation (\ref{e:bi graded net prod}), explicitly obtained from the crossed product construction of Definition \ref{d:rapo K on comlem}, the peculiarities about the locality Property \textbf{3\hspace{0,5pt}b.} and the graded duality Property  \textbf{5\hspace{0,5pt}b.} are collected in the following result, using notations as in Definition \ref{d: ktwisted net}
\bprop\label{o:grad prod}
Let $\cF_\cP$ a net defined as in equation (\ref{e:bi graded net prod}) for $\cP=\cI, \cD$, with $\cN_\cP$ local, irreducibly represented and satisfying Haag duality on $\cI$ and the net $\widehat{K}_{l,\cP}$ local. Then 
\bdes
\item[i)]
for any $I_l,\,I\in \cI$ with $I_l<I$ we have $\cF_l(I_l)\bigcap\cF_l(I)'=\cN(I_l)^K$, where the fixed point subalgebra $\cN(I_l)^K$ is considered under the adjoint action of any element in $\widehat{K}_{l,\cI}(I)$;       
\item[ii)]
for any $I\in \cI$ it holds $\cF_l^g(I')= \bigvee_{I_l<I}\cN(I_l)^K \,\bigvee_{I<I_r}\cF_r(I_r)$ and 
\beqn
\cF_l^{gd}(I)= \bigcap_{I_l<I}\Big(\cN(I_l)^K\Big)' \,\bigcap_{I<I_r}\cF_r(I_r)'\bigcap \cF_l^\cI(\bR)\,. 
\eeqn
In particular, if the additivity of $\cF_\cI$ and $\cN_\cI^K$ on the half lines holds, we have 
\beqn
\cF_l^g(I')=\cN(J_l)^K\bigvee \cF_r(J_r)\,,\qquad \textrm{and}\qquad 
\cF_l^{gd}(I)=\big(\cN(J_l)^K\big)'\bigcap \cF_r(J_r)'\bigcap \cF_l^\cI(\bR)\,.
\eeqn
\item[iii)]
for any $E\in \cI_2$, it holds $\cF_l^g(E')= \bigvee_{I_l<I}\cN(I_l)^K \,\bigvee_{I<I_r}\cF_r(I_r)\bigvee \big(\bigcap_{I_0\bot I_3} \cF_l(I_0)'\big)$ and 
\beqn
\cF_l^{gd}(E)= \bigcap_{I_l<I}\Big(\cN(I_l)^K\Big)' \,\bigcap_{I<I_r}\cF_r(I_r)'\,\bigcap 
\big(\bigvee_{I_0\bot I_3} \cF_l(I_0)\big)\bigcap \cF_l^\cI(\bR)\,. 
\eeqn
In particular, if the additivity of $\cF_\cD$ and $\cN_\cD^K$ on the half lines holds, we have 
\beqan
&&\cF_l^g(E')=\cN(J_l)^K\bigvee \cF_r(J_r)\bigvee \cF_l(I'_3)'\,,\quad \textrm{and}\\ 
&&\cF_l^{gd}(E)=\big(\cN(J_l)^K\big)'\bigcap \cF_r(J_r)'\bigcap \cF_l(I'_3)\bigcap \cF_l^\cI(\bR)
=\cF_l^{gd}(I)\bigcap \cF_l(I'_3)\,.
\eeqan
\edes
\eprop
\prf
i) is obtained by a calculation on the generators of the cross product algebras of the relative commutant $\cF_l(I_l)\bigcap\cF_l(I)'$. In fact, for any elements $A\in \cN(I)$, $B\in \cN(I_l)$ and $V\in \widehat{K}_{l,\cI}(I)$, $W\in \widehat{K}_{l,\cI}(I_l)$ we have $AV(h)\,BW(k) =A\a_V(h)(B)V(h)W(k)=\a_V(h)(B)W(k)AV(h)$ for any $h,k\in K$, where the Definition \ref{d:rapo K on comlem} of the net  $\widehat{K}_{l,\cP}$ and its locality has been used. Observe that because of the Haag duality of $\cN_\cI$ and point ii) of Lemma \ref{l:local grading}, this construction do not depend on the representations used.\\
ii) and iii) come from a straight application of i) to the Definition \ref{d: ktwisted net}. 
\qed
Complying with the Remarks \ref{r:obs graded prop}, in the crossed product case the intersection defining the twisted algebras $\cF^g_l(I')$ and the graded dual algebras $\cF^{gd}_l(I)$ reduces on the left complement to the fixed point algebra of the identity subnet, under the action of the represented grading group. Whereas, on the right complement, through the cited minimal choice of the right algebras and according to the graded locality, just exclude the elements of $\widehat{K}_{l,\cI}(I_r)$ for $I<I_r$, that actually do not commute with $\cF_l(I)$.

\smallskip  
Similar definitions and results as in Proposition \ref{o:grad prod}, may be given for a trivially graded net, i.e.\ graded through $\widehat{K}_\cP\equiv\widehat{K}$, and for a globally graded net, i.e.\  graded through $\widehat{K}_{\bot, \cP}$. We shall refrain from giving a general approach here, but will treat an explicit example in the Streater and Wilde case, in the Proposition \ref{o:properties} and in equation (\ref{e: global  k duality}).

\smallskip
In the general construction of physical models, the representations $\widehat{K}_{c,\cP}$ 
may be implemented through a quantization procedure of smearing test functions, for example by the Weyl or loop group quantization.
In this case, we may choice test functions that define smooth partitions of the unity, having (derivatives with) support on the elements of the index set $\cP$, see Subsection \ref{ss:Other properties of fields subnets} for an explicit example. 

Focusing on the case $\cP=\cI$, a relevant property of this choice is the following: 
for any $I\in \cI$, there exist two representations $V_c\in\widehat{K}_c(I)$ for $c=l,r$, and a faithful unitary representation $V$ of $K$, acting by adjoint action on all the net $\cN_\cI$, such that it holds
\footnote{\label{e:strong of K}
In many cases, as the Streater and Wilde model below, the representation $V$ of $K$ may be strained to be a strongly continuous representation of $K$, albeit $V_r$ and $V_l$ are not.
}  
\beq\label{e:comm lr}
V_l(k)V_r(k)\,=\,V_r(k)V_l(k)\,=\,V(k)\,, \qquad k\in K\,.
\eeq 
Moreover, we may distinguish in notation the isomorphic discrete groups grading on the components $c$ and $Sc$ and identify the copy $K_{Sc}$ with the group $K$ itself, obtaining the obvious isomorphism of groups 
\beq\label{e:iso KK}
K_c\times K_{Sc}\longrightarrow K_c \times K \qquad \textrm{such that}\qquad (h, k)\longmapsto (hk^{-1},k)\in K_c\times K\,.
\eeq
Then, we obtain via the representation $V_c\times V$ and for any $\cF_\cI(I)$ in equation (\ref{e:bi graded net prod}), the isomorphism
\beq\label{e: Vin F}
\cF_\cI(I)\cong\cN_\cI(I)\rtimes_{\a_c}V_c(K)\rtimes_\a V(K)\,, \qquad I\in \cI\,.
\eeq 
We derive hence the following hints, that will be also a guide for the model in Section \ref{s:completenes SW}.

\smallskip
Let $\cF=\cF_\cI$ be a $K\times K$-graded net, with grading flipped by $S$, represented on the Hilbert space $\cH_f$ and defined as in equations (\ref{e:bi graded net prod}) and (\ref{e: Vin F}), such that there exists a compact symmetry gauge group $G$ acting on it, with $K\subset Z(G)$, the center of $G$. 

Suppose that $G$ is represented on $\cH_f$ by the strongly continuous extension of the representation $V$ of $K$ given in equation (\ref{e:comm lr}), and that it acts locally on $\cF$ by the adjoint action $\a$. We denote by $\cF^G$ the fixed point subnet of $\cF$ under the action $\ad V$ of $G$, defined by $\cF^G(I):=\cF(I)^G$; by $\cA$ the fixed point net of $\cF_c$ under the same action of $G$, i.e.\ $\cA(I)=\cF_c(I)^G$; and by $\cK:=V(K)''$ the von Neumann algebras generated by $K$ in the representation $V$ on $\cH_f$. Hence, for $c=l,r$ and $I\in \cI$, we have
\beqa\label{e: decomps fields}
\cF^G(I) &=&\big(\cN(I)\rtimes_{\a_c}V_c(K_c)\rtimes_\a V(K)\big)^G = \big(\cN(I)\rtimes_{\a_c}V_c(K_c)\big)^G\rtimes_\a V(K)\nonumber\\
				&=&\big(\cN(I)\rtimes_{\a_c}V_c(K_c)\big)^G\bigvee V(K)'':=\cF_c^G(I)\bigvee \cK =: \cA(I)\bigvee \cK\,.
\eeqa
Here has been supposed and used the fact that $K$ acts trivially on $\cF_c^G(I)=\big(\cN(I)\rtimes_{\a_c}V_c(K)\big)^G$, 
through the action $\a=\ad V$. 

\smallskip
We denote as usual by $\repb\cN_\cP$ the category of the representations of the net $\cN_\cP$, satisfying \emph{the DHR superselection criterion}, i.e.\ such that for $\p\in\repb\cN_\cP$ it holds $\p\rest o^\bot \cong \p_0\rest o^\bot$, where $\p_0$ is the defining (vacuum) representation and $\cong$ means unitary equivalence.

\smallskip
If the net $\cF_{c\cI}$ (and a fortiori $\cF_\cI$) is the \emph{minimal} net with \emph{full Hilbert spectrum} for the group $G$  (see \cite[Definition 7.3]{rob04}) i.e.\ in the terminology of the Hopf - von Neumann algebras of \cite{nt},  $\cF_c$ is given as the fixed point net $\cA$ acted on by the duals action $\b$ of the dual $\widehat{G}$, respectively dual to $\a$ and $G$, then for the DHR representations of the fixed point net, we have $\repb \cF_c^G \cong\widehat{G}$ and furthermore
\footnote{\label{f:charge cont}
If $\cF_c$ has full Hilbert $G$-spectrum, for both $c=l,r$, and we are interested in the \emph{charge contents} of $\cF$ as a fields net, i.e.\ in the superselection sectors of the net $\cA$ with gauge group $G$, we may use any of the subnet $\cF_c$.
}
\beqn
\cF_c(I)=\cA(I)\rtimes_\b \widehat{G}\qquad \textrm{and}\qquad \cF_c^G(I)=\big(\cA(I)\rtimes_\b \widehat{G}\big)^G\cong \cA(I)\,.
\eeqn
The grading group has a further role in the superselection sectors category of the theory, suggested comparing this approach with the (more general) one of the braided $G$-Categories in \cite{mug05}.

\smallskip
For this purpose, and with the same notations as above, we suppose that $\repb \cF_c^G$ is a rigid generally \emph{braided} tensor category, and the relation $\repb \cF_c^G \cong\widehat{G}$ holds only as a tensor category and not as a \emph{symmetric} one, i.e.\ on the same objects and morphisms but with a defined braid. 

\smallskip
For $K$ as in the equation (\ref{e:iso KK}), we may consider the $K$-category $K-\Loc \cN$ of the $K$-twisted representations of the net $\cN$, see for details the categorial-oriented definition \cite[Definition 2.6]{mug05} similar to the net-oriented Definition \ref{d:rapo K on comlem}. 
The identity grade subcategory of $K-\Loc \cN$ is then equivalent to the category $\repb \cN$. 

Denoted by $bK$ the Bohr compactification of $K$, similarly to \cite{mug05} we also have the following equivalence of braided (eventually symmetric) categories 
\beq\label{e:fix cat}
\repb\cN^{bK}\,\cong\, \big(K-\Loc \cN\big)^K\,.
\eeq 
Furthermore, if there exists a full symmetric tensor category $C\subset \repb \cN^{bK}$ such that $C$ is contained in the center $\cZ_2\big(\repb \cN^{bK}\big)$, a further result from \cite{mug05} is that $\repb \cN^{bK}\rtimes C$, \emph{the Galois extension of the braided tensor categories} $\repb \cN^{bK}$ by the category $C$, is a braided category, see \cite{mug00,mug04,mug05} for precise definitions and results. 

\smallskip
If we consider $C\cong\widehat{bK}$, i.e.\ the representations contained in the vacuum representation of the net $\cN$, then from the results in \cite{mug05} about finite or compact infinite group, we also derive the following equivalence of categories
\beqn
K-\Loc \cN\,\cong\,\repb \cN^{bK}\rtimes C\,.
\eeqn
That said, returning to the equation (\ref{e: decomps fields}), if $\cF_c^G=\cN^{bK}=\cA$ we also have
\beq\label{e:dual K}
\cF_c(I)\cong \big(\cA(I)\rtimes_\b C\big)\rtimes_\b\widehat{G}/ C \qquad \textrm{and}\qquad 
\cF_c^{bK} (I)=\cA(I)\rtimes_\b \widehat{G}/C\,.
\eeq
Hence we obtain that $\repb \cA\cong (K-\Loc \cN\big)^K\cong\widehat{G}/C\rtimes C$ and $\widehat{G}/C\cong K_c$, where $K_c$ is the discrete grading group as in the equations (\ref{e:iso KK}) and (\ref{e: decomps fields}). This means that $K_c$ represents the DHR representations of $\cA$, obtained as restriction of twisted representations of $\cN$, and  the subcategory $C$ instead is composed by the untwisted ones.

\smallskip
In the Section \ref{s:completenes SW}  we shall see that this is the case of the Streater and Wilde model, where $G\cong b\bR^2$ is the global, compact gauge group, described in \cite{fabio3}, and $\widehat{G}/C\rtimes C \cong \bR^2_d$ is the associated charge braided category, with symmetric subcategories $\widehat{G}/C$ and $C$ whose objects are labeled by $\bR_d$. 

\smallskip 
We end the section mentioning some examples of inversion covariant net, apart from the one treated in details in Section \ref{s:completenes SW}.
The Buchholz, Mack and Todorov field net of the $U(1)$-current on the light-line, before its compactification to the circle, see \cite{bmt88} and \cite[Subsection 4.2]{fabio3} for details, may be seen as a net with grading group $K=\bR$.
Moreover, some chiral models, for example the loop group model with value in a lattice, may be studied in a similar fashion, see \cite{fabio7}.

\smallskip
Finally, this approach will be used in a forthcoming paper (see \cite{cr}) to describe explicitly the topological nature of the sectors introduced by Brunetti and Ruzzi in \cite{br08} for quantum fields theories on the circle $S^1$, their relation with the $G$-category description of M\"uger in \cite{mug05} and the subfactors approach, for example in \cite{ckl08}.
\section{Net cohomology and Superselection Sectors}\label{s: net coom model rep}
In this section we briefly review the Roberts' theory of net cohomology presented in \cite{rob90, rob04}. 
This is the cohomology of a poset with coefficients in a net of von Neumann algebras. Its main applications is the characterization of the set of the DHR superselection sectors for a net of observables. In what follows we are interested in the problem of the completeness of a set of sectors for specific model, and present some generalizations of the net cohomology in this direction.

\smallskip
Recalling the definition of a simplicial set $\S_*(\cP)$ associated to a poset $\cP$ and other notations from Subsection \ref{ss:Nets}, it is possible to give the following 
\footnote{
Actually we consider here only $0,1-$order net cohomology. For higher-order cohomologies with non Abelian coefficients  see \cite{rob77}. 
} 
\bdfn\label{d:cocycle}(\cite{rob90}) 
A \emph{0-cocycle} in the net of von Neumann algebras $\cN_\cP$ over the index set $\cP$ is a map $z:\S_0(\cP)\to \cN_\cP$ satisfying the cocycle identity
\beqn
z(\partial_0 b) \,=\,z(\partial_1 b)\,,\qquad b\in \S_1(\cP)\,.
\eeqn 
and the locality condition 
\beqn
z(a)\in \cN_\cP(|a|)\,,\qquad a\in \S_0(\cP)\,.
\eeqn
A \emph{1-cocycle} in the net $\cN_\cP$ is a map $z:\S_1(\cP)\to \cU(\cN_\cP)$, the unitary group of  $\cN_\cP$, satisfying the cocycle identity
\beqn
z(\partial_0 c)\,z(\partial_2 c)\, =\,z(\partial_1 c)\,,\qquad c\in \S_2(\cP)
\eeqn 
and the locality condition 
\beqn
z(b)\in \cN_\cP(|b|)\,,\qquad b\in \S_1(\cP).
\eeqn
\edfn    
The 1-cocycles in the net $\cN_\cP$ are objects of a $W^*$-category denoted by $Z^1(\cN_\cP)$, where an arrow from the cocycle $z'$ to the cocycle $z$ is a map $w:\S_0\to \cN_\cP$ satisfying 
\beqn
z(b)\,w(\partial_1 b)\, =\,w(\partial_0 b)\,z'(b)\,,\qquad b\in \S_1(\cP)\,,
\eeqn 
and the locality condition
\beqn
w(a)\in \cN_\cP(|a|)\,,\qquad a\in \S_0(\cP)\,.
\eeqn
Let $Z_t^1(\cN_\cP^d)$ denotes the full subcategory  in $Z^1(\cN_\cP^d)$ of the 1-cocycles of $\cP$ in $\cN_\cP^d$ defined by the two further properties: 
\bitem
\item
any object $z\in Z_t^1(\cN_\cP^d)$ is trivial in $\cN^\cP$, i.e.\ for every $a\in \S_0(\cP)$ there exists a unitary $V_a\in \cN^\cP$ such that $z(b)=V_{\partial_0 b}V_{\partial_1 b}^*$, for $b\in\S_1(\cP)$; and
\item
for any path $p$ in $\cP$ we have $z(p)Az(p)^*=A$, if $A\in \cN_\cP(o)$ with $\partial_0 p,\, \partial_1 p \in |o|^\bot$.
\eitem
A first relevant result of net cohomology is the cohomological interpretation of the superselection sectors given in \cite[Theorem 26.2]{rob04}:
if $\cP$ is connected, the $W^*$-categories of the DHR representations $\repb\cN_\cP$ and $Z_t^1(\cN_\cP^d)$ are equivalent.

\smallskip
In the sequel we also use the result \cite[Theorem 30.1]{rob04} about superselection sectors and the change of the index set:
let $\cL\subset\cP$ be an inclusion of connected posets with a common causal disjointness relation $\bot$, and $\cN_{\cL}\subset \cN_\cP$ an inclusion of nets of von Neumann algebras obtained by restriction from $\cP$ to $\cL$, in the same reference representation.  Then, if the two generated von Neumann algebras associated to the nets are equal, i.e.\ $\cN^{\cL}=\cN^\cP$, restricting representations from $\cP$ to $\cL$ and acting identically on the intertwiners defines a faithful functor from $\repb\cN_\cP$ to $\repb\cN_{\cL}$. Moreover, if $\cN_\cP$ is additive and $\cL$ generates $\cP$, then this functor is an equivalence.
An analogous theorem  for the corresponding net 1-cohomology holds, see \cite[Theorem 30.2]{rob04}. 
\subsection{Computing net cohomologies}\label{ss:comp cohom}
The computation of the net $0$-cohomology is based on the connectedness of the index set and on the properties of locality and irreducibility of  $\cN_\cP$, see \cite[Section 3.4.4]{rob90}.

\smallskip
Computing the net 1-cohomology is usually based on a general \emph{referring model} presented in \cite[Section 3.4.5]{rob90}: here a net of von Neumann algebras on the (finite) index set associated to the poset of subsimplices of the standard $2$-simplex $\D_2$ is constructed. In this model, the subalgebras of each sub-simplex is supposed to be canonically isomorphic to the tensor product of the algebras associated with its faces. 
For a local net of von Neumann algebras $\cN_\cP$, the above condition for disjoint localized vertices of the 2-simplex, coincides with the so called \emph{quasi-split property}, see \cite{dl84} and references therein,  i.e.\  for any $o_1\bot o_2$ there exists a canonical isomorphism such that 
$\cN(o_1)\bigvee\cN(o_2)\cong\cN(o_1)\otimes \cN(o_2)$. 

\smallskip
If this strong structural property do not hold in a specific model, we cannot proceed  to discuss the local $1$-cohomology along these lines.
\footnote{\label{f:quasi split}
We recall that the validity of the (quasi-)split property for a net that is a fixed point net under the action of a global gauge group, may be lifted to the bigger net only if the group is \emph{finite} and Abelian, see for example \cite[Corollary 9.9]{dl84}.
}

\smallskip
As a first step, we hence generalize the toy model on $\D_2$ when the split (and quasi-split) property is lacking. Using an obvious notation, denote by $\D_2=c$ the standard 2-simplex, by $\partial_0 c,\,\partial_1 c,\,\partial_2 c$ its 1-subsimplices and by $\partial_{01} c,\,\partial_{02} c,\,\partial_{12} c$ its 0-subsimplices. 
We define a net $N=N_{\D_2}$ of von Neumann algebras on $c$ letting $N=N(|c|)=N^{\D_2}$ denote the (quasilocal) von Neumann algebra associated to the 2-simplex $c$. We also write $N_i$ and $N_{ij}$ for the von Neumann algebras  $N(|\partial_i c|)$ and $N(|\partial_{ij} c|)$ respectively. Moreover, for any 1-cocycle $z$ with values in $N$ we use the notation $z_i:= z_{\partial_i c}\in N_i$, for $c$ the standard 2-simplex.

\smallskip
Remembering the definition of a Hilbert space inside a von Neumann algebra, see footnote \ref{f: hilbert in vna},  we give the following
\blemma\label{l:cohotoy}
Let $N,\, N_{i}$, $N_{ij}$  and $z\in Z^1(N)$ as above. If  
\bdes
\item[a)]
$N_0=N_{01}\bigvee N_{02}$, $N_1=N_{01}\bigvee N_{12}$, $N_2=N_{02}\bigvee N_{12}$ and $N=N_{01}\bigvee N_{02}\bigvee N_{12}$, i.e.\ additivity on subsimplices;
\item[b)]
$N_{0}\bigwedge N_{1}=N_{01}$, $N_{0}\bigwedge N_{2}=N_{02}$, $N_{1}\bigwedge N_{2}=N_{12}$ and $N_{01}\bigwedge N_{02}=N_{01} \bigwedge N_{12} =N_{02} \bigwedge N_{12}=\bC$, i.e.\ minimality of the intersections on subsimplices;
\item[c)]
for any $0\neq z_2\in N_2$ there exists a conditional expectation $\F:N\to N_0$ such that $\F(z_2)\neq 0$,  $\F(N_2)\subset N_{02}$ and $\F(N_1)\subset N_{01}$,
\edes    
then  there are unique Hilbert spaces $H_{ij}\subset N_{ij}$ with support $I$ such that 
\beqn
z_2 H_{12}=H_{02},\quad z_1 H_{12}=H_{01}\quad \text{and}\quad z_0 H_{02}=H_{01}.
\eeqn
\elemma
\prf
The proof is very close to the one in \cite[Section 3.4.5, Lemma 1]{rob90}, to which we refer for details.
Because of points a) and b), the net $N$ is a lattice homomorphism from the lattice of the subsimplices of $\D_2$ into the lattices of the subalgebras of $N$. Setting $H_{12}:=\{n\in N_{12}: z_2 n \in N_{02}\}$, $H_{02}:=\{n\in N_{02}: z_0 n \in N_{01}\}$ and $H_{01}:=\{n\in N_{01}: z_1^* n \in N_{12}\}$ we define the norm-closed linear subspaces, mapped in the right way by the unitary operators of the 1-cocycle. The conditional expectation required in c) and the minimality of the intersection on subsimplices, assure that there are exist scalar products on the above linear subspaces, with supports equal $I$. The uniqueness of these Hilbert spaces follow from the construction.   
\qed
It follows that, as in the  Roberts' original formulation, there exist isometries $\psi^\a_{ij}\in N_{ij}$ such that $\S_\a \psi^\a_{ij} \psi^{\a*}_{ij} =I$ and 1-cocycle $z$ such that $z_2\psi^\a_{12}=\psi^\a_{02}$, $z_1\psi^\a_{12}=\psi^\a_{01}$ and $z_0\psi^\a_{02}=\psi^\a_{01}$. 
Hence, any $z\in Z^1(N)$ is a direct sum of trivial $1$-cocycles and the $1$-cohomology is said to be \emph{quasitrivial} in this case.

\smallskip
Secondly, we want to generalize the following result of Roberts, in the case of lacking of the (quasi-)split property and when  dealing with a net defined from a Weyl algebra, even if in a non-regular representation, i.e.\ on a non-separable Hilbert space: 
\bthm\label{th:robcoho trivial}(\cite[Theorem 2, Section 3.4.5]{rob90})
Let $\cN_\cP$ a local net of von Neumann algebras over the connected index set $\cP$, such that $\cG_\cP^\bot$ has only one connected component. If 
\bdes
\item[a)]
for every $b\in \S_1(\cP)$ and every path $p=\{b_1,b_2,\dots,b_n\}$ such that\\ $\partial p:=\{\partial_0 b_1,\partial_1 b_n\}=\{\partial_0 b,\partial_1 b\}=:\partial b$ we have 
\beq\label{e: aa)}
\bigcap_{\partial p =\partial b} \cN_\cP(|p|)\,=\,\cN_\cP(|\partial_0 b|)\bigvee \cN_\cP(|\partial_1 b|)\,;
\eeq 
\item[b)] for $o_1,o_2 \in \cP$ and $o_1\bot o_2$ it holds
\beq\label{e:qsplitN}
\cN_\cP(o_1)\,\bigvee\, \cN_\cP(o_2)\,\cong \,\cN_\cP(o_1)\,\otimes\, \cN_\cP (o_2)\,,
\eeq
\edes
then $Z^0(\cN_\cP)=\bC$ and for given any object $z\in Z^1(\cN_\cP)$ there are unique Hilbert spaces with support $I$, $H(a)\subset \cN_\cP(|a|)$ such that for $b\in \S_1(\cP)$ it holds $z(b)H(\partial_0 b)=H(\partial_1 b)$, i.e.\ $Z^1(\cN_\cP)$ is quasitrivial.
\ethm
The result of Lemma \ref{l:cohotoy}, about the  quasi-triviality on $\D_2$ without (quasi-)split property, is hence used to give the following  
\bprop\label{p:trivial coho}
Let $\cN_\cP$ a local net of von Neumann algebras over the connected index set $\cP$ such that $\cG_\cP^\bot$ has only one connected component. If 
\bdes
\item[a)]
condition (\ref{e: aa)}) holds; 
\item[b)]
for every $c\in \S_2(\cP)$ we have $\cN_\cP(\partial_0 \partial_1 c)= \cN_\cP(\partial_0 c)\bigwedge\cN_\cP(\partial_1 c)$, and if 
$\partial_0 c\bot \partial_1 c$ we have $\cN_\cP(\partial_0 c)\bigwedge\cN_\cP(\partial_1 c)=\bC$; 
\item[c)]
picked $c\in \S_2(\cP)$ such that $|\partial_{01}c|\bot|\partial_{02}c|$, $|\partial_{01}c|\bot|\partial_{12}c|$ and $|\partial_{02}c|\bot|\partial_{12}c|$
and setting $N_{ij}:=\cN_\cP(|\partial_{ij}c|)$ there exists a conditional expectation $\F$ as in point c) of Lemma \ref{l:cohotoy} above,  
\edes
then $Z^0(\cN_\cP)=\bC$ and $Z^1(\cN_\cP)$ is quasitrivial.
\eprop
\prf
The proof is obtained rephrasing the one in Roberts' Theorem \ref{th:robcoho trivial} and is based on Lemma \ref{l:cohotoy}. We give a sketch of it for completeness. 
The triviality of $Z^0(\cN_\cP)$ is directly obtained from the connectedness of the index set and condition b): chosen $b\in \S_1(\cP)$ such that $|\partial_0 b|\bot|\partial_1 b|$, for any $w\in Z^0(\cN_\cP)$ we have $w(\partial_0 b)=w(\partial_1 b)\in \cN_\cP(|\partial_0 b|)\cap \cN_\cP(|\partial_1 b|)=\bC$. 
Because of the $1$-cocycle identity, for every $z\in Z^1(\cN_\cP)$ we have $z(p)=z(b)$ if $\partial p = \partial b$. Thus $z(b)\in \cN_\cP(|\partial_0 b|)\bigvee \cN_\cP(|\partial_1 b|)$ because of a).
For given $a\in \S_0(\cP)$, pick $b\in \S_1(\cP)$ with $|a|=|\partial_0 b|\bot |\partial_1 b|$ and define
\beq\label{e:Hilbert space in vna}
H(a):=\{F \in \cN_\cP(|a|) : z(b)^*F \in \cN_\cP(|\partial_1 b|)\}.
\eeq
Choosing $c\in \S_2(\cP)$ with $\partial_0 c=b$ and $|b|=|\partial_0 c|\bot |\partial_1\partial_2 c|$ and using the Lemma \ref{l:cohotoy} with $N_{ij}=\cN_\cP(|\partial_i\partial_j c|)$, it is possible to show that $H(a)$ is a Hilbert space in $\cN(|a|)$, independently of $b$. This because of  b) and c) above.
Finally, for any vertex in a $2$-simplex with the above choice of $b\in \S_1(\cP)$ and $c\in \S_2(\cP)$, the definition as in equation (\ref{e:Hilbert space in vna}) gives an Hilbert space $H(\partial_i \partial_j c)\subset N_{ij}$, i.e.\ $H(a) \subset \cN(|a|)$ for $a\in \S_0(\cP)$, with $z(b) H(\partial_1 b) =H(\partial_0 b)$.
\qed
As noticed in \cite[Section 3.4.5]{rob90}, the hypothesis (\ref{e: aa)}) is the true model-dependent cohomological condition giving the triviality of $Z^1(\cN_\cP)$. 

\smallskip
The technical (quasi-)split property required for the net $\cN_\cP$ in the original Roberts' formulation is replaced by the conditions b) and c). The trivial intersection requirement  in point b) is  widely satisfied, for example if the net $\cN_\cP$ satisfies the Schlieder condition, see for example \cite{dan90}, or if the stronger Borchers or type III factor property holds. 

\smallskip
The condition b) of Proposition \ref{p:trivial coho} is tailored for local nets, but we need of a generalization to the case of graded nets, specializing to the case of grading by a crossed product as in Subsection \ref{ss:cross prod graded net}, that we give in the sequel.

\smallskip
Suppose that $K$ is a discrete Abelian group, and $\cF_\cP$ is a net on $\cP=\cI, \cD$, with $K\times K$-grading flipped by $S$, in the sense of Definition \ref{d:graddnetdef}, and satisfying $K$-graded locality, according to the equation 
(\ref{e:K graded locality}). 
Suppose moreover that there exists for every $o\in \cP$ and $s=l,r$ two representations $V_s\in\widehat{K}_s(o)$, so that $\cF_\cP$ is obtained through crossed products as in equations (\ref{e:bi graded net prod}), and a unitary faithful representation $V$ of the grading group $K$ such that also the equation (\ref{e:comm lr}) holds. 
Hence, according to the equation (\ref{e:net grad o}) we have $\cF_{s\cP}(o)=\cN_\cP(o)\rtimes_{\a_s} V_s(K)$, for $s=l,r$,  and because of (\ref{e: Vin F}) we also have $\cF_\cP=\cN_\cP\rtimes_{\a_l}V_l(K)\rtimes_\a V(K)$.
Recalling that we denoted by $\cK: = V(K)''$ the group von Neumann algebra of $K$ in the representation $V$, we may state the following
\bcor\label{c:cohograd} 
Let $\cF_\cP$ and $\cF_{s\cP}$, for $s=l,\,r$  as in the notation above. If 
\bdes
\item[a)]
condition (\ref{e: aa)}) holds for $\cF_{s\cP}$; 
\item[b)]
for every $c\in \S_2(\cP)$ we have $\cF(\partial_0 \partial_1 c)= \cF(\partial_0 c)\bigwedge\cF(\partial_1 c)$, and for $o_1\bot o_2$ it holds\\ $\cF(o_1\cap o_2)= \cF(o_1)\bigwedge \cF(o_2)= \cK$ and $\cF_s(o_1)\bigwedge \cF_s(o_2)= \bC$;
\item[c)]
condition c) of Proposition \ref{p:trivial coho} holds for the nets $\cF_s$,
\edes
then 
$Z^0(\cF_\cP)=\cK$, $Z^0(\cF_{l\cP})=Z^0(\cF_{r\cP})=\bC$, and $Z^1(\cF_l)$ and $Z^1(\cF_r)$ are quasitrivial.
\ecor 
\prf
The result for $Z^0 (\cF_\cP)$ is obtained from the condition b) as in Proposition \ref{p:trivial coho}; the same holds for 
$Z^0 (\cF_{s\cP})$ with $s=l,\,r$. Notice that if $\cF_{s\cP}$ satisfies the condition (\ref{e: aa)}), also $\cF_\cP$ does. 
Then it is possible to use the Proposition \ref{p:trivial coho} for the nets $\cF_{s\cP}$ obtaining the result.
\qed
\brem\label{r:z mod}
If in point b) of the previous Lemma \ref{l:cohotoy} the requirement $N_{01}\bigwedge N_{02} =\bC$ and similar is replaced by $N_{01}\bigwedge N_{02} =\cZ$ and similar, where $\cZ$ is a common (Abelian) von Neumann subalgebra of the $N_{ij}$, then the $1$-cocycles result to be associated to Hilbert right-$\cZ$-modules, instead of Hilbert spaces, i.e.\ $\psi^* \psi'\in \cZ$ for $\psi, \psi'\in H\subset N_{ij}$. 

The result of Proposition \ref{p:trivial coho} change consequently, giving in the $K\times K$-graded case of the Corollary \ref{c:cohograd}, $\cZ=\cK=V(K)''$.
Because this is not relevant for the model we are going to treat in details below, we refrain from further comments.
\erem
\subsection{Net cohomology results for a net of Weyl algebras}\label{ss: coho weyl}
In this subsection we show that a net of von Neumann algebras derived from a Weyl algebras (local, graded local and, in the general case,  non-regularly  represented), satisfies the hypothesis b) and c) of Proposition \ref{p:trivial coho} or Corollary \ref{c:cohograd}. 
The stronger hypothesis a) has instead to be verified for any specific model.

\smallskip
We refer to the notation of Sections 2.1 of \cite{fabio3}, recalling some general requirements and results about the twisted crossed product of Weyl algebras and their representations. 
Notice that the nets of the Streater and Wilde model fulfill these requirements, that may be easily adapted to the case of finite or denumerable sectors for a net derived from a Weyl algebra. Hence, furnishing detailed references to \cite{fabio3}, we consider and suppose to have:
\footnote{\label{f:dicho}
We remember however the dichotomy established in \cite{lx04} about conformal fields theory: if all the irreducible sectors of a model have a conjugate sector, then either the model is completely rational (and with a finite number of irreducible sectors) or it has uncountably many different irreducible sectors. 
}
\bitem
\item 
a net of von Neumann algebras $\cF_\cP$ on index set $\cP=\cI, \cD$, defined from a (non-regular) representation $\p_f$ of a Weyl algebra over a symplectic space $V_f=V_a\oplus L$, such that $L=C\oplus N$. Here, $C\cong N\cong \bR^s_d$ for some $s\in \bN$, is the (non-regular) symplectic subspace of (the sets of) additive charge, see \cite[Subsections 2.3, 3.2 and 4.1]{fabio3}. 
Notice that, any set of charge $c\in C$ is composed by $s\in \bR^s_d$ reals, representing charges also of different nature (as in the case of the Streater and Wilde model), and that $L, C$ and $N$ may be replaced by more general symplectic (sub-)spaces, for example  symplectic (sub-)spaces of functions;
\item 
for any $o\in \cP$, the local symplectic space decomposition of $V_f(o)$ in $V_a\oplus L$ depends only on $V_a$, i.e.\ $V_f(o)\subset V_a(o)\oplus L$. 
This means that,  defined for any $o\in \cP$ the algebras $\cB(o):= \p_b(\cW(V_a(o)\oplus N,\s_f))''$ with $\p_b=\p_f\rest \cW(V_a\oplus N)$, then the algebras $\cF(o):= \p_f(\cW(V_f(o),\s_f))''$ can be written as a crossed product
\beq\label{e:action sympl}
\cF(o)\,=\, \cB(o)\rtimes_{\s_L}\cU(C)\,.
\eeq
Here the crossed product is in general obtained by the symplectic form $\s_L$ as a twisted cross product of Weyl algebras, for a proper representation $\cU$ of $C$ on $\cH_b$, see \cite[Subsections 2.1 and 4.1]{fabio3}; 
\item
denote the GNS vector of $\p_f$ by $\O_f\cong\O_a\otimes \O_L= \O_a\otimes |0\rangle$, we suppose that it is cyclic and separating for every local algebras $\cF(o):=\p_f(\cW(V_f(o)))''$. We assume that $\O_a$ is cyclic and separating for the local algebras (of observables) $\cA_\cP (o):=\p_a(\cW(V_a(o)))''$ and that $\O_L$ is cyclic for $\p_L (\cW(L))''$, 
see \cite[Subsections 2.3 and 3.2]{fabio3}. 
The separating property of $\O_f$ for $\cF(o)$ do not implies the corresponding property of $\O_L$ for $\p_L (\cW(L))''$;
\item
there exist a map $c:F \mapsto c(F)\in C$ giving the charges of $F\in V_f$, such that if $c(F)=0$ then $F\in V_a\oplus 0\oplus N \cong V_a\oplus N$, and a second map $n:F \mapsto n(F)\in N$, so that we have $F=F_a\oplus c(F)\oplus n(F)\in V_a\oplus C\oplus N$. Both such maps result to be additive; and
\item
the (non-separable) Hilbert space $\cH_f$ of the representation $\p_f$ admits a decomposition in \emph{charge subspaces} for the subnet $\cA_\cP$. 
If $\cH_a$ is the separable Hilbert space the of representation for the net $\cA$ and $\cH_c \cong \cH_a$, this means that $\cH_f= \oplus_{c\in C}\cH_c\cong \cH_a \otimes \cH_L$. Here $\cH_L\cong \ell^2 (C)$, and we are actually identifying $\cH_c\cong \cH_a\otimes |c\rangle$, for every $c\in C$, see \cite[Subsections 3.2, 4.1 and 4.3]{fabio3}.
\footnote
{\label{f:prod rep}
As in \cite{fabio3}, by $\p_f \cong\p_a \otimes \p_L$ we mean the representation $\p_f$ for the decomposition of the Hilbert space $\cH_f$ as above and not a tensor product of representations of $\cF_\cP$. 
}
\eitem
Under these assumptions, we first show the existence of a conditional expectation for the algebras of the net $\cF$, as in point c) of Proposition \ref{p:trivial coho} and Corollary \ref{c:cohograd}. 
The existence of such a conditional expectation is obtained in the case of the (quasi-)split property of the net, through the Tomiyama slide map.
In the setting we are working, one of the difficulty is hence the non-separability of the Hilbert space of representation of the net $\cF$. 

\smallskip  
We begin defining for every (set of) charge $c\in C$, the following linear weakly closed local subspace, of weakly convergent series 
\beqn
\sF_c(o):=\{ \S_{i\in I}\, a_i \p_f(W(F_i)), \, a_i \in \bC,\, F_i\in V_f(o),\, c(F_i)=c\in C,\, o\in \cP\}^-.
\eeqn
Notice that, because of the definition of $\sF_c(o)$, any element $A\in\cF(o)$ is a weak-convergent net over finite subsets $\L\subset C$, where $A_c\in \sF_c(o)$, for $c(F_i^c)=c$, i.e.\
\beq\label{e:decompos aaa}
A={\rm w}-\lim_{\L\subset C}\S_{c\in\L}\,A_c\,=\, 
{\rm w}-\lim_{\L\subset C}\S_{c\in\L}\,\S_{i\in I}\, a_i^c\, \p_f(W(F_i^c))
\eeq
We shall use this shorthand sum notation to mean the convergence on finite subsets of $C$.
For the spaces $\sF_c(o)$ it holds 
\blemma\label{l:space}
With the assumptions and notations as above we have:
\bdes
\item[i)] 
$\sF_c(o)$ is  a weakly closed linear space and a von Neumann algebra iff $c=0\in C$. We call $\sF_0(o)$ the zero charge local von Neumann algebra on the index $o\in \cP$, it coincides with $\cB(o)$ and $\cA(o)=\sF_0(o)\rest \cH_a$; 
\item[ii)] 
the local von Neumann algebra $\cF(o)$ is generated by the disjoint sum of the subspaces $\sF_c(o)$, i.e.\ $\cF(o)=\bigvee_{c	\in C} \sF_c(o) = \big(\S_{c	\in C} \,\sF_c(o)\big)''$;
\item[iii)] 
if $\e\in \cH_{c'}$ then $(\sF_c(o)\e)^-=\cH_{c+c'}$. In particular, choosing  $\e=\O_f$, $(\sF_c(o) \O_f)^-=\cH_c$ and we have
\beqn
\sF_c(o)= \{A \in \cF(o):  A\O_f\in \cH_c\}\,;
\eeqn
\item[iv)] 
the restriction of the representation $\p_f$ to the subspace $\sF_c(o)$ is regular; more precisely for every $F\in V_f(o)$ and $F=F_a\oplus c(F)\oplus n(F) \in V_a\oplus C\oplus N$, the maps 
\beqn
\bR\ni\l\to \p_f(W(\l\,F_a\oplus c\oplus n(F)))\quad \textrm{and}\quad
\bR\ni\m\to \p_f(W(F_a\oplus c\oplus \m\, n(F)))
\eeqn
are weakly (and strongly) continuous;
\item[v)] 
the von Neumann algebra $\sF_0(o)$ is contained in any subspace $\sF_c(o)$ for every $c\in C$ and $o\in \cP$, i.e.\  $\sF_0(o)\O_f\subset \bigcap_{c\in C}\cH_c$. Moreover, for any chosen $F\in V_f(o)$ with $c(F)=c$, we have $\sF_c(o)=\sF_0(o)\, \p_f(W(F))$. 
\edes
\elemma  
\prf 
The easy proof of all the statements follows from the theory of non-regular representations, treated  in Sections 2.3 of \cite{fabio3}. In particular, the disjoint decomposition of the local algebras $\cF(o)$ in ii) follows from the non-regularity of the representation $\p_f$ that does not permit taking weak limits along the subspace $C$. The same is true for the representation $\p_L$, along the subspace $C$.
\qed
From the above assumed decomposition of the symplectic space, we define for every $o\in \cP$ the common von Neumann subalgebra $\cN:=\p_f(\cW(N))''\subset \cF(o)$. We suppose that the symplectic form $\s_f\rest \cN$ is trivial, so that $\cN$ is Abelian, and that $\cN\rest \cH_0=\bC I$. 
Assume that $\cN$ commutes with $\sF_0(o)$, for any $o\in \cP$, so that from the above assumptions it is possible to write 
\beq\label{e:decomp f0}
\cB(\cH_f)\supset\sF_0(o)=\cA(o)\bigvee \cN\cong\cA(o)\otimes \cN\subset \cB(\cH_a)\otimes \cB(\cH_L)\,,\qquad o\in \cP.
\eeq
Here we identified $\cN$ with $I_a\otimes\cN \subset \cB(\cH_a)\otimes \cB(\cH_L)$.
\footnote{Referring to the Streater and Wilde model (as formulated in \cite{fabio3} ad recalled below in Section \ref{s:completenes SW}), the field net $\cF$ defined from the symplectic space $V_f:=\cS\oplus\duS$ is not contemplated in this simplified situation; instead the net $\cC$, defined from the symplectic space $V_c:=\cS\oplus\duzS$, well suits the requirements above, see \cite[Subsection 3.1]{fabio3}. We shall return on these conditions discussing the local $1$-cohomology of the model, see also footnote \ref{f:diag}.
}

\smallskip
Assume that $\cA$ satisfies the split property, see \cite[Section 9]{rob04} for a review and details. Hence given $o_1\,,o_2\in \cP$ with $o_1\bot o_2$, there exists $\widetilde{o_1}\in \cP$ such that $o_1\subset \widetilde{o_1}$ and $\widetilde{o_1}\bot o_2$, and  also exists an intermediate type I factor $\cM$ such that $\cA(o_1)\subset \cM\subset\cA(o_2)'$. This entails that $\cA(o_2)\subset \cM'$ and we can even choose $\widetilde{o_1}\supset o_1$ such that $\cM\subset \cA (\widetilde{o_1})$. 
From the split property we deduce the quasi-split property, i.e.\ the existence of an isomorphism of von Neumann algebras $\Psi_a$ such that
\beq\label{e:qsplit}
\Psi_a\big(\cA(o_1)\bigvee \cA(o_2)\big) \,=\,\cA(o_1)\otimes \cA(o_2) \,.
\eeq
We extend now the isomorphism $\Psi_a$ to the tensor product of the local algebras of the net $\sF_0$, and some preliminary observations about the role of $\cN$ are in order. 
By the decomposition in equation (\ref{e:decomp f0}), the isomorphism $\Psi_a$ easily extends to an isomorphism of von Neumann algebras from $\sF_0(o_1)\bigvee \sF_0(o_2)$ to a \emph{central $\cN$-relative tensor product}, see \cite{sau80} for basic definitions, as
\beq\label{e:qsplitiso}
\Psi_0\big(\sF_0(o_1)\bigvee \sF_0(o_2)\big) \,=\,\sF_0(o_1)\otimes_\cN \sF_0(o_2) \,.
\eeq
Here we mean that the von Neumann algebra generated by the two commuting von Neumann algebras $\sF_0(o_1)$ and $\sF_0(o_2)$ is isomorphic to the von Neumann algebra tensor product $\sF_0(o_1)\otimes \sF_0(o_2)$ provided with the following property
\beq\label{e:relative tensor}
A_1N\otimes A_2 =A_1\otimes NA_2
\eeq 
for all $A_i\in \sF_0(o_i),\;i=1,2$ and $N\in \cN$.
If $\cN$ is in the center of any $\sF_0(o)$, i.e.\ $\cN\subset Z(\sF_0(o))$, this relative tensor product is central, and we have 
\beq\label{e:qsplitz}
NA_1\otimes A_2=A_1N\otimes A_2=A_1\otimes NA_2 =A_1\otimes A_2 N.
\eeq
For any $o,\, o_1,\,o_2\in \cP$ with $o_1,\,o_2 \subset o$, if we denote by $\sF_c(o_1)\bigvee \sF_{-c}(o_2):=\{\sF_c(o_1)\bigcup \sF_{-c}(o_2)\}^-$, we have $\sF_c(o_1)\bigvee \sF_{-c}(o_2) \subset \sF_0(o)$  and it is isomorphic to a $\cN$-relative tensor product but not central, i.e.\  the property described by the equation (\ref{e:qsplitz}) does not hold in this case.\smallskip

We prefer however not to follow this approach that use the relative tensor products (see also Remark \ref{r:z mod}), but switch to a more physical description, distinguishing two cases, both useful in the sequel: 
\bitem
\item 
the local algebras $\cF_\cP(o)$ commute with $\cN$, for every $o\in \cP$ (for instance if the net $\cF_\cP$ is purely Bosonic); or
\item 
the net $\cF_\cP$ is a net with $N\times N$-grading, flipped by the space inversion $S$, according to the Definition \ref{d:graddnetdef}, and satisfying $N$-graded locality, as in equation (\ref{e:K graded locality}).
\eitem
The first case is easier, and we treat it in the Lemma \ref{l:coditexp1} below. 

For the second case, we assume that the left and the right $N$-gradings of the net $\cF_\cP$ are obtained by crossed products, assumption already used in the general Corollary \ref{c:cohograd} and in view of the models we shall treat in Section \ref{s:completenes SW} constructed as twisted crossed product of Weyl algebras. 

Hence, in the notation of Subsection \ref{ss:cross prod graded net}, $N$ is a discrete additive group, replacing $K$ of that subsection, isomorphic to the Abelian Weyl subgroup $\cU(N)$ into the algebras $\cW(V_f)$, see \cite[Subsection 2.1]{fabio3} for details. According to the Definition \ref {d:rapo K on comlem}, the interesting representations of $N$ are the ones generating the nets $\widehat{N}_{s, \cP}$ for $s=l,r$ and, if the further property of equation (\ref{e:comm lr}) holds, we may identify the representation $\p_f\rest \cW(N)$ with $V:=V_l\cdot V_r$, for appropriate $V_s\in \widehat{N}_s(o)$, with $s=l,r$ and any $o\in \cP$. Finally, $\a:=\ad V$ is a (local) automorphic action of $N$ on \emph{all} the net $\cF_\cP$. 

\smallskip
On the other hand, we define the left $\cF_{l, \cP}$ and right $\cF_{r, \cP}$ subnets of $\cF_\cP$, requiring that there exist two different symplectic subspaces $V_{fl},\,V_{fr}\subsetneq V_f$ such that $V_f=V_{fl}\oplus N=V_{fr}\oplus N$, giving the algebras $\cF_s(o):= \p_f(\cW(V_{fs}(o),\s_f))''= \cB(o)\rtimes_{\a_s}V_s(N)$, and that the larger net is given by  $\cF(o)=\cF_s(o)\rtimes_{\a_{Ss}} V_{Ss}(N)$ for $s=l,\,r$. 
The adjoint actions of the group $N_s$ for $s=l,r$ via the representations $V_s\in \widehat{N}_{s \cP}(o)$ and of the group $N$ via $V$, are all defined by the symplectic form $\s_L$, similarly to the equation (\ref{e:action sympl}).

\smallskip
Turning to the existence of a conditional expectation for this case, we define the left/right fixed charge, weakly closed linear spaces by
\beqn
\sF_{c,s}(o):= \sF_c(o)\bigcap \cF_s(o), \qquad o \in \cP,\quad c\in C,\quad s=l,\,r\,. 
\eeqn
Observe that, depending on the nature of the charge $c$, if the group $N$ for $s=l$ or $s=r$ is equivalent to a subcategory of the local superselection sectors of $\cA$ (recall the comments after the equation (\ref{e:dual K}) distinguishing twisted and untwisted sectors), then the space $\sF_{c,s}(o)$ contains, i.e.\ is in general larger, than $\widehat{N}_s(o)$.
Moreover, because $N\nsubseteq V_{fs}$, through the representation $V$ we have $\cN\nsubseteq\cF_s(o)$ and $\cN\nsubseteq\sF_{c,s}(o)$ for all $o\in \cP$ and $c \in C$. From the $N$-graded locality of $\cF_\cP$, we hence obtain for any $o_1, o_2\in \cP$ with  $o_1<o_2$ 
\beq\label{e: propgradede}
\sF_{c,l}(o_1) \,\bot \,\sF_{q,r}(o_2)\qquad \textrm{for all}\quad c, q\in C\,.
\eeq
We may then state the following lemma and proposition about the existence of a useful conditional expectation, both for the local and graded-local case: 
\blemma\label{l:coditexp1} 
Suppose that  $\cF_\cP$ is a local net, additive on $\cP$, i.e.\ such that for $o,\,o_1,\, o_2\in \cP$ and $o=o_1\cup o_2$ it holds $\cF_\cP(o)=\cF_\cP(o_1)\bigvee\cF_\cP(o_2)$, that satisfies the above assumptions, where the net $\sF_0$ is quasi-split and 
\beq\label{e:condition}
\sF_0(o)=\S_{c\in C}\,\sF_c(o_1)\,\bigvee\,\sF_{-c}(o_2)\,,
\eeq 
where $\bigvee$ means the weakly continuously generated subspace, then
\bdes
\item[i)]
there exists an isomorphism $\Psi$ such that $\Psi\big(\sF_0 (o)\big)\,=\,\S_{c\in C}\,\big(\sF_c(o_1)\otimes \sF_{-c}(o_2)\big)$; 
\item[ii)] 
$\sF_c(o)\cong \S_{q\in C}\, \big(\sF_q(o_1)\otimes \sF_{c-q}(o_2)\big)$.
\edes
\elemma
\prf
Preliminary, we notice that for any $o=o_1\cup o_2\in \cP$ with $o_1\bot o_2$ and for every $c\in C$, we may choice two elements $F_c\in V_f(o_1)$ and $G_c\in V_f(o_2)$ with $c(F_c)=c$ and $c(G_c)=-c$ and define $W_c(1):=\p_f(W(F_c))$  and $W_{-c}(2):=\p_f(W(G_c))$.
Hence, from the result in point v) of Lemma \ref{l:space} and the decomposition of equation (\ref{e:condition}), any element $A\in\sF_0(o)$ is obtained by a weak-convergent net, taking a limit on finite subsets $\L\subset C$ such that $c\in \L$ iff $-c\in \L$, i.e.\ by  
\beq\label{e:decompos aa}
A={\rm w}-\lim_{\L\subset C}\S_{c\in\L}\,(\S_j \,A_c^j(1)A_{-c}^j(2))\,W_c(1)W_{-c}(2)\,.
\eeq
Choosing $\L\subset C$ closed under charge conjugation, i.e.\ under the transformation $c\mapsto -c$, allows the weak convergence to the element $A$, always remaining in $\sF_0(o)$.
Here $\S_c\S_j \,A_c^j(1)A_{-c}^j(2)$ is a weakly convergent series in $\sF_0(o)$ with  $A_c^j(i)\in \cA(o_i)$, $i=1,2$.
\\
i)
the result is obtained extending the isomorphism $\Psi_0$ in equation (\ref{e:qsplitiso}) from the subalgebra $\sF_0(o_1)\bigvee \sF_0(o_2)\subset \sF_0(o)$ to an isomorphism $\Psi$ from  \emph{all} the algebra $\sF_0(o)$. This is possible using the decomposition in fixed charge linear spaces and the split property for the local (observable) net $\cA$, extended to the net $\sF_{0\cP}$. 

From v) in Lemma \ref{l:space}, we can write the fixed charge subspaces on the single $o_1$ or $o_2\in \cP$ as $\sF_c(o_1)=\sF_0(o_1)W_c(1)$ and $\sF_{-c}(o_2)=\sF_0(o_2)W_{-c}(2)$. 
Moreover, to handle with the fixed charge subspaces, we recall that given $T\in V_f$ with charge $c(T)\neq 0$ 
it is possible to construct a symplectic space isomorphism $\psi_T$ that exponentiates to a Weyl and von Neumann algebras isomorphism $\Psi_T$ called  \emph{a regularizing isomorphism} and defined in \cite[Propositions 3.2 and 4.1]{fabio3}, to witch we refer for details. 

We define  $\psi_T(V_f(o))=\psi_T(V_a(o))\oplus L= \psi_T(V_a(o))\oplus(C\oplus N)$ and obtain the algebras isomorphism $\Psi_T$ applying the Weyl functor as in \cite[Remark 3.3]{fabio3}. Hence, for every $o\in \cP$ it holds
\beqn
\Psi_T (\cF(o))\subset \cB(\cH_a)\otimes\cB(\cH_L) \quad\textrm{and}\quad 
\Psi_T (\cA(o))=\cA(o)\otimes I_L\,.
\eeqn
Now we observe that the split inclusion for the net $\cA_\cP$ is written as $\cA(o_1)\subset\cM\subset \cA(o_2)'$ where $\cM$ a type $I$ factor; moreover $\cM'\subset \cA(o_1)'$ and $\cM\subset \cA(\widetilde{o_1})$ for $o_1\subset \widetilde{o_1}$ and $\widetilde{o_1}\bot o_2$. 

Being  the vector $\O_a$ cyclic and separating for the local algebras $\cA(o)$, we may extend the isomorphism $\Psi_T$ to $\cM'$; this is obtained identifying the factor $\cM$ with $\cM\otimes \bC I_L$ and defining $\Psi_T(\cM')\cong\cM'_a:=\cM'\otimes \bC I_L\subset \cB(\cH_a)\otimes\cB(\cH_L)$. This means that $\cM'_a$ is isomorphic, under $\Psi_T$, to the subalgebra of the elements in $\cB(\cH_f)$ acting as the commutant of $\cM$ on $\cH_0\cong \cH_a$ and as the scalars on $\cH_c$, for every $0\neq c\in C$.

Being  also $\cA(o_2)\cong \cA(o_2)\otimes \bC I_L\subset\cM'_a$, using the identification $\cN\cong I_a\otimes \cN \cong \Psi_T(\cN)$, we also have the following inclusions 
\beqan
&&\sF_c(o_1)=\sF_0(o_1)\,W_c(1)\,= \,\cA(o_1)\bigvee \cN \, \,W_c(1)\,\subset\, \cM \bigvee \cN\, \,W_c(1)\,,\\
&&\sF_{-c}(o_2)=\sF_0(o_2)\,W_{-c}(2)\,=\,\cA(o_2)\bigvee\cN\,\,W_{-c}(2)\,\subset\,\cM'_a \bigvee\cN\, W_{-c}(2)\,.
\eeqan
The last terms of the above inclusions become under the isomorphism $\Psi_T$,
\beq\label{e:formaiso}
\cM\otimes \cN \,\Psi_T(W_c(1)) \qquad \textrm{and}\qquad \cM'\otimes \cN\, \Psi_T(W_{-c}(2)).
\eeq
The proof proceed generalizing a quite general argument that allows to show the quasi-split property from the split property of a net, see for example \cite[Section 9]{rob04}. 

\smallskip
Let $E\in \cM$ be a minimal projection such that $E\O_f=\O_f$, so that we have $E=[\cM'_a\O_f]$ and we may set $E'=[\cM\O_f]\in \cM'_a$. 
Observe that $\cM \bigvee \cN \, W_c(1)\O_f= \cH_c$ and $\cM'_a\, W_{-c}(2) \O_f =\cH_{-c}$, $\O_f$ being  cyclic for $\cM\bigvee \cN$ and $\cM'\bigvee \cN$ and $\O_a$ cyclic and separating for $\cM$ in $\cH_0\cong\cH_a$. 
The weakly closed linear spaces in equation (\ref{e:formaiso}) commute elementwise. In fact, being the net $\cF$ local, from the tensor product form of equation (\ref{e:formaiso}), the only thing  really to prove is that $W_c(1)$ commutes with $\cM'_a$. We choose to this purpose an element $T\in V_f(o_1)$ such that $W_f(T)=W_c(1)$ and $\Psi_T(W_c(1))=W_a(1)\otimes W_L(c)$, where $W_a(1)\in \cA(o_1)$, that clearly commutes with $M'\otimes \l I_L$, for every $M'\in \cM'$ and $\l\in \bC$ because $\cA(o_1)'\supset \cM'$. 

Now we observe that the closed linear space spanned by the vectors of the form 
\beq\label{e: choice}
M N_1 M'N_2W_c(1)W_{-c}(2)\,\O_f\,,\qquad \textrm{with} \quad M\in \cM,\,M'\in \cM'_a \quad\textrm{and}\quad N_1,\,N_2\in \cN\,, 
\eeq
contains $\cM\bigvee\cM'_a\bigvee\cN\, W_c(1)W_{-c}(2)\,\O_f=\cH_0$.
We deduce that the map of dense subspaces 
\beq\label{e:iso space} 
MN_1M'N_2W_c(1)W_{-c}(2)\,\O_f \longmapsto MN_1W_c(1)\,\O_f \otimes M'N_2W_{-c}(2)\,\O_f
\eeq
extends to a isomorphism identifying $\cH_0$ with $\cH_c\otimes \cH_{-c}$.
The above results are sufficient to prove both that
\beqan 
&&\sF_c(o_1)\subset \cM\bigvee\cN\, W_c(1) \,\cong\, \S_{q\in C}\,\cB(\cH_q, \cH_{q+c}) \otimes I\,, \\
&&\sF_{-c}(o_2)\subset \cM'_a\bigvee\cN\, W_{-c}(2) \,\cong\, I\otimes\S_{q\in C}\,\cB(\cH_q, \cH_{q-c})\,.
\eeqan
Hence, the subspaces  $\sF_c(o_1)$ and $\sF_{-c}(o_2)$ generate a von Neumann algebra $\sF_{-c}(o_2)\bigvee \sF_{-c}(o_2)$ on the Hilbert space $\cH_0$. If we denote by $\Psi_0$ the isomorphism obtained from the equation (\ref{e:iso space}), we have that  $\Psi:= \Psi_T\circ\Psi_0$ realizes the immersion of $\sF_{-c}(o_2)\bigvee \sF_{-c}(o_2)$ in $\S_{q\in C}\,\cB(\cH_q, \cH_{q+c}) \otimes \cB(\cH_q, \cH_{q-c})$. Then, in any concrete representation of the net $\cF$ and because of the linearity of the spaces $\sF_c(o_i)$, we obtain the result summing up on finite subsets of $C$ and taking the weak limit.\\
ii)
Picking an element $F\in V_f(o_2)$ such that $c(F)=c$ and identifying  under the above isomorphism $\Psi_T$ the  element $\p_f(W(F))$ with the element $I\otimes \p_f(W(F))=:W_c$, we have from i) that 
\beqn
\sF_c(o) =\sF_0(o) W_c \cong\S_{q\in C}\,\sF_q(o_1)\otimes \sF_{-q}(o_2)\,(I\otimes W_c)
= \S_{q\in C}\, \sF_q(o_1) \otimes \sF_{c-q}(o_2).
\eeqn
\qed
Denoting by $\Psi$ also the isomorphism in point ii) of the above Lemma, we have the following 
\bprop\label{o:coditexp2} 
Under the hypothesis of Lemma \ref{l:coditexp1}, we also have 
\bdes
\item[i)] 
if $\cF_\cP$ is a local net, for every $o_1\bot o_2$ with $o_1,o_2\in \cP$, and $0\neq A\in \cF(o)=\cF(o_1)\bigvee \cF(o_2)$ there exists a conditional expectation $\F: \cF(o)\to \cF(o_1)$ such that $\F(A)\neq 0$; and  
\item[ii)]
if  $\cF_\cP$ is a net with $N\times N$-grading flipped by $S$ and $N$-graded local, for every $o_1,o_2\in \cP$  with $o_1< o_2$, the result of point i) holds.
\edes
\eprop
\prf
i) 
from equation (\ref{e:decompos aaa}), any element $A\in \cF(o)$ can be written as $A={\rm w}-\lim_{\L\subset C}\S_{c\in\L}\,A_c$. Here $A_c$ is an element in $\sF_c(o)$, and for $o=o_1\cup o_2$ and $o_1\bot o_2$, the isomorphism $\Psi$ in ii) of the above Lemma \ref{l:coditexp1}, maps $A_c$ as 
\beq\label{e:A_c}
A_c\mapsto \Psi(A_c)={\rm w}-\lim_{\L\subset C}\S_{q\in\L}\,\S_j A_q^{c,j}(1)\, W_q(1)\otimes A_{c-q}^{c,j}(2)\,W_{c-q}(2)\,.
\eeq
Hence, for the element $A$, decomposed by $A_q^{c,j}(i)\in \sF_q(o_i)$, $i=1,2$, we have  
\beq\label{e:decomcq}
A\mapsto \Psi(A)={\rm w}-\lim_{\L\subset C}\S_{c,q\in\L}\,\S_j A_q^{c,j}(1)\, W_q(1)\otimes A_{c-q}^{c,j}(2)\,W_{c-q}(2)\,.
\eeq
For any such an element $A$ as in equation (\ref{e:decomcq}) and for any state $\o\in \cF(o_2)_*$ the map $\F:\cF(o)\to \cF(o_1)$ defined by 
\beq\label{e:condexprea}
\F\big(\Psi(A)\big):={\rm w}-\lim_{\L\subset C}\S_{c,q\in\L}\,\S_j A_q^{c,j}(1)\, W_q(1)\otimes \o(A_{c-q}^{c,j}(2)W_{c-q}(2))
\eeq
is a normal conditional expectation on $\cF(o)$, relative to the charge decomposition, that turns out \emph{not} to be  weakly continuous along $C$.

The definition of the expectation $\F$ is consistent because the set of the product states on the tensor product of the local algebras $\cF(o_1)\otimes \cF(o_2)$ is total in the set of the normal functionals on any $\sF_q(o_1)\otimes \sF_{c-q}(o_2)$.
Hence, as in the case of the  Tomiyama slice map for spitting von Neumann algebras, it is always possible to choose the state $\o$ above, such that $\F(A)\neq 0$ if $A\neq 0$, because of the assumed cyclicity of $\O_f\in \cH_f$ for any $\cF(o)$.

We can easily verify this last claim on the dense set of the vector states, and for a simple non vanishing element $A\in \cF(o)$ with only $A_c(2)\neq 0$ for one $c\in C$. We mean here that 
\beqn
A={\rm w}-(\lim_{\L\subset C}\S_{q\in\L}\,\S_j A_q^j(1)\, W_q(1))\otimes\S_i\, A^i_c (2)W_c(2)\,.
\eeqn
Referring to notation in \cite[Subsection 2.3]{fabio3},  we can choose a vector $x	\otimes|0\rangle + y\otimes|c\rangle\in \cH_a\otimes \cH_L$ with $c\neq 0$ and call $\x\in \cH_f$ the vector it gives under the isomorphism from $\cH_a\otimes \cH_L$ to $\cH_f$, obtained from the regularization by the element $W_c(2)$. We can use now the vector state $\o_\x$ of $\x$ to define a conditional expectation as in equation (\ref{e:condexprea}). Then, evaluating it on the element $A\cong \Psi(A)$, we have  
\beqan
\F(\Psi(A))&=& {\rm w}-(\lim_{\L\subset C}\S_{q\in\L}\,\S_j A_q^j(1)\, W_q(1))\otimes\S_i\, \o_\x(A^i_c (2)W_c(2))\\
&=& {\rm w}-(\lim_{\L\subset C}\S_{q\in\L}\,\S_j A_q^j(1)\, W_q(1))\otimes\S_i\, (x,\,A^i_c (2) x)\langle 0|c\rangle +\\
&&\qquad+ (x,\,A^i_c (2) y)\langle 0|2c\rangle +(y,\,A^i_c (2) x)\langle c|c\rangle +(y,\,A^i_c (2) y)\langle c|2c\rangle)\\
&=& {\rm w}-(\lim_{\L\subset C}\S_{q\in\L}\,\S_j A_q^j(1)\, W_q(1))\otimes\S_i\, (y,\,A^i_c (2) x))\,.
\eeqan
As $x,y\in \cH_a$ can be freely chosen, the result follows.\\
ii)
If the net $\cF$ is $N$-graded local, it suffices to choose for $o_1<o_2$  the elements in equation (\ref{e: choice}) as $W_f(T)=W_c(1)\in \cF_l(o_1)$ and $W_{-c}(2)\in \cF_r(o_2)$. 
Then, the conditional expectation $\F$ is now constructed as above, but using the isomorphism $\Xi_{o_2}$ defined in Lemma \ref{l: stat net implement}. 
Hence, for $o_1<o_2$ and $c\in C$ we substitute  $\sF_c(o_1)$ by $\sF_{c,l}(o_1)$ and the subspaces $\sF_{-c}(o_2)$ by $\Xi_{o_2}(\sF_{-c,l}(o_2))=\sF_{-c,r}(o_2)$. 
Notice that, because of Remark \ref{r: xi and cofinality}, it suffices to give the proof for index elements $o_i\in \cI$, for $i=1,2$.
\qed
After these results on the existence of the necessary conditional expectations for a net defined from Weyl algebras, we obtain the following results about the minor hypothesis b) and c) of the Proposition \ref{p:trivial coho} and Corollary \ref{c:cohograd}:
\bprop\label{o:cohoweyl}
Let $o\mapsto \cF_\cP(o)=:\p_f(\cW(V(o)))''$ be a net of von Neumann algebras defined over the connected index set $\cP$, from the Weyl algebra of the symplectic space $(V_f,\s_f)$, where the representation $\p_f$ decomposes as $\p_f=\p_a\otimes\p_L$ (in the sense of footnote \ref{f:prod rep}). Then
\bdes
\item[i)] 
if the symplectic space is not degenerate and the net $\cF_\cP$ is a local net, then the conditions b) and c) of the Proposition \ref{p:trivial coho} hold; and 
\item[ii)] 
if $\cF_\cP=\cF_{s\cP}\rtimes_{\a_{Ss}}V_{Ss}(N)\cong\cF_{s\cP}\rtimes_{\a}V(N)$, with $s=l,r$ is a $N\times N$-graded net flipped by $S$ and $N$-graded local, then for the subnets $\cF_{s,\cP}$ obtained from the non degenerate symplectic subspaces $(V_s, \s_f)\subset(V_f,\s_f)$, the conditions b) and c) of Corollary  \ref{c:cohograd} hold.
\edes
\eprop
\prf
i) the first part of the cited condition b) holds because for a Weyl algebra net $\cF$ in a Fock regular representation $\o$, we have  $\cF(\bigwedge_j V_j)=\bigwedge_j \cF(V_j)$ for $V_j$ any closed real subspaces of the symplectic space $V$, see for example \cite{lrt}. If the net is given in a (non trivial) non-regular representation $\p_f$ with the stated properties, the same result holds because of the normality of this representation on the regular part. The non-degeneracy (i.e.\, the property that given $v,\,v'\in V$,  if $\s(v,v')= 0$  $\forall\, v'\in V$ then $v=0$) of the symplectic space assures that the intersection of local algebras with disjoint supports trivialize.
The existence of a conditional expectation required in the cited condition c), follows from i) in Proposition \ref{o:coditexp2}, choosing $c\in \S_2(\cP)$ such that $|\partial_{01}c|\bot|\partial_{02}c|, \, |\partial_{01}c|\bot|\partial_{12}c|$ and $|\partial_{02}c|\bot|\partial_{12}c|$,  such that 
\[
|\partial_{01}c|\cup|\partial_{02}c| = o_1\,\bot\,o_2=|\partial_{12}c|\,.
\]  
Hence we choose  the algebras $\cF_f(o_1)=\cF_f(|\partial_{01}c|)\bigvee \cF_f(|\partial_{02}c|)$ and  $\cF_f(o_2)=\cF_f(|\partial_{12}c|)$.\\
ii) 
The proof is similar to point i). We use ii) of Proposition \ref{o:coditexp2}, choosing the disjoint supports such that  $|\partial_{01}c|<|\partial_{02}c|<|\partial_{12}c|$. The algebras are hence defined, for $o_1=|\partial_{01}c|\cup|\partial_{02}c|<|\partial_{12}c|=o_2$, by $\cF_{lf}(o_1)=\cF_{lf}(|\partial_{01}c|)\bigvee \cF_{lf}(|\partial_{02}c|)$ and $\cF_{lf}(o_2)$. The presence of $\Xi_{o_2}$ in the definition of the conditional expectation $\F$, transforms this last algebra to $\cF_{rf}(o_2)$, so that  $\cF_{lf}(o_1) \bot\cF_{rf}(o_2)$. 
Also here it suffices to give the proof for the index elements $o_i\in \cI$ for $i=1,2$, because of the Remark \ref{r: xi and cofinality}.
\qed
\section{Sectors for the Streater and Wilde model}
\label{s:completenes SW}
In this section we discuss the question of the completeness of the superselection sectors of the Streater and Wilde model presented in 
\cite{fabio3}, via the net $1$-cohomology theory of the index sets $\cI$ and $\cD$ with values in the graded net of the fields, of witch we shall establish the quasi triviality and study the concerns for different fixed-point nets. 

\smallskip
For this purpose, we use notation and results of the previous sections and those established in \cite{fabio3}. 
In this paper, we denoted by $\cS$ the Schwartz space of real-valued rapidly decreasing test functions on $\bR$; by $\dS:=\{ f \in \cS : f=\partial g ,\  g \in \cS\}$ its subspace of functions with vanishing  Fourier transforms at zero momentum, i.e.\ with vanishing integrals on $\bR$; and by $\duS:=\{ f \in C^\infty (\bR) :\, \partial f \in \cS \}$ the space of functions having derivatives in $\cS$, in particular by $\duzS$ and $\duqS$ respectively the elements  $f\in \duS$ such that $\lim_{x\to -\infty} f(x)=\pm\lim_{x\to +\infty} f(x)$. 

Through these functions spaces we defined in \cite[Subsection 3.1]{fabio3} the time-zero data of the model. For example the element $F=\fzu \in \cS\oplus \duS=:V_f$, and using the symplectic form $\s_f (F, G) = \int_\bR \, (f_0 g_1 - f_1 g_0 )\, dx$ we obtain the symplectic spaces of the model
\begin{small}
\beq\label{e:def inclusion symplectic}
\barr{ccccc}
V_a:=\dS\oplus\cS&\subset&V_b:=\dS\oplus\duzS&\subset&V_c:=\cS\oplus\duzS\\\\
 \cap&&\cap & & \cap \\\\
 V_q:=\dS\oplus\duqS &\subset&V_e:=\dS\oplus\duS&\subset&V_f:=\cS\oplus\duS\,.
\earr
\eeq
\end{small}
The local structure of these spaces comes from the definition of the \emph{localization} of an element on $\cP$: 
\beq\label{e:localization}
F=\fzu \in V_f(o) \quad \textrm{iff}\quad \loc F:=\supp f_0 \cup \supp \partial f_1\, \subseteq o\in \cP\,.
\eeq 
A full description of the known superselection sectors of the local net of observable algebras $\cA_\cI$, defined on the open intervals index set $\cI$, has been given in terms of some auxiliary nets $\cB_\cI,\,\cC_\cI,\,\cQ_\cI$ and $\cE_\cI$ and of a larger (putative) field net $\cF_\cI$. These nets of von Neumann algebras are obtained from the respective symplectic spaces recalled in (\ref{e:def inclusion symplectic}), in the standard way of equation (\ref{e:nets von weyl}). 
We only remind from \cite[Subsections 3.1 and 3.2]{fabio3} that the quasi-free state that defines the Fock (vacuum) representation $\p_a$ of the Weyl algebras $\cW(V_a,\s_a)$ is given, for $F=\fzu\in V_a$ and $\o=|p|$ the module of the energy-momentum $p$, by 
\beqn\label{e:Fock state}
(\O_a,W_a(F)\O_a)\,=\,\o_a (W_a(F)):=\exp\Big\{\,- \frac{1}{4}\,\int\,( \o^{-1} |\tilde{f}_0|^2 + \o |\tilde{f}_1 |^2 )\,dp \,\Big\}\,,
\eeqn
and the non regular representation $\p_f$ of $\cW(V_f,\s_f)$ is obtained as the GNS representation associated to the non regular state
\beqn\label{e:ams state}
(\O_f,W_f(F)\O_f)\,=\,\o (W_f(F)) := \exp \Big\{- \frac{1}{4} \, q(F) \Big\}, \qquad F \in V_f\,,
\eeqn
where $q(F)$ is a so called \emph{generalized quadratic form}, defined by
\beqn\label{e: g.quadratic}
q(F) = \left\{ \barr{ll}
\int \,( \o^{-1} |\tilde{f}_0|^2 + \o |\tilde{f}_1 |^2 )\, dp \qquad &\textrm{if}\quad F=\fzu \in V_a \,,\\
+\infty \qquad &\textrm{if}\quad F \notin V_a\,.
\earr
\right.
\eeqn
The following diagram presents the von Neumann algebras nets of the model, together with the actions of the gauge and charge groups, see \cite[Subsection 4.4]{fabio3} for details
\begin{small}
\beq\label{e:SW inclusion of net}
\barr{ccccc}
&&\cA\otimes \Zb&&\cB\rtimes \cU(C)\\
&&\Vert &&\Vert\\
 \cA=\cQ^{\cG_q}& \subset & \cB=\cE^{\cG_q}=\cC^{\cG_c}=\cF^\cG&\subset&\cC=\cF^{\cG_q}\\\\
 		\cap	&& \cap& & \cap \\\\
		\cQ&\subset&\cE=\cF^{\cG_c}& \subset&\cF\\
\Vert&&\Vert&&\Vert \\
\cA\rtimes \cU(Q)&&\cB\rtimes\cU(Q)&&\cB\rtimes\cU(Q)\rtimes\cU(C)\,.
\earr
\eeq
\end{small}

\noindent
The net $\cB_\cI$ is the fixed point subnet of $\cF_\cI$ under the action of a global compact group of gauge symmetries $\cG=\cG_c\times \cG_q$, isomorphic to the Bohr compactification of $\bR^2$, and whose dual additive group $\dG=C\oplus Q\cong \bR_d\oplus\bR_d$, in the discrete topology, is the charge group of the sectors of  $\cA_\cI$. 

\smallskip
Moreover we define the Abelian discrete group $N:=\duzS /\cS \cong \bR_d$ that has been shown in \cite[Subsection 3.1]{fabio3} to be a subnet of $\cG_c$. The group Abelian von Neumann algebras of $N$ in representation $\p_f$ is $\cZ_b:=\p_f(\cW(N))''$ and result to be the center of $\cB(I)$, for every $I\in \cI$.

\smallskip
We also recall that, denoted by $\cH_L:=\ell^2(C)\cong \ell^2(\bR_d)\cong \ell^2(Q)=:\cH_M$, the Hilbert representation spaces of the nets are: $\cH_a,\,\cH_b\cong\cH_c\cong\cH_a\otimes \cH_L,\,\cH_q\cong\cH_e\cong\cH_a\otimes \cH_M$ and $\cH_f\cong\cH_a\otimes \cH_L\otimes\cH_M$, see e.g.\ \cite[Proposition 3.5]{fabio3}. All of them \emph{but} $\cH_a$ of $\cA_\cI$, are non-separable and carry a non-regular representation of the respective Weyl algebras defining the nets.
\subsection{More properties for the Streater and Wilde model}
\label{ss:Other properties of fields subnets}
In order to discuss the triviality of the net 1-cohomology along the lines previously indicated, we have to understand better the structural properties of all the nets on $\cI$ recalled in diagram (\ref{e:SW inclusion of net}), and their extensions on $\cD=\cI\cup \cI_2$. 

\smallskip
The definition of the nets on $\cD$ need some attention: in fact, as an example of great relevance, the net $\cA_\cD$ is given in the representation $\p_a$ by 
\beq\label{e:nets von weylA}
 \cA_\cD :D\longmapsto \cA_\cD(D):= \p_a (\cW (V_a(D),\s_a))'', \qquad D \in \cD\,,
\eeq
but it is easy to show that for any proper double interval $E=I_1\cup I_2\in \cD$ we have 
\beq\label{e:easy inclusion}
\cA_\cD^\cI(E):= \cA_\cI(I_1)\,\bigvee \, \cA_\cI(I_2)\,\subsetneq\,\cA_D(E)\,,
\eeq
where the algebra $\cA_\cD^\cI(E)$ is defined by additivity as in equation (\ref{e:N(o)}).
This simple specimen, is based on the following observation: given functions $f^1\in \cS(I_1)$ and $f^2\in \cS(I_2)$ such that $\int_\bR f^1\,dx=-\int_\bR f^2\,dx \neq 0$, then $f=f^1+f^2\in\dS(E)$, i.e.\ $\int_\bR f\,dx=0$, but $f^1\notin\dS(I_1)$ and $f^2\notin\dS(I_2)$.\smallskip

\smallskip
As another example, and to fix notation, the net $\cF_\cI$ is easily shown to have a $N\times N$-grading flipped by $S$ and to be $N$-graded local. In fact, the  net of symplectic subspaces $\cI\ni I\mapsto V_{f\cI}(I)$, has left subnet defined by 
\beq\label{e:Vfl}
V_{fl}(I):=\{F=\fzu\in V_f(I): F_{+\infty}:= \lim_{x\to+\infty}f_1(x)=0\}\,,
\eeq
and for any $n\in N$, it has $n$-homogeneous left subspaces, defined at $-\infty$ by 
\beq\label{e:homo left Vf}
V_{fl}(I)_n:=\{F=\fzu\in V_{fl}(I): F_{-\infty}:= \lim_{x\to-\infty}f_1(x)=n\in N\}\,.
\eeq
These subsets are linear subsets iff $n=0$ and moreover $V_{fl}(I)=\bigcup_{n\in N} V_{fl}(I)_n$,  $V_{fl}(I)_n \cap V_{fl}(I)_m = \emptyset$ if $n\neq m$ and $V_{fl}(I)_n \bigcup V_{fl}(I)_m = V_{fl}(I)_{n+m}$, where the last subspace is defined as in equation (\ref{e:homo left Vf}), for $n+m\in N$. 
This suffices to show that $V_{fl,\cI}$ is a  symplectic spaces net graded by $N$ at $-\infty$. 
Replacing  $+\infty$ by $-\infty$, we define the symplectic spaces net $V_{fr,\cI}$, with  $N$-grading at $+\infty$. Notice that the definition of the right subnet may be equivalently obtained using the action the space inversion map $S$ on the symplectic space, i.e.\ through the automorphism defined in Remark \ref{r:sympl xi}.
These two subnets generate $V_f$ as a net with $N\times N$-grading flipped by $S$, similarly to Definition \ref{d:graddnetdef}. 

From the symplectic homogeneous  subspaces, the homogeneous space of von Neumann algebras are defined for the component $s=l,r$, by $\cF_s(I)_n:=\p_f\big(\cW(V_{fs}(I)_n)\big)''$, so that it holds 
\beqn
\cF_s(I)\,=\,\bigvee_{n\in N}\cF_s(I)_n\,=\,\p_f\big(\cW(V_{fs}(I))\big)''\,.
\eeqn
This nets graded by $N$, generates $\cF_\cI$ as a von Neumann algebras net with  $N\times N$-grading, flipped by $S$, as in Definition \ref{d:graddnetdef}, that results to be $N$-local in the sense of the equation (\ref{e:K graded locality}) by a easy calculation on the symplectic form.

\smallskip
The $N$-grading of the $s=l,r$ algebras $\cF_s(I)$, is equivalently obtained through the crossed products of an identity subnet we are going to construct $\cF_{\cI,0}\subsetneq\cF_{s,\cI}$, and by the left and right representations of $N$, i.e.\ respectively of the groups $N_l\cong N_r\cong N$, as in Subsection \ref{ss:cross prod graded net}, in particular equation (\ref{e:iso KK}). 

In what follows we analyze this and some related constructions in details, also to exhibit a concrete example of  trivially and globally graded net, introduced in Subsection \ref{ss:cross prod graded net}, see in particular Proposition \ref{p: product grad1}, and its specific locality and duality features.

\smallskip
The net $\cB_\cI=\cA_\cI\otimes \cZ_b$ is obviously a trivially graded net, being $n\mapsto V(n):=\p_f\big(W (0,n)\big)$ a faithful, strongly continuous (because $N$ sets in a regular symplectic subspace for $\p_f$, see \cite[Proposition 3.5]{fabio3}), unitary global representation of the discrete group $N$ on the Hilbert space $\cH_f$, such that $\cZ_b=V(N)''$, see to the diagram (\ref{e:SW inclusion of net}).

\smallskip
The net $\cF_{\cI,0}$ is defined in representation $\p_b$, starting from the symplectic subspaces obtained from equation (\ref{e:homo left Vf}) at $n=0$, i.e.\, for any $I\in \cI$ by  
\beq
V_f(I)_0\,:=\,\{F=\fzu\in V_f(I): F_{\pm\infty}= \lim_{x\to\pm\infty}f_1(x)=0\}\,=\, \cS(I)\oplus \cS(I)\,.
\eeq
Notice that the net $\cF_{\cI,0}$ is a local net, represented on the Hilbert space $\cH_b=\cH_a \otimes \cH_L\cong \cH_a \otimes \ell^2(C)$ and, in the Hilbert space $\cH_f$, it holds the following inclusion of net $\cA_\cI\subsetneq\cF_{\cI,0}\subsetneq \cC_\cI$, and $\cB_\cI\nsubseteq \cF_{\cI,0}$ nor $\cB_\cI\nsupseteq\cF_{\cI,0}$. 
We shall see in point ii) of Proposition \ref{o:properties} below, that the net $\cF_{\cI,0}$ satisfies the Haag duality on $\cI$, hence may be used to define a globally graded net up to elements in $\cF_{\cI,0}$, as the net $\cN_\cI$ in 
Lemma \ref{l:local grading},see also point 1) of Remark \ref{r:obs grading}. 

\smallskip
Namely we have $\cC_\cI=\cF_{\cI,0}\rtimes_\a V(N)$, i.e.\ the net $\cC_\cI$ in the diagram (\ref{e:SW inclusion of net}), is obtained as the discrete crossed product of $\cF_{\cI,0}$ by the global action $\a:=\ad V$ of the group $N$, hence it is globally $N$-graded, according to  equation (\ref{e:product grad}) and the Proposition \ref{p: product grad1}, and represented on the Hilbert space $\cH_f$. 

\smallskip
The $N$-locality up to elements of $V(N)''$ of $\cC_\cI$ is given by definition, being $\cF_{\cI,0}$ local and $V(N)$ are the elements that break the locality of $\cC_\cI$. 

Similarly, to Proposition \ref{o:grad prod}, with the same notations and using the additivity on the net on the intervals respect to the half lines, we may treat the global graded duality: for the globally $N$-graded net $\cC_\cI$ and for any $I\in \cI$ we have $\cC^g(I')=\bigvee_{I_0\bot I}\cC(I_0)^N$ according to Definition \ref{d: ktwisted net}, and hence
\beq\label{e: global  k duality}
\cC^g(I')=\bigvee_{I_0\bot I}\cB(I_0)= \cB(J_l)\bigvee\cB(J_r)=
\p_c\big(\cW(\dS(J_l)\bigcup\dS(J_r)\oplus \duzS(I'))\big)''.
\eeq
This requires that  $\cC^{gd}(I) := \cC^{g}(I')'\bigcap \cC^\cI(\bR)$ if generated by the Weyl elements with function $G=\gzu\in V_c$ such that $\s(G,F)=0$ for any $F=\fzu\in V_b(I')=\dS(I')\oplus \cS(I')$. This entails $\int f_0 g_1\,dx =0=\int f_1 g_0\,dx$ hence 
$G\in \cS(I)\oplus\duzS(I)=V_c(I)$, i.e. the result (notice that the constant values of $g_1$ on $I'_l$ and $I'_r$ have to be equal, because of the intersection with $\cC^\cI(\bR)$ in the definition).

\smallskip
We may pass now to examine the grading on the two connected components of the net $\cF_\cI$.
To define the left and right gradings, it is  possible to choose for any $I\in\cI$, two elements $F_l=(0,f_{1l})\in V_{fl}(I)$ and $F_r=(0,f_{1r})\in V_{fr}(I)$ such that the couple $f_{1l},f_{1r}$ form a partition of the unity with derivative supported on $I$, i.e.\ $f_{1l}(x)+f_{1r}(x)\equiv 1$. This entails that for $s=l,r$ it holds $f_{1s}\rest I^\bot_s=1$ and $f_{1s}\rest I^\bot_{Ss}= 0$, and we obtain the following representations of $N_s\cong N$, in the net of representations $\widehat{N}_s(I)$ giving the grading:
\beq\label{e:costruz rapo complem N}
V_s:N_s \longrightarrow \cU(\cH_f)\, \qquad \textrm{such that} \qquad n\longmapsto V_s(n):=\p_f\big(W(n\, F_s)\big)\,. 
\eeq
These representations are not strongly continuous, because the elements $F_s$ sit in a non regular symplectic subspace for $\p_f$, see \cite[Proposition 3.5]{fabio3}.
The results of Lemma \ref{l:local grading} are fulfilled and the left and right nets $\cF_{s,\cI}$ are uniquely defined for $s=l,r$ through the equation (\ref{e:net graded c}). These two nets are homogeneously conjugated by the space inversion operator $S$, see equation 
(\ref{e:S action}), so the net they generate is $\cF_\cI$ as in equation (\ref{e:bi graded net prod}). Equivalently, the identity subnet $\cF_{\cI,0}$ is $S$-covariant and for $s=l,r$, the two nets of representations $\widehat{N}_{s,\cP}$ are local, $S$-conjugated up to elements in $\cF_{\cI,0}$ as in Lemma \ref{l:local grading}, with the components flipped by $S$, 
see equation (\ref{e:conj for l r}).

\smallskip
Moreover, because of the above choice of a partition of unity, the equation (\ref{e:comm lr}) holds, so that for any $n\in N$, we have  $V(n)=V_l(n)V_r(n)= \p_f\big(W((0,n))\big)$, hence the same representation of $N$ defined above. 
Notice that, in obvious notation, the isomorphism in the equation (\ref{e:iso KK}) gives in this case
\beq\label{e:iso KK NN}
N_l\times N_r\longrightarrow Q \times N  \qquad \textrm{such that}\qquad (n_l,n_r)\longmapsto (n_r-n_l,n_l)\in Q\times N\,.
\eeq
Using the representation $V_s \times V: Q\times N\to \cB(\cH_f)$, also an equation similar to (\ref{e: Vin F}) holds, and considering the \emph{discrete} action $\a_s$ of $N_s$ for $s=l,r$, and $Q$, and the action $\a$ of $N$, we have
\beqa\label{e: Vin F SW}
\cF_\cI&=&\cF_{\cI,0}\rtimes_{\a_s}V_s(N_s)\rtimes_{\a_{Ss}} V_{Ss}(N_{Ss})\cong\big(\cF_{\cI,0}\rtimes_\a V(N)\big)\rtimes_{\a_s} V_s(Q)\nonumber \\
&\cong& \cC_\cI\rtimes_{\a_s} V_s(Q) \cong \cF_{s\cI}\rtimes_\a V(N)\,.
\eeqa
Having at this point both the direct symplectic and the graded formulation of the nets of the the Streater and Wilde model, in the rest of this subsection we establish some of their structural results, also useful for the cohomological interpretation. 
In Subsection \ref{ss: reps} we shall return on the relation between the approach of the graded net and the one of the $G$-category of M\"uger in \cite{mug05}, already recalled in Subsection \ref{ss:cross prod graded net}. 
We begin from the following 
 
\bprop\label{o:properties}
In the notation above, in particular referring to diagram (\ref{e:SW inclusion of net}), we have
\bdes
\item[i)]
all the nets $\cA_\cI,\,\cB_\cI,\,\cC_\cI,\,\cQ_\cI,\,\cE_\cI,\,\cF_{\cI,0}$ and $\cF_\cI$ are additive;
\item[ii)]
in their defining (vacuum) representation, the  nets $\cA_\cI,\,\cB_\cI,\,\cQ_\cI,\,\cE_\cI$ and $\cF_{\cI,0}$ are local and satisfy Haag duality. 
The net $\cC_\cI$ is globally $N$-graded and satisfy $N$-graded locality, up to elements in $\cZ_b=V(N)''$, and $N$-graded duality.
The net $\cF_\cI$ has $N\times N$-grading flipped by $S$, satisfy  $N$-graded locality and $N$-graded duality;
\item[iii)]
\emph{Reeh-Schlieder Property:} for  any $I\in \cI$, the observable vacuum vector $\O_a$ is cyclic and separating for any  local algebra $\cA_\cI(I)$;  the field vacuum vector $\O_f$  is cyclic for any local algebra $\cF_\cI(I),\, \cF_{l \cI }(I)$ and $\cF_{r \cI}(I)$, and separating  for the algebras $\cF_{l \cI }(I)$ and $\cF_{r \cI}(I)$, but $\cF_\cI(I)$. The vector $\O_{f,0}:=\O_a\otimes \O_L$, i.e.\ the vector  $\O_f$ projected on $\cH_a\otimes \cH_L$, is cyclic and separating for any local algebra $\cF_\cI(I)_0$. 
\edes
\eprop 
\prf
i) Being the elements of $\cI$ connected, the  additivity of these nets is generally given because they are derived from Weyl algebras, see \cite[Section 7]{rob04}.\\
ii)
Locality or $N$-graded locality easily results from calculation on the symplectic spaces, by the definition of the symplectic form $\s_f$  above, i.e.\ by the definition of the graded nets obtained through crossed products as seen above.
The duality of the net $\cA_\cI$ in its vacuum representation $\p_a$, recalled in \cite[Subsection 4.2]{fabio3}, is well established in \cite{hl82}. 

To show the duality for the other nets, we use directly the classical result in \cite[Proposition 6.2]{dhr69II}: we just observe that for every couple of charges $(c,q)\in \dG$ and any arbitrary small non-void interval $I\in \cI$, there exists an element $F=\fzu\in V_f (I)$ such that $\r^I_F:=\ad \p_f(W(F))$ furnishes a sector automorphism (or solitonic automorphism of the net $\cC_\cI$) carrying these charges. This has been established in \cite[Proposition 4.5]{fabio3} and for a solitonic automorphism of $\cC_\cI$ the \emph{charge} $q$ has to be understood as the difference of the limit values of the function $f_1$ at $+\infty$ and $-\infty$, i.e.\ as an element in $Q$, referring  to the isomorphism of equation (\ref{e:iso KK NN}).

The net $\cA_\cI$ is an Haag dual net in its vacuum representation and also with respect to the representations $\p_a \circ\r^I_F$ for any automorphism $\r^I_F$ because of \cite[Lemma 2.2]{dhr71}, so the nets $\cF_\cI$ and $\cQ_\cI$ are Haag dual by \cite[Proposition 6.2]{dhr69II}. Moreover, the nets $\cB_\cI$, $\cC_\cI$ and $\cE_\cI$, are the fixed point nets of $\cF_\cI$ under the action of the compact group $\cG$, $\cG_q$ and $\cG_c$ respectively. Hence, because of the just cited result, we can use the Haar integral mean over these groups  and, performing a little \emph{diagram chase} in the diagram (\ref{e:SW inclusion of net}), we obtain Haag duality for all the nets on the index set $\cI$. As an example, consider  $\cC_\cI^{\cG_c}=\cB_\cI$. Here $\cB_\cI=\cA_\cI\otimes \cZ_b$ so that being $\cA_\cI$ Haag dual so are $\cB_\cI$ and $\cC_\cI$, because of  \cite[Lemma 2.2]{dhr71}.
The results for $\cC_\cI$ have been presented above.\\
iii) The result for the observable net $\cA_\cI$ is classical: passing to its corresponding 2-dimensional local net $\widetilde{\cA}$ (see \cite[subsection 4.2]{fabio3} for details) it follows from the additivity of the net and the positivity of energy. For the nets with grading group, we adapt the classical argument for a $\bZ_2$-grading.  
Consider for example the subnet $\cF_l:=\cF_{l\cI}$. For any $I \in\cI$ we have $\cF_l(I)\subset \cF(I)$ and $\cF_l(I)$ contains fields of any charge $(c,q)\in \dG$. This suffices to have the cyclicity of $\O_f$ for $\cF_l(I)$ and a fortiori for $\cF(I)$. If we take now $I,\,I_1\in \cI$ with $I<I_1$,  we have $\cF_l(I)\subset \cF_r(I_1)'$. So the vector $\O_f$, being cyclic for $\cF_r(I_1)$, is also separating for any $\cF_l(I)$, and vice versa. 
Because the local algebras $\cF(I)$ equates $\cF_l(I) \bigvee \cZ_b$, where  $\cZ_b$ fixes the vector $\O_f$, see \cite[Subsection 4.4]{fabio3}, so that for $X\in \cZ_b$ we have  $X\O_f=\O_f$ and $\O_f$ is not separating for the net $\cF_\cI$.
The result for the net $\cF_{\cI,0}$ is similar.
\qed
Apart from the abstract argument used above, for all the nets, it is possible to verify the duality Property \textbf{5.} and the $N$-graded duality Property \textbf{5\hspace{0,5pt}b.} for the net $\cF_\cI$, directly from the definitions by simple calculations on the symplectic spaces, using  the Definitions \ref{d: ktwisted net} of the $N$-twisted and $N$-graded dual algebras.
Consider for example the net $\cF_{l,\cI}$; for any $o\in \open(\bR)$ we introduce the notation 
\beqan
&&\dulS(o):=\{f\in \duS(o): \lim_{x\to +\infty}f(x)=0\}\,\quad \textrm{and}\\
&&\durS(o):=\{f\in \duS(o): \lim_{x\to -\infty}f(x)=0\}\,.
\eeqan
Hence, recalling the Definitions \ref{d: ktwisted net}, we have 
\beqa\label{e: gr f I}
\cF^g_l(I')&=&\bigvee_{I_0\bot I}\cF_l(I_0)^N\bigvee_{I<I_r}\cF_r(I_r)\nonumber\\
					 &=&\p_f\big(\cW(\dS(J_l)\oplus \partial^{-1}_l\cS(J_l))\big)''\bigvee \p_f\big(\cW(\cS(J_r)\oplus \partial^{-1}_r\cS(J_r))\big)''.
\eeqa
Then $\cF^{gd}_l(I):=\cF^g_l(I')'\bigcap \cF_l^\cI(\bR)$ is obtained from the elements $G=\gzu\in V_{fl}$ such that $\s_f(G,F)=0$ for any 
$F=\fzu\in \big(\dS(J_l)\bigcup \cS(J_r)\big)\oplus \big(\partial^{-1}_l\cS(J_r)\bigcup \partial^{-1}_r\cS(J_r)\big)$, that entails 
$G\in \cS(I)\oplus\dulS(I)=V_{fl}(I)$, i.e. the result.

\smallskip
Notice that the Reeh-Schlieder property assures that, for any $I_1,\,I_2\in \cI$ such that $\overline{I}_1\subset I_2$, the triples $(\cF_{\cI,0}(I_1),\,\cF_{\cI,0}(I_2),\,\O_{f,0})$ and for $s=l,\,r$, $(\cF_{s\cI}(I_1),\,\cF_{s\cI}(I_2),\,\O_f)$ are \emph{standard $W^*$-inclusions}. Hence, the \emph{split property} and the \emph{quasi-split property} for the nets $\cF_{\cI,0}$ and $\cF_{s\cI}$ are equivalent, see \cite{dl84} for details. From this observation it follows
\bprop\label{o:split}
The net $\cF_\cP$ and its subnets $\cF_{\cP,0}$, $\cF_{l\cP}$ and $\cF_{r\cP}$, for $\cP=\cI, \cD$, do not satisfy neither the split property nor the quasi-split property. 
\eprop
\prf
We know from \cite[Proposition 1.6]{dl84} that the split property for $\cF_{\cI,0}$ and $\cF_{s\cI}$, for $s=l,\,r$, and the standardness for the inclusion as above, assured by Reeh-Schlieder property of $\cF_{\cI,0}$ and $\cF_{s\cI}$, are sufficient conditions for the separability of the representation Hilbert space. The result hence follows from the non-regularity of the representation $\p_b$ of $\cF_{\cI,0}$ and $\p_f$ of $\cF_{s\cI}$ and $\cF_\cI$.  
\qed
A further easy consequence of diagram (\ref{e:SW inclusion of net}) is the following: $\cF_\cI (I)$ being  the crossed product of the type III$_1$ factor $\cA_\cI(I)$ by the discrete group of charges $\dG$, is itself a factor of type III, and the weaker \emph{Property B} of Borchers also holds for the net $\cF_\cI$. Actually this result is also a consequence of the positivity of the energy, see \cite{fabio3} for details.\smallskip

\smallskip
We pass to list the relevant properties of the nets on the index set $\cD$, in particular in relation to the extension by additivity,  by the following
\bprop\label{o:propert}
It holds
\bdes
\item[i)]
the net $\cA_\cD$ is a local, non-additive extension of the net $\cA_\cI$. Moreover, it equals the dual of the additively extended canonical net $\cA_\cD^{\cI d}$, i.e.\ 
\beq\label{e:a dual}
\cA_\cD (D)= \cA_\cD^{\cI d}(D)\,, \qquad \textrm{for every}\quad D\in \cD\,.
\eeq
This implies that the double interval duality, for the additively extended net $\cA_\cD^\cI$ is not satisfied, i.e.\ 
\beq\label{e:D duality}
\cA_\cD^\cI\, \subsetneq\,\cA_\cD^{\cI d}\,=\,\cA_\cD;
\eeq   
\item[ii)]
the nets  $\cF_{\cD,0}$ and $\cF_\cD$ equal respectively the nets $\cF_{\cD,0}^\cI$ and $\cF_\cD^\cI$, extended by additivity. The net $\cF_{\cD,0}$ is local and the net $\cF_\cD$ is $N$-graded local on the index set $\cD$, i.e.\
$\cF_l(D_1)\,\bot\,\cF_r(D_2)$ for $D_1,D_2\in \cD$ and $D_1< D_2$;
\item[iii)]
the net $\cF_\cD$ satisfies  $N$-graded duality, i.e.\  for any $D\in \cD$ it holds $\cF_{l\cD}(D)=\cF_{l\cD}^{gd}(D)$;
\item[iv)] 
the results of point i) hold for the nets $\cB_\cD$,\,$\cQ_\cD$ and $\cE_\cD$. The net $\cC_\cD$ equals the additivity extended net $\cC_\cD^\cI$, is globally $N$-graded and satisfy $N$-graded locality up to elements in $V(N)''$, and $N$-graded duality in the global grading sense.
\edes
\eprop
\prf
i)
We just have to show the equality (\ref{e:a dual}) for a generic double interval $E=I_1\cup I_2\in \cI_2$. In the fixed notations and omitting to write the intersection with the quasi-local algebras $\cA^\cI(\bR)\equiv\cA_\cD^\cI(\bR)$ of the additive nets, we have  
\beqa\label{e:hat cA}
\cA_\cD^{\cI d}(E) &:=& \cA_\cD^\cI(E')'= \big(\cA_\cI(I_3 ) \bigvee \cA_\cD^\cI(I' ) \big)'= \cA_\cI(I_3 )' \bigwedge \cA_\cD^\cI( I' )' \nonumber\\
     &=& \cA_\cI(I_3 )' \bigwedge \cA_\cI( I ) = \cA_\cD^\cI(I_3 ') \bigwedge \cA_\cD^\cI( I' )'
\eeqa
where we used $\cI$-duality (Haag duality) and the additivity of $\cA_\cI$. 
From these equalities and by a direct calculation on the generators of the algebras involved, i.e.\ on the elements in the local symplectic subspaces, we verify (\ref{e:a dual}). 

The inclusion $\subseteq$ is proved from the first equality in the second line of (\ref{e:hat cA}). The generators of $\cA_\cD^{\cI d}(E)$ are given by elements $G = \gzu\in V_a (I)$ such that $\s_a (F, G) = 0$, for $F = \fzu \in V_a (I_3 )$, i.e.\ such that $\int \, f_0\, g_1 \, dx =0$ and $\int \, f_1\, g_0 \, dx =0$. 
These give $g_0 \rest I_3 = 0$ and $g_1 \rest I_3 = const$, so that $g_0 \in \dS (E)$ and $g_1 \in \cS (I) : g_1 \rest I_3 = const$.
Such a set of functions contains $V_a (E)$, and the inclusion $\subseteq$ of (\ref{e:a dual}) hence holds by definition of $\cA_\cD (E)$.

The inclusion $\supseteq$ of (\ref{e:a dual}) follows from the second equality in the second line of  (\ref{e:hat cA}). $\cA_\cD^\cI(I'_3)$ being defined by additivity, the generators of $\cA_\cD^{\cI d} (E)$ are obtained from elements $G = \gzu\in V_a (I_0 )$, with $I_0\in \cI$ and $I_0 \subset I_3^\bot$, such that
$\s_a (F, G) = 0$ for $F=\fzu \in V_a (I_4 )$, for every $I_4 \subset I'$ and  $I_4\in \cI$.
Hence we have: $g_0 =0$ on $I_3$ and on every $I_4$ as above, i.e.\ $g_0 \in \dS(E)$; $g_1 = const$ on $I'$, vanishing on $I_3$. As $g_1 \in \cS$ and  $I'$ is not bounded, such constants on  $I'$ have to vanish, so we have  $g_1 \in \cS(E)$ and  $G \in V_a (E)$.\\
ii)
The additivity on $\cD$ is obtained by a calculation on the Weyl generators of the algebras as in above point i).
The locality and $N$-graded locality is a consequence of the cofinality of $\cI$ in $\cD$.\\
iii)
From point iii) of Proposition \ref{o:grad prod}, and with the same notation, we have that for a generic additive, $N$-graded local net $\cF_l$ it holds  $\cF_l^{gd}(E)= \cF_l^{gd}(I)\bigcap \cF_l(I'_3)$; then using the result following equation (\ref{e: gr f I}) for the graded dual algebra $\cF^{gd}_l(I)$, we have that the generators of the algebras $\cF_l^{gd}(E)$ derive from elements in  $V_{fl}(I)\bigcap V_{fl}(I_3')$, i.e.\ $\big(\cS(I)\oplus\dulS(I)\big)\bigcap\big(\cS(I'_3)\oplus\dulS(I'_3)\big)=\cS(E)\oplus\dulS(E)=V_{fl}(E)$, yielding the result.\\ 
iv) Follows as in the previous results, by calculating on the symplectic spaces. In particular, for the globally graded net $\cC_\cD$, we have $\cC^{gd}(E)=\cC_\cD(I)\bigcap \cC(I_3')= \cC_\cD(E)$, where the additivity of $\cC_\cI$ on $\cD$ and on $\cJ$ has been used, i.e.\ $\cC(I_3')=\bigvee_{E_0\in \cD\cap I_3^\bot} \cC(E_0)$. 
\qed
The above results may provide insights for more general theories, at least for Weyl algebras models or loop groups model, see e.g.\ \cite{fabio7} for partial work in this direction, hence we make the following remarks
\bitem
\item
The non-triviality of the inclusion in equation (\ref{e:D duality}), i.e.\ the absence of the duality on double intervals for the net $\cA_\cD^\cI$, implies that there are non-trivial DHR sectors for the net $\cA_\cI$, according to a classical argument: the larger algebras $\cA_\cD^{\cI d}(E)$ contains non-trivial charge carrying observable operators, relative to localized sector endomorphisms supported on $I_1 \cup I_2$, that are not present in $\cA_\cI(I_1)\bigvee\cA_\cI(I_2)=: \cA_\cD^\cI(E)$.
\item
Observe that the above argument directly justifies condition (\ref{e: aa)}) as implying the triviality of superselection sectors. In fact, the intersection of the algebras associated with the path $p\in \S_1(\cP)$ is as small as possible.
\item
The equality in equation (\ref{e:a dual}) means that for given two sectors automorphisms of the net $\cA_\cI$, indicated by $\r_i$ for $i=1,2$ and localized in the interval $\cI_i$, the bilocal intertwining operator $U\in (\r_1,\r_2)$, i.e.\ $U\in \{ \cA_\cI(I'_1)\cap \cA_\cI(I'_2)\}'$ is actually contained in $\cA_\cD(I_1\cup I_2)=\cA_\cD(E)$. 
This gives a characterization of a general condition on the sector intertwiner formulated in \cite{dhr71}, that allows to predict that all the observable data defining the superselection sectors of the net $\cA_\cI$ are just encoded in the nets $\cA_\cI$ and $\cA_\cD$.
\item
In other words, recalling the definitions in diagram (\ref{e:def inclusion symplectic}), if $V_b(E)\rest I_1$ denotes the restriction of  $V_b(E)$ to the subspace of elements localized in the interval $I_1$, where for an element $F=\fzu$ the function $f_1$ is prolonged continuously by different constants out of $I_1$, we have $V_b(E)\rest I_1=V_f(I_1)$ and $V_a(E)\rest I_1=V_{fr}(I_1)$. 
This implies that $\p_f(\cW(V_b(E)\rest I_1))''=\cF_\cI(I_1)$ and $\p_f(\cW(V_a(E)\rest I_1))''=\cF_{r\cI}(I_1)$, i.e.\ the complete algebras of fields and its right graded subalgebra localized in $I_1$ is associated to $V_b(E)\rest I_1$ and $V_a(E)\rest I_1$ respectively.
\item
We recall that the study of double interval duality for a net $\cA_\cD^{\cI d}$ plays a prime role when investigating  the superselection sectors of the net $\cA_\cI$. See also \cite{dri79}, and \cite{klm01} for the study of low dimensional conformal, \emph{completely rational models} using the subfactor theory of von Neumann algebras. 
\item
Finally, we notice that the implication of iii) of the above Proposition \ref{o:propert}, i.e.\ duality for $\cF_\cI$ implies duality for $\cF_\cD$, is a specific feature of the model and not a general result, see also \cite[Section 30]{rob04}.
The existence of general conditions giving this implication is an interesting topic for further research.
\eitem
We end this subsection commenting about the non-trivial centers, or relative commutants, of the quasi-local algebras of observables in the model at hand. 

\smallskip
We recall that this structural topic of a net of observables, was discussed at the beginning of AQFT, as traces of classical global observable, and has been proposed again in a series of paper by Baumgaertel and Lled\'o, culminating in \cite{bl03}, and have given rise to some interesting mathematical counterparts, see for example M\"uger in \cite{mug04a} and Vasselli in \cite{vas07}.
We recall that the presence of such a center is also a feature of the general theory of constrains by Baumgaertel, Grunling and Lled\'o, see for example \cite{gl00,bg05, gru06}, that also accounts  for electromagnetic and QED models.\smallskip

In the Streater and Wilde model, we may discuss this issue referring to diagram (\ref{e:SW inclusion of net}) and to \cite[Sections 3 and 4]{fabio3}. 
As $\cB_\cI(I)=\cA_\cI(I)\otimes \cZ_b$ for $I\in \cI$, the relative commutant of the quasilocal algebra $\cA=\cA^\cI(\bR)$ in all other quasilocal algebras but $\cQ$ is non-trivial and equals $\cZ_b$, i.e.\ 
\beq\label{e:relative comm}
\cA^c:=\cA'\cap\cF=\cA'\cap\cB=\cA'\cap\cE=\cA'\cap\cC=\cZ_b.
\eeq
On the other hand, we have trivial relative commutants of $\cA$ in the nets that do not contain elements of the representation $V$ of the group $N$, see Subsection \ref{ss:Other properties of fields subnets}, i.e.\  $\cA'\cap\cQ =\cA'\cap\cF_0 =\cA'\cap \cF_s=\bC$, for $s=l,r$.

The center of $\cA$ is trivial, as $\cA$ is irreducible in $\cH_a$, i.e.\ $Z(\cA):=\cA\cap\cA'=(\p_a,\p_a)=\bC$ so, from the physical point of view, we are in the usual situation of DHR charge. 
The centers of $\cQ$, $\cC$ and $\cF$ are also trivial. 
On the other hand, the action of the group $\cG$ gives $\cF^\cG=\cB$, and $\cB$ is not irreducibly represented on $\cH_b\cong\cH_a\otimes\cH_L$, since $Z(\cB):=\cB'\cap\cB=(\p_b,\p_b)=\cZ_b$. Similarly  $Z(\cE)=\cZ_b$.
However, although the elements of $\cZ_b$ appear as fixed points under the action of the gauge group $\cG$, it is not proper to think of them as local observables for various reasons:
\bitem
\item
they are trivially represented in the vacuum representation  space of the observables $\cH_a$, where they are multiples of the identity; 
\item
they have void localization region, according to the definition of localization in equation (\ref{e:localization}); and
\item
they do satisfy the DHR superselection criterion only respect to the sector automorphism with charge $q\in Q$, in fact for any $F\in V_f(I)$ with charge $(F_c,F_q)\in C\times Q$ and for a generator $\p_b(W(0,n))$ of $\cZ_b$, we have 
\beqn
\r^I_F(\p_b(W(0,n)))=e^{-i F_c n}\p_b(W(0,n))\,.
\eeqn
\eitem
We remember  that the algebra $\cZ_b$ is the Abelian von Neumann algebra generated by the elements of the discrete subgroup $N\subset\cG_c\subset\cG$, strongly represented on the Hilbert space $\cH_b\subset \cH_f$, i.e.\ of the discrete group $N$ in a non regular representation strongly continuous along the element of $N$ (that means  $\l \mapsto \p_f(W(0, \l n))$ is strongly continuous for any $n\in N$). Actually, this representation of $N$ also furnishes a global grading for the net $\cC_\cI$, in the sense of the equation (\ref{e:product grad}) and as seen before Proposition \ref{o:properties}, that trivialize on $\cB_\cI$.

\smallskip
We could have choice to not contemplate the elements of $\cZ_b$ as part of the auxiliary nets $\cB, \cC, \cE$ or $\cF$.
\footnote{\label{f:diag}
In fact, we could have considered from the beginning, instead of the six terms diagram of symplectic space (3.4) of \cite{fabio3}, giving rise to the diagram (\ref{e:SW inclusion of net}), the following four terms one 
\beq\label{e:def 1 inclusion symplectic}
\barr{ccc}
V_a:=\dS\oplus\cS&\subset&\cS\oplus\cS=:V_{f0}\\\\
 \cap& & \cap \\\\
 V_q:=\dS\oplus\duqS &\subset&\cS\oplus\duqS.
\earr
\eeq
But in this way the role played by the Weyl elements associated to the subspace $N$ would not be clarified at all.
}
However, a further meaning of $\cZ_b$ is presented in the Remark \ref{r:z mod3} below.
\subsection{Net $1-$cohomology for the Streater and Wilde model}\label{ss:Net cohomology for the SW}
After these results about the properties of the nets of the model under study, we turn back to the problem of its superselection theory, carrying out the analysis in two steps:
\bitem
\item[1)] 
show the (quasi-)triviality of the $1$-cohomology for the $N\times N$-graded net $\cF_\cD$, where the index set $\cD$ is chosen to simplify the calculations for verifying the condition (\ref{e: aa)}) in Corollary \ref{c:cohograd}. Then, derive the same result for the net $\cF_\cI$; and 
\item[2)] 
use the action of the global gauge symmetry (sub-)group on $\cF_{s\cI}$, for $s=l,\,r$, to investigate the  $1$-cohomology of the fixed point subnets and the nature of their superselection sectors. In particular for the observable net $\cA_\cI$.
\eitem
We begin by computing the 0-cohomology for all the nets of the model defined in diagram (\ref{e:SW inclusion of net}), both on the index sets $\cI$ and $\cD$, stating the following
\blemma\label{e:0 cohom}
For the index sets $\cP=\cI,\,\cD$ we have
\bdes
\item[i)]\qquad $Z^0(\cA_\cP)\,=\,Z^0(\cQ_\cP)\,=\,Z^0(\cF_{\cP,0})\,=\,Z^0(\cF_{l\cP })\,=\,Z^0(\cF_{r\cP})\,=\,\bC$; 
\item[ii)]\qquad $Z^0(\cB_\cP)=Z^0(\cE_\cP)=Z^0(\cC_\cP)=Z^0(\cF_\cP)=\Zb$.
\edes
\elemma
\prf
Being $\cP$ is connected, it is a standard  result of \cite{rob90} that for a net $\cN_\cP$ we have $Z^0(\cN_\cP) =\cap_{o\in \cP} \cN_\cP(o)$.\\
i)
Locality and the triviality of the center give the result for $\cA_\cP$, $\cQ_\cP$ and $\cF_{\cP,0}$.
For $\cF_{s\cD}$, $s=l,\,r$,  the result follows from the triviality of the intersection; for example $\cF_{l\cI}(D_1) \cap \cF_{l\cI}(D_2) =\bC$ if $D_1 \bot D_2$, for $D_1,D_2 \in \cD$.\\
ii)
Locality and the equality $\cZ(\cB_\cP)=\cZ(\cE_\cP)=\Zb$, give the result for $\cB_\cP$ and $\cE_\cP$. $N$-graded locality and condition $\cZ_b\subset \cC_\cP(D)$ for any $D\in \cD$, give the result for $\cC_\cP$ and for $\cF_\cP$.  
\qed
\brem\label{r:z mod3}
Along the lines of the Remark \ref{r:z mod} and the comments about the trivial centers at the end of Subsection \ref{ss:Other properties of fields subnets}, we notice that the item ii) in the above Lemma indicates a slight difference from the usual case of trivial $0$-cohomology of a net $\cN$, where the set of arrows from the trivial 1-cocycle $\1$ to itself , i.e.\ for the 1-cocycle $\1(b)= I$ for every  $b\in \S_1(\cP)$, coincides with $Z^0 (\cN)=\bC$. 
For the nets in point ii) instead, such a set of arrows is  identified with $\cZ_b=V(N)''$, giving a further cohomological meaning for this Abelian algebra.  
\erem
We pass to consider step 1), the $1-$cohomology on the net $\cF_\cP$, stating the following
\bprop
The net 1-cohomology $Z^1(\cF_{s\cP})$ is quasitrivial, for both $\cP=\cI,\,\cD$ and for $s=l, \, r$.
A fortiori, $Z^1_t(\cF_{s\cP})$ is also quasitrivial. 
Moreover, the DHR representation category $\repb \cF_{s\cP}$ have no nontrivial irreducible objects. The same hold for the net $\cF_\cP$.
\eprop
\prf
Observe that for the net $\cF_\cD$ with grading group $N\times N$, all the conditions of the Corollary \ref{c:cohograd} are satisfied. The condition (\ref{e: aa)}) is fulfilled by the net property, just because of the choice of the index set $\cD$. In fact it suffices to show that it is true for $\partial_0 b, \partial_1 b \in \cI \subset \cD$ because of the additivity of $\cF_\cD$ and $\cF_{s\cD}$. 
Hence the Corollary \ref{c:cohograd} gives the triviality of $Z^1(\cF_{s\cD})$.
The net $\cF_\cD$ is $N$-graded dual, so the quasitriviality of $Z^1(\cF_{s\cD})$ is equivalent, up to direct sums, to that of $\repb (\cF_{s\cD})$, see the Roberts equivalence theorem \cite[Theorem 26.2]{rob04}. For this observe that $\cD$ is connected because $\cI$ is cofinal in $\cD$ and connected, Lemma \ref{l:cofinal}.

The results on the index set $\cI$ holds because all the conditions of Roberts' result on the change of index set \cite[Theorem 30.1]{rob04} are fulfilled. In fact $\cI$ being  cofinal in $\cD$, the nets $\cF_{s\cI}$ and $\cF_{s\cD}$ generate the same quasilocal von Neumann algebras, the net $\cF_{s\cD}$ is additive by iii) of Proposition \ref{o:propert} and $\cI$ generates $\cD$.

The results for the net $\cF_\cP$ follow as it is a represented discrete crossed product of $\cF_{s\cP}$ by the Abelian discrete group $N$ in the representation $V$, as sees in equation (\ref{e: Vin F SW}).
\qed
The above enounced step 2) is discussed in the following subsection.
\subsection{The set of sectors of $\cA_\cI$ associated to $\cF_\cI$}\label{ss: reps}
Here we answer the question about the completeness problem for the DHR superselection sectors of the Streater and Wilde model. 
Namely, we show that the set of sectors described in Subsection 4.3 of \cite{fabio3} are \emph{all} the DHR sectors of the model, so that the net $\cF_\cI$ may be called \emph{the} field net of the observable net $\cA_\cI$.

\smallskip
According to Definition \ref{d:rapo K on comlem}, starting from an element $F_s=(0,f_{1s})\in V_{f,s}(I)$ with $f_{1s}\in \dulS (I)/\duzS(I)$, we constructed in Subsection \ref{ss:Other properties of fields subnets} for the net $\cF_{\cI,0}$ of the Streater and Wilde model, the net of representations $\widehat{N}_{s,\cI}$ for the discrete grading groups $N_s\cong\bR_d$, for both $s=l,r$. For any $I\in \cI$, through the choice of the elements $F_l$ and $F_r$ forming a partition of unity at $I$, we also have a global representation for the discrete group $N\cong \bR_d$. 

Moreover, $N_s\cong Q$ is a subgroup of the charge group $\widehat{\cG}=C\times Q$, so that the crossed product definition of the net $\cF_\cI$ in equation (\ref{e: Vin F SW}) holds. Furthermore, referring to diagram (\ref{e:SW inclusion of net}), we also have that  $N\subset \cG_c$ is a discrete subgroup of the gauge group $\cG=\cG_c\times \cG_q$. 

Using  the results of \cite[Sections 4.3 and 4.4]{fabio3}, we obtain by duality from the net of representations $\widehat{N}_{s,\cI}\cong\widehat{Q}_\cI$, the net of representations $\widehat{\cG_q}_\cI$, of the compact gauge subgroup $\cG_q$.
 
Similarly, starting from an element $F=(f_0,0)\in V_{f,l}(I)$ with $f_0\in \cS(I)/\dS(I)$, it is possible to construct the Abelian net $\widehat{C}_\cI$, the net of the representations of the charge group $C$ on the bounded open intervals, and  for $V\in\widehat{C}_\cI$ it holds $\cF_{\cI,0}=\cA_\cI\rtimes V(C)$. 
As above, we have by duality the net of representations of the compact gauge subgroup $\widehat{\cG_c}_\cI$. 

\smallskip
We notice that both the nets $\widehat{\cG_q}_\cI$ and $\widehat{\cG_c}_\cI$ are generated by represented Weyl unitaries, i.e.\ one dimensional representations, contained in the net $\cF_{s\cI}$. 

\smallskip
Hence we are in the particular case of a net of von Neumann algebras $\cN_\cP$ with gauge automorphisms group $G$, and a net $\widehat{G}_\cP$ of classes of Hilbert spaces with support identity, in any local algebra $\cN_\cP(o)$, for any $o\in \cP$.
In this case, we may invoke the following general cohomological classification result \cite[Theorem 4, Section 3.4.5]{rob90}: if $Z^0 (\cN_\cP)$ is trivial  and $Z^1 (\cN_\cP)$ is quasitrivial and if $\cN^G_\cP$ is the net of fixed point of $\cN_\cP$ under the action of a group $G$ of gauge automorphisms, then the unitary equivalence classes of $Z^1 (\cN^G_\cP)$ are in 1-1 correspondence with $Z^0(\widehat{G})$.\smallskip

\smallskip
For the net $\cF_\cI$, its subnets and the group $\cG$ described in diagram (\ref{e:SW inclusion of net}), we have the following
\bprop\label{o:subnet coho}
Let the net $\cF_\cI$ and the group $\cG:=\cG_c\times \cG_q$ be defined as above. Then 
\bdes
\item[i)]
$\cC_\cI=\cF_\cI^{\cG_q}$ and the objects of $Z^1_t(\cC_\cI)$ are in 1-1 correspondence with those of $Z^0(\widehat{\cG}_q)$. The net $\cC_\cI$ has no non-trivial DHR sectors but only $N$-solitonic sectors labeled by the charges $q\in Q$, given by $N$-solitonic automorphisms.
The net $\cF_{\cI,0}$ has the same sectors as the net $\cC_\cI$; 
\item[ii)]
$\cE_\cI=\cF_\cI^{\cG_c}$ and the objects of $Z^1_t(\cB_\cI)$ are in 1-1 correspondence with those of $Z^0(\widehat{\cG}_c)$. The net $\cE_\cI$ has only DHR sectors labeled by charges $c\in C$.
The net $\cQ_\cI$ has the same sectors as the net $\cE_\cI$;
\item[iii)]
$\cB_\cI=\cF_\cI^{\cG}$ and the objects of $Z^1_t(\cB_\cI)$ are in 1-1 correspondence with those of $Z^0(\widehat{\cG})$. The net $\cB_\cI$ has only DHR sectors labeled by charges $(c,q)\in\dG\cong C\oplus Q$, and  the net $\cA_\cI$ has the same sectors as the net $\cB_\cI$.
\edes
\eprop
\prf
The results about the fixed point are established in \cite[Subsection 4.4]{fabio3} and reported in the diagram (\ref{e:SW inclusion of net}).
Notice that, according to the Proposition \ref{o:properties}, any of the above listed subnet $\cN_\cI\subset \cF_\cI$ is Haag dual, or $N$-graded dual (globally or on the connected component), in its defining representation. Hence  $Z^1_t(\cN_\cI^d)=Z^1_t (\cN_\cI)$ and because of this, $\repb \cN_\cI$ and $Z^1_t(\cN_\cI)$ are equivalent, see \cite[Theorem 30.1]{rob04} recalled after Definition \ref{d:cocycle}. The superselection sectors of these subnets are obtained as in the above reminded 
\cite[Theorem 4, Section 3.4.5]{rob90}.
The results on the nature of the sector automorphisms are given in \cite[Proposition 4.5]{fabio3}, using the fact that $\cZ_b$ is Abelian.
\qed
We notice that the obtained classification of $\repb \cA_\cI$, through a net cohomological computation as above, excludes the existence of sectors of infinite statistical dimension, because the result is acquired in terms of $Z^0(\dG)$, the net 0-cohomology of the charge group $\dG$, i.e.\ of the dual of a compact gauge group.
For this reason, the described representations are \emph{all} the (finite statistical dimension) DHR representations of net $\cA_\cI$ and, by the discussion in \cite[Subsection 4.2]{fabio3}, \emph{all} the positive energy M\"{o}bius covariant DHR representations of the corresponding 1+1-dimensional net $\widetilde{\cA}_\cK$, for $\cK$ the causal index set of the double cones on the 2-dimensional Minkowski spacetime.  

\smallskip
Further considerations about the category $\repb \cA_\cI$ are now in order. Is is well known (see for example \cite{rob04}) that
$\cI$ being a directed index set, we may describe $\repb \cA_\cI$ by the equivalent tensor $W^*$-category of the localized transportable endomorphisms $\cT_t$ of the net $\cA_\cI$, where the vacuum representation corresponds to the identity morphism, the tensor unit $\i$ of $\cT_t$, and the set of the arrows between $\r, \t\in \cT$ is denoted by $(\r, \t)$. 
Moreover, as a general feature, because $\cG$ is Abelian, the irreducible elements in $\cT_t$ have statistical dimension $1$, and  any irreducible object in $\cT_t$ is the composition of two objects from the following two full subcategories, faithfully embedded in $\cT_t$
\bitem
\item
$\cT_{t,C}$, whose objects are the automorphisms implemented by Weyl operators in $\cC_\cI$, with zero $Q$-charge and the arrows are given by elements in $\cB_\cI$;
\item
$\cT_{t,Q}$, whose objects are automorphisms implemented by  Weyl operators in  $\cE_\cI$, with zero $C$-charge and and the arrows are given by elements in $\cQ_\cI$.
\eitem  
Returning to the discussion about the approach with the $G$-categories at the end of the Subsection \ref{ss:cross prod graded net}, we can equivalently characterize the subcategory $\cT_{t,C}$ as the DHR representations of the net $\cA$ that result from the restriction of the (untwisted) representations contained in the vacuum representation of the net $\cF_0$. This subcategory also corresponds to the subcategory $C$ appearing in the equations (\ref{e:dual K}). The subcategory $\cT_{t,Q}$ instead, corresponds to the subcategory $\widehat{G}/ C$ in the same equations. 

Moreover, according to the equation (\ref{e:fix cat}), reading $\cN=\cF_0$ and $K=N$, then the fixed point braided $N$-category  $\big(N-\Loc \cF_0)^N$ of the $N$-twisted representations $N-\Loc \cF_0$, corresponds to all the DHR representations $\repb \cA$, hence is equivalent to $\cT_t$.    

\smallskip
Through the category $\cT_t$ we express other properties of these equivalent categories, mainly about their braiding symmetry, collecting them in the following 
\bprop\label{o:braid} 
Let $\r$ and $\t$ be two objects in the tensor $W^*$-category $\cT_t$, localized in $I_\r$ and  $I_\t$ respectively, with $I_\r,\,I_\t\in \cI$ and $I_\r < I_\t$. Then 
\bdes
\item[i)]
is defined a statistical operator $\ep(\r,\t) \in (\r \t,\t \r)$ such that  
\beq\label{e:eps}
\ep (\r,\t)=e^{-i(C_\r Q_\t +Q_\r C_\t)}\,\p_f\big(W_f(F_\t)W_f(F_\r)W_f(F_\t)^*W_f(F_\r)^*\big)\,,
\eeq
where $\p_f(W_f(F_\r))$ and $\p_f(W_f(F_\t))$ are  the represented Weyl operator implementing the automorphisms $\r$ and $\t$ respectively, for $F_\r,\,F_\t\in V_f$, with any localization; 
\item[ii)]
if $\r\cong\t$ then $\ep(\r,\t)=e^{-2iC_\r Q_\r }\,\p_f\big(W_f(F_\t)W_f(F_\r)W_f(F_\t)^*W_f(F_\r)^*\big)$, where the Weyl operators are as in point i);
\item[iii)]
$\cT_t$ is a braided tensor $W^*$-category with $(\i,\i)_{\cA_	\cI}=\bC$. The subcategories $\cT_{t,C}$ and $\cT_{t,Q}$ are symmetric tensor $W^*$-categories, i.e.\ if $\r, \t\in \cT_{t,C}$ or $\r, \t\in \cT_{t,Q}$ then the phase $e^{-i(C_\r Q_\t +Q_\r C_\t)}=1$. A fortiori  $(\i,\i)_{\cA_	\cI}=\bC$ for both of them.
\edes
\eprop
\prf
The equality (\ref{e:eps}) follows from a routine calculation, see e.g. \cite{frs89}, that we report for completeness.
We use two equivalent auxiliary automorphisms $\widetilde{\r}\cong \r$ and $\widetilde{\t}\cong \t$ that are also localized in $I_\r$ and $I_\t$ respectively. These are chosen such that they are implemented by Weyl elements $F_{\widetilde{\r}}\in V_f(I_1)$ and $F_{\widetilde{\t}}\in V_f(I_2)$, and with $F_{\widetilde{\r}\ +\infty}=F_{\widetilde{\t}\ -\infty}=0$. In fact this choice may be freely done, up to  the an element in $N$. Then, if $U_1\in (\r,\widetilde{\r})$ and $U_2\in (\t,\widetilde{\t})$ are unitary (Weyl) intertwiners,  we have equation (\ref{e:eps}) from the usual definition
\beqn
\ep (\r,\t)=\t(U_1^*)U_2^*U_1 \r(U_2)\,.
\eeqn
The other results simply derives from the expression of the statistical operator $\ep$.
\qed
It is to observe that we may obtain the results of the above Proposition using the chiral approach, passing to the corresponding sectors and charges by D'Alembert formula, see \cite[Subsection 4.2]{fabio3}.

\smallskip
The category $\cT_t$ hence turns out to be equivalent as a tensor $W^*$-category to the category of the representations of the Bohr compactification of the Abelian group $\bR^2$, in symbols $\repb b\bR^2$. 
However, as expressed in equation (\ref{e:eps}), the category $\cT_t$ is braided whereas $\repb b\bR^2$ is symmetric. This is in  accord with the approach of graded category recalled in Subsection \ref{ss:cross prod graded net}.

\smallskip
This occurrence is similar to the one in \cite{ik02} where two different groups are shown to possess the same abstract representation  tensor category. By the Doplicher-Roberts duality in \cite{dr89}, this means that this last category reveals at least two different symmetries on its objects, as a manifestation of the \emph{isocategoriticity problem} in a physical model. New results on the mathematical side of this topic, have recently been announced by M\"uger.
\section{Conclusions and Outlook}
To demonstrate the completeness of a known set of DHR superselection sectors of the observable net of the Streater and Wilde model, we improve a strong result of Roberts on the triviality of the $1$-cohomology of a causal poset with values in a net of von Neumann algebras. 

\smallskip
Namely,  for a large class of Weyl algebra models,  we exploited condition (\ref{e: aa)}) as the main net $1$-cohomology triviality condition, i.e.\ as a condition for trivial sectors, that may be more easily established by a proper choice of the causal poset. 
Moreover we showed that the absence of the rather technical (quasi-)split property for a (field) net, may be overcome, at least in Weyl algebra models, once it holds for a fixed point (observable) subnet under the action of a gauge compact group. This is possible even in the unfavorable case of the non-separable Hilbert space of a non-regular representation for the larger (field) net. 

\smallskip
The formalism of a net with a grading group has also been introduced to describe the commutation relations of the algebras of a (field) net, localized on disjoint elements of the index set, and to generalize the locality and Haag duality for (field) nets on low dimensional spacetimes. 
This equivalently means that it is possible to describe the same usual relevant properties for the observable nets in a non vacuum representation, both in the twisted and untwisted case. 

\smallskip
In view of \cite{glrv} and \cite{br08, bfm}, the acquired analysis on models may be of interest on different spacetimes, both for the DHR or topological sectors. In particular, it is to better establish the relation between the following different approaches to the nature of the twisted sector: the above proposed graded net approach; the $G$-categories one in \cite{mug05}; the superconformal net case in \cite{ckl08} and the topological description in  \cite{br08}. Some work in this direction has been done in \cite{cr}.

\smallskip
From the model-oriented point of view, we expect that these hints from the Streater and Wilde case may help to describe the superselection structure, at least in presence of an underlying crossed product of Weyl algebras, see \cite[Subsection 2.1]{fabio3}. 
Some first results have been carried out by the author for the St\"uckelberg-Kibble $QED_2$ interacting model, in a non-regular representation, where the charge symplectic subspace and its symplectic complement are spaces of test functions, see \cite{fabio6}. In this case, non sharply localized sectors are expected.

\smallskip   
Finally, we stress that the methods developed in this series of two papers, may  result useful in other (simple) currents (extension) theories, such as  orbifold, coset, loop groups and vertex algebras models. Some first results in this direction have been also performed in \cite{fabio7}. 

\smallskip
Actually, all this twisted crossed algebras model we investigated, seems to be a version of a more general theory of superselection sectors of models, based on Galois Theory and Group Extension.
\bigskip

\noindent {\bf Acknowledgements.\,}
I'm grateful to John Roberts for drawing my attention on the problem, deep discussions and his constant interest on the aim of this work.  
I also like to thank  L\'aszl\'o Zsid\'o for hints about conditional expectations and Sebastiano Carpi and Giuseppe Ruzzi for many conversations about the argument of this paper. 
%

{\footnotesize
%
}

\begin{thebibliography}{alpha}




\bibitem{bg05}
Baumgaertel, H., Grundling, H.: Superselection in the presence of constraints. J. Math. Phys. \textbf{46},  34 pp. (2005).


\bibitem{bl03}
Baumgaertel, H., Lled\'o, F.: Duality of compact groups and Hilbert C*-systems for C*-algebras with a nontrivial center. Internat. J. Math. \textbf{15}, 759-812 (2004).




\bibitem{bfm}
Brunetti,R. , Franceschini, L.,  Moretti, V.: Topological features of massive bosons on two dimensional Einstein space-time. I: Spatial approach. Ann. Henri Poincar\'e. \textbf{10}, 1027-1073 (2009).	


\bibitem{br08}
Brunetti, R., Ruzzi, G.: Quantum charges and spacetime topology: The emergence of new superselection sectors. Commun. Math. Phys. \textbf{287},  523-563(2009). 
 


\bibitem{bmt88}
Buchholz, D., Mack, G., Todorov, I.: The current algebra on the circle as a germ of local field theories. Nucl. Phys. B. (Proc. Suppl.) \textbf{5B}, 20-56 (1988).

\bibitem{bdlr92}
Buchholz, D., Doplicher, S., Longo, R., Roberts, J.E.: A new look at Goldstone's theorem. Rev. Math. Phys. \textbf{Special Issue}, 49-83 (1992).


\bibitem{ckl08}
Carpi, S., Kawahigashi, Y., Longo, R.: Structure and classification of superconformal nets. 
Ann. Henri Poincar\'e. \textbf{9}, 1069-1121 (2008).




\bibitem{fabio3} 
Ciolli, F.: Massless scalar free Field in 1+1 dimensions I: Weyl algebras Extensions and Superselection Sectors.  Rev. Math. Phys. \textbf{21}, 735-780 (2009). ArXiv math-ph/0511064.

\bibitem{fabio6}
Ciolli, F.: The St\"uckelberg-Kibble $QED_2$ model in a net cohomology approach. (In preparation).

\bibitem{fabio7}
Ciolli, F.: Net Cohomology for conformal nets of lattices (In preparation).

\bibitem{cr} 
Ciolli, F., Ruzzi, G.: (In preparation).

\bibitem{cc05}
Carpi, S., Conti, R.: Classification of subsystems for graded-local nets with trivial superselection structure. Commun. Math. Phys. \textbf{253}, 423-449 (2005).

\bibitem{cdr01}
Conti, R., Doplicher, S., Roberts, J.E.: Superselection theory for Subsystems. Commun. Math. Phys. \textbf{218}, 263-281 (2001).


\bibitem{dan90}
D'Antoni, C.: Technical properties of the quasi-local algebra. In: \cite{Kas}.




\bibitem{dhr69I}
Doplicher, S., Haag, R., Roberts, J. E.: Fields, observables and gauge transformation I. Commun. Math. Phys. \textbf{13}, 1-23 (1969).

\bibitem{dhr69II}
Doplicher, S., Haag, R., Roberts, J. E.: Fields, observables and gauge transformation II. Commun. Math. Phys. \textbf{15}, 173-200 (1969).

\bibitem{dhr71}
Doplicher, S., Haag, R., Roberts, J. E.: Local observables and particle statistics I. Commun. Math. Phys. \textbf{23}, 199-230 (1971).

\bibitem{dl}
Doplicher, S. (Editor), Longo, R. (Editor): \emph{Noncommutative Geometry.} Martina Franca, Italy 2000. Lecture Notes in Mathematics \textbf{1831}. Springer-Verlag, 2004.

\bibitem{dl84}
Doplicher, S., Longo, R.: Standard and split inclusions of von Neumann algebras, Invent. Math. \textbf{73}, 493-536 (1984).

\bibitem{dr89}
Doplicher, S., Roberts, J.E.: A new duality theory for compact groups. Invent. Math. \textbf{98}, 157-218 (1989).


\bibitem{dri79}
Driessler, W.: Duality and Absence of Locally Generated Superselection Sectors for CCR-Type Algebras. Commun. Math. Phys.  \textbf{70},  213--220 (1979).


\bibitem{frs89}
Fredenhagen, K., Rehren, K.-H., Schroer, B.: Superselection sectors  with braid group statistics and exchange algebras I. General theory.  Commun. Math. Phys. \textbf{125}, 201-226 (1989).









\bibitem{gru06}
Grundling, H.: Quantum constraints. Rep. Math. Phys. \textbf{57}, 97-120 (2006). 

\bibitem{gl00}
Grundling, H., Lled\'o, F.: Local quantum constraints. Rev. Math. Phys. \textbf{12}, 1159-1218 (2000). 

\bibitem{gl}
Guido, D., Longo, R.: Relativistic invariance and charge conjugation in quantum field theory. Commun.
Math. Phys. \textbf{148}, 521–551 (1992).

\bibitem{glrv}
Guido, D., Longo, R., Roberts, J. E., Verch, R.: Charged sectors, spin and statistics in quantum field theory on curved spacetimes.  Rev. Math. Phys.  \textbf{13}, 125-198 (2001).






\bibitem{hl82}
Hislop, P. D., Longo, R.: Modular Structure of the Local Algebras Associated with the Free Massless Scalar Field Theory. Commun. Math. Phys. \textbf{84}, 71-85 (1982).

\bibitem{ik02}
Izumi, M., Kosaki, H.: On a subfactor analogue of the second cohomology, Rev. Math. Phys. \textbf{14}, 733-757 (2002). 


\bibitem{klm01}
Kawahigashi, Y., Longo, R., M\"uger, M.: Multi-interval Subfactors and Modularity of Representations in Conformal Field Theory. Commun. Math. Phys. \textbf{219}, 631-669 (2001).

\bibitem{kl04}
Kawahigashi, Y., Longo, R.: Classification of Two-dimensional Local Conformal Nets with $c<1$ and 2-cohomology Vanishing for Tensor Categories. Commun. Math. Phys. \textbf{244}, 63-97 (2004).



\bibitem{Kas}
Kastler, D. (ed.): \emph{The algebraic theory of superselection sectors: introduction and recent results}. World Scientific, 1990.



\bibitem{lrt}
Leyland, P., Roberts, J.E., Testard, D.: Duality for Quantum Free Fields. Preprint 78/o.1016, C.N.R.S. Marseille 1978.


\bibitem{lon}
Longo, R. (ed.): \emph{Mathematical Physics in Mathematics and Physics. Quantum and Operator Algebraic Aspects}. Fields Institute Communications, Volume \textbf{20}. American Mathematical Society, 2001.


\bibitem{lx04}
Longo, R., Xu, F.: Topological sectors and a dichotomy in conformal field theory. Comm. Math. Phys. \textbf{251} (2004), 321-364 . 









\bibitem{mug00}
M\"uger, M.: Galois theory for braided tensor categories and the modular closure. Adv. Math. \textbf{150}, 151–201 (2000).

\bibitem{mug01}
M\"uger, M.: Conformal Field Theory and Doplicher-Roberts Reconstruction. In: \cite{lon}. 

\bibitem{mug04}
M\"uger, M.: Galois extensions of braided tensor categories and braided crossed G-categories. J. Alg. \textbf{277}, 256–281 (2004).

\bibitem{mug04a}
M\"uger, M.: On the center of a compact group. Int. Math. Res. Not. \textbf{51}, 2751-2756 (2004). 

\bibitem{mug05}
M\"uger, M.: Conformal Orbifold Theories and Braided Crossed G-Categories. Commun. Math. Phys. \textbf{260}, 727-762 (2005).



\bibitem{nt}
Nakagami, Y., Takesaki, M.: \emph{Duality for Crossed Products of von Neumann Algebras.} Lecture Notes in Mathematics \textbf{731}. Springer-Verlag, 1979.
 





\bibitem{rob76}
Roberts, J. E.: Local cohomology and superselection rules. Comm. Math. Phys. \textbf{51}, 107-119 (1976).

\bibitem{rob77}
Roberts, J. E.: Mathematical Aspects of Local Cohomology. Proceedings of the Colloquium on Operator Algebras and their Applications to Mathematical Physics, 321-332, CNRS, Marseille, 1977.

\bibitem{rob90}
Roberts, J.E.: Lectures on algebraic quantum field theory. In: \cite{Kas}.


\bibitem{rob04}
Roberts, J.E.: More lecture on Algebraic Quantum Field Theory. In \cite{dl}. 

\bibitem{rr06}
Roberts, J. E., Ruzzi, G.: A cohomological description of connections and curvature over posets. Theory Appl. Categ. \textbf{16}, 855-895 (2006).

\bibitem{ruz05}
Ruzzi, G.: Homotopy of posets, net-cohomology and superselection sectors in globally hyperbolic space-times. Rev. Math. Phys. \textbf{17}, 1021-1070 (2005).

\bibitem{sau80}
Sauvageot, J. L.: Produits tensoriels de $\cZ$-modules. Publ. Univ. P. \& M. Curie \textbf{23}, 468-485 (1980).


\bibitem{sw70}
Streater, R. F., Wilde, I. F.: Fermion states of a Boson field. Nucl. Phys. \textbf{B24}, 561-575 (1970).

\bibitem{Str}
Stratila, S.: \emph{Modular theory in operator algebras}. Abacus Press, 1981.



\bibitem{vas07}
Vasselli, E.: Some remarks on group bundles and $C^\ast$-dynamical systems. Comm. Math. Phys. \textbf{274}, 253-276 (2007). 



\end{thebibliography}
\end{document}